\documentclass[a4paper,11pt]{article}
\usepackage{pos}
\usepackage{amsmath}
\DeclareMathOperator{\sinc}{sinc}
\usepackage[dvipsnames,svgnames]{xcolor}
\usepackage{hyperref}
\usepackage[%
        capitalise
        ]{cleveref}
\crefmultiformat{figure}{Figs.~#2#1\xdef\mycreffirstarg{#1}#3}%
 { and~#2{\crefstripprefix{\mycreffirstarg}{#1}}#3}{, #2#1#3}{ and~#2#1#3}
\PassOptionsToPackage{demo}{graphicx}
\hypersetup{%
        colorlinks=true,%
        linktocpage=true,%
        pdfstartpage=1,%
        pdfstartview=FitV,%
        breaklinks=true,%
        pdfpagemode=UseNone,%
        pageanchor=true,%
        pdfpagemode=UseOutlines,%
        plainpages=false,%
        bookmarksnumbered,%
        bookmarksopen=true,%
        bookmarksopenlevel=1,%
        hypertexnames=true,%
        pdfhighlight=/O,%
        urlcolor=Cerulean,%
        linkcolor=Cerulean,%
        citecolor=Cerulean,%
}

\title{Training School Lecture Notes: Probing ultralight axion-like particles with quantum technology}
\ShortTitle{Probing ultralight ALPs with quantum technology}

\author*{Sreemanti Chakraborti}

\affiliation{Institute for Particle Physics Phenomenology,\\ Department of Physics,\\ Durham University,\\ Durham, UK}

\emailAdd{sreemanti.chakraborti@durham.ac.uk}

\abstract{We review the physics of ultralight axion-like particles (ALPs) as dark matter candidates and the experimental strategies used to search for them with precision and quantum technologies. In the ultralight regime, the enormous occupation number of the dark matter field motivates a classical description in terms of a coherently oscillating background, leading to distinctive, time-dependent signatures in laboratory observables. We discuss the effective
field theory framework governing ALP interactions with Standard Model fields, and show how different operators give rise to qualitatively different experimental signals. The lecture notes cover both conversion-based searches enabled by the
axion-photon coupling, such as haloscopes and helioscopes, and precision experiments sensitive to oscillations of fundamental constants and material properties. These include atomic and nuclear clocks, optical cavities, laser and unequal time-delay interferometers, and mechanical or solid state resonators. Emphasis is placed on the physical origin of the sensitivity of each platform,
the role of coherence, bandwidth, and noise, and the complementarity between different technologies across a wide range of ALP masses. Together, these approaches provide broad and overlapping coverage of ultralight dark matter parameter space and define a rapidly evolving experimental programme with strong discovery potential.}

\FullConference{3rd Training School of the COST Action COSMIC WISPers (CA21106) (COSMICWISPers2025)\\
16–19 Sept 2025\\ Annecy, France\\}

\tableofcontents

\begin{document}
\maketitle
\tableofcontents

\section{Introduction}

These lecture notes provide an introduction to the physics of ultralight axion-like particles (ALPs) and to the modern experimental strategies used to search for them using precision and quantum technologies. The central theme is that, for sufficiently small masses, dark matter behaves not as a collection of individual particles but as a coherently oscillating classical field. This field-like nature leads to characteristic, time-dependent signatures that are best accessed through high-precision measurements of frequencies, phases, and length scales.

The notes begin by motivating ultralight dark matter from both observational and theoretical perspectives, emphasising the broad range of possible masses and interactions encompassed by ALP models. In the ultralight regime, the enormous occupation number of the dark matter field justifies a classical description, with a well-defined oscillation frequency set by the ALP mass and a finite coherence time determined by the galactic velocity distribution. This framework
provides the basis for understanding how ALPs can induce coherent, oscillatory signals in laboratory experiments.

A first major focus of the notes is on dimension--5 ALP interactions, in particular the axion-photon coupling, which enables conversion-based searches. The working principles and scaling laws of haloscopes and helioscopes are discussed, highlighting how resonant enhancement, magnetic fields, coherence, and noise considerations determine experimental sensitivity. These examples illustrate how traditional particle physics couplings manifest as measurable electromagnetic signals when probed with carefully designed apparatus.

The notes then broaden the scope to precision experiments sensitive to oscillations of fundamental constants and material properties induced by ultralight fields. This includes atomic and nuclear clocks, optical cavities, laser interferometers, unequal time-delay interferometers, and mechanical or solid-state resonators. Particular emphasis is placed on how different platforms probe different frequency (mass) ranges, why certain technologies dominate at low or high masses, and how comparisons between clocks, cavities, and interferometers allow one to disentangle genuine new physics signals from systematic effects.

Overall, the aim of these lecture notes is to provide a coherent, model-aware overview of the ultralight axion and ALP search programme, bridging effective field theory descriptions of ALP interactions with the concrete observables of precision experiments. By highlighting both the complementarity and the shared physical principles underlying diverse experimental approaches, the notes are intended to equip the reader with the conceptual tools needed to understand, evaluate, and contribute to ongoing and future searches for ultralight dark matter.

The lecture notes are organised as follows. In \cref{sec:dm}, we briefly revisit the dark matter problem from a particle physics perspective, summarising the current state of the field. In \cref{sec:uldm}, we introduce the theoretical and observational motivation for ultralight dark matter, with particular emphasis on axions and ALPs, their role as dark matter candidates, and the classical field description appropriate in the ultralight regime. We also present the EFT framework for ALPs, detailing their dimension--5 and higher dimensional interactions with Standard Model fields and the resulting phenomenology relevant for laboratory searches. \cref{sec:probe-dim5} focuses on conversion-based searches enabled by the axion–photon coupling, including the operating principles and sensitivity scaling of haloscopes and helioscopes. In \cref{sec:probe-dim6}, we shift to precision experiments probing scalar-type effects of ultralight ALPs, covering atomic and nuclear clocks, optical cavities, interferometers, and mechanical resonators, and explaining how oscillations of fundamental constants and material-dependent observables arise and are measured. Finally, in \cref{sec:sum}, we summarise the current experimental landscape and projected future reach of laboratory searches, highlighting the strong complementarity between different technologies across many orders of magnitude in ALP mass. 

Throughout the notes, the emphasis is placed on connecting effective theoretical descriptions to concrete experimental observables, with the aim of equipping the reader to understand and critically assess ongoing and future searches for ultralight ALPs.

\section{Dark Matter Basics from a Particle Physics Perspective}
\label{sec:dm}

\subsection{The Observational Evidence}

A wide range of astrophysical and cosmological observations provide compelling evidence for the existence of \emph{non-luminous matter} that interacts predominantly through gravity~\cite{Zwicky:1933gu,Zwicky:1937zza}. Importantly, all currently established evidence for dark matter is \emph{gravitational only}; no confirmed non-gravitational interaction with Standard Model particles has yet been observed. The evidence for dark matter arises consistently across very different physical scales:

\paragraph{Galactic rotation curves:} Observations of the rotational velocity of stars and gas in spiral galaxies reveal a striking discrepancy with the rotation curves predicted by standard gravitational dynamics applied to the observed luminous mass distribution. In galaxies such as NGC~7541~\cite{Rubin:1978kmz}, the observed circular velocity remains approximately constant at large radii, instead of falling off as $v(r) \propto \frac{1}{\sqrt{r}}$, as would be expected if the mass distribution followed the luminous disk and gas components. This flat rotation curve implies the presence of an extended, approximately spherical dark matter halo whose mass dominates at large radii.

\paragraph{Galaxy clusters and gravitational lensing:} In galaxy clusters, the gravitational potential inferred from galaxy dynamics and from weak and strong gravitational lensing exceeds that expected from luminous matter alone~\cite{Bartelmann:2010fz,Hoekstra:2013via}. In merging clusters, such as the Bullet Cluster~\cite{Clowe:2006eq}, multi-wavelength observations show that the dominant mass component is spatially offset from the baryonic gas, providing strong evidence for a collisionless dark component.

\paragraph{Cosmic Microwave Background (CMB):}
Precision measurements of temperature anisotropies in the cosmic microwave background, as quantified by the angular power spectrum, require a substantial dark matter component to reproduce the observed peak structure~\cite{Sunyaev:1970eu,Gelmini:2015zpa,PhysRevLett.49.1110}. The CMB data are well described by a cosmological model containing cold dark matter (CDM), and cannot be explained by baryonic matter alone.

The present-day energy composition of the Universe is inferred to be approximately $68.5\%$ dark energy, $26.6\%$ dark matter, and $4.9\%$ ordinary (baryonic) matter~\cite{Bennett_2013,Planck:2018vyg}. This highlights that dark matter constitutes the dominant form of matter in the Universe.

In the Milky Way, dark matter is believed to form an approximately spherical halo enveloping the visible galactic disk and bulge. The Sun is located within this halo, and therefore resides in a local dark matter environment. The key parameters relevant for laboratory and astrophysical searches are the local energy density and the dark matter velocity in the solar neighbourhood ~\cite{Bovy:2012tw,Bertone:2004pz,Kamionkowski_2010}:
\begin{align}
\rho_{\mathrm{DM,local}} &\simeq 0.4~\mathrm{GeV/cm^3}, \\
v_{\mathrm{DM}} &\sim 300~\mathrm{km/s}.
\end{align}

From the strong gravitational evidence accumulated so far, one can infer several robust and essentially model-independent properties of dark matter~\cite{Kolb:1990vq}. First, dark matter has a well-measured cosmic abundance, as established by cosmological observations, and a local abundance in the Milky Way halo that can be inferred from galactic dynamics. Second, on cosmological scales, it behaves as an approximately pressureless component, consistent with the success of the cold dark matter paradigm in describing large-scale structure and the cosmic microwave background. Third, whatever constitutes dark matter must be sufficiently long-lived to survive over the age of the Universe, since it continues to dominate the matter density today. Finally, dark matter must be electrically neutral, or at least extremely weakly charged, because even a small electric charge would typically lead to strong electromagnetic interactions that are
tightly constrained by astrophysical and cosmological observations.

Despite these robust inferences, the fundamental nature of dark matter remains unknown. In particular, we do not yet know what the underlying particle (or field) identity of dark matter is, nor do we know its spin or its mass. It is also unknown whether dark matter possesses non-gravitational interactions with Standard Model particles, and if so, what the relevant coupling structures and strengths are. On the cosmological side, it remains unclear whether dark matter was ever in thermal equilibrium with the Standard Model plasma, i.e., whether it is a thermal relic or whether it was produced through a non-thermal mechanism. More broadly, we do not know when dark matter was generated in the early Universe, nor the detailed physical process by which its present-day abundance
was established.

These open questions motivate a broad experimental programme to explore dark matter across a wide range of masses and interaction strengths. This encompasses laboratory-based experiments spanning over diverse energy scales, space-based searches for astrophysical phenomena that may signal interactions between dark matter and Standard Model particles, and cosmological probes in which dark matter interactions leave observable imprints, such as modifications to the light-element abundances produced during the Big Bang Nucleosynthesis~\cite{RevModPhys.88.015004}.

\subsection{Dark Matter across Different Mass Scales }

DM candidates span an extraordinary range of masses, extending over more than \(\sim 50\) orders of magnitude~\cite{Baryakhtar:2022hbu,Drlica-Wagner:2022lbd}. This wide range reflects the fact that gravitational evidence alone does not fix the microscopic nature of DM, allowing for very different theoretical realisations with distinct production mechanisms and experimental search strategies.
\begin{figure}[!ht]
    \centering
    \includegraphics[width=0.9\linewidth]{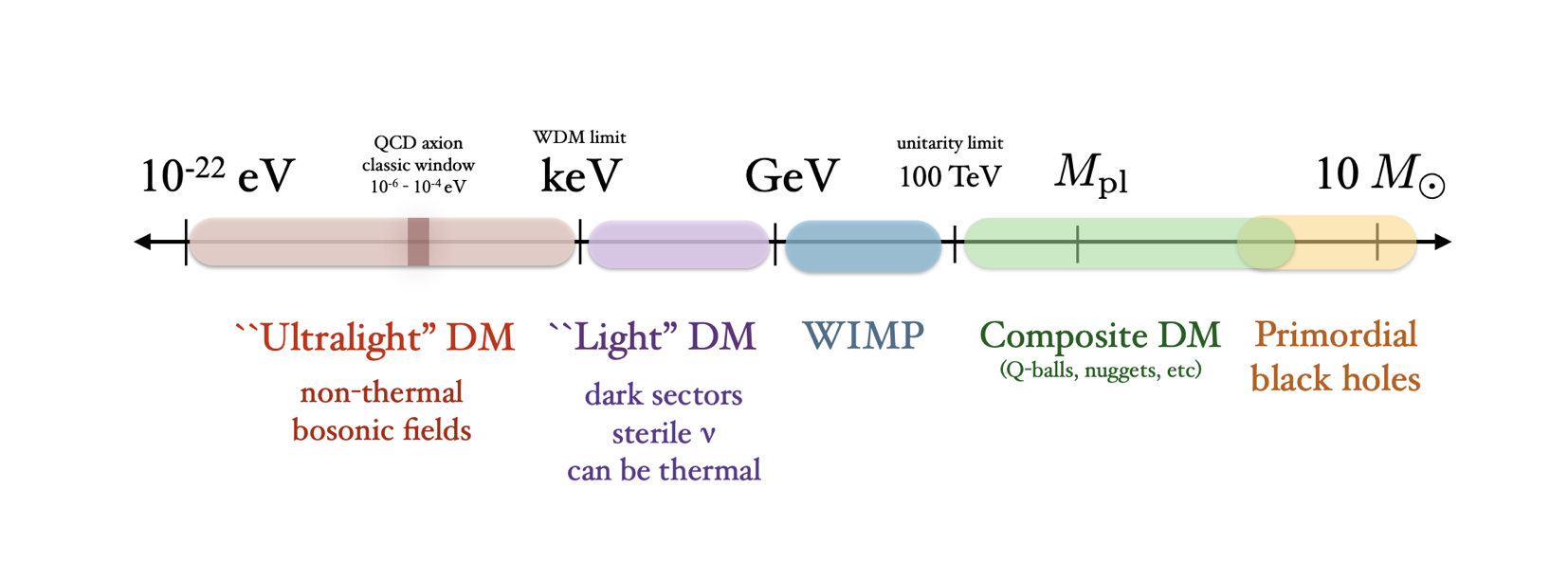}
    \caption{A schematic overview of the possible mass ranges of dark matter candidates considered in the literature, illustrating the broad landscape from ultralight fields to macroscopic objects.}
\end{figure}

\paragraph{Ultralight dark matter:} At the lowest end of the mass spectrum, one finds \emph{ultralight dark matter} (ULDM)~\cite{Hui:2016ltb} with masses as small as $m_{\rm DM} \sim 10^{-22}\ \mathrm{eV}$. Candidates in this regime are typically \emph{bosonic fields} rather than individual particles. Because of their extremely small mass, their occupation number in galactic halos is enormous, and DM behaves as a coherent, classical field rather than a collection of particles. Such candidates are generally produced through \emph{non-thermal mechanisms}, such as vacuum misalignment~\cite{PhysRevD.28.1243}. 

A particularly well-motivated example in this category is the QCD axion~\cite{Preskill:1982cy,Abbott:1982af,Dine:1982ah}, whose ``classic window'' lies approximately in the mass range $m_a \sim 10^{-6} \text{--} 10^{-4}\ \mathrm{eV}$. 

Ultralight dark matter motivates a fundamentally different experimental paradigm based on precision measurements and quantum sensors, which probe coherent, time-dependent effects imprinted on physical observables.

\paragraph{Light dark matter:} Moving to higher masses, the \emph{light dark matter} regime typically spans masses from the keV scale upward. This region includes candidates such as sterile neutrinos~\cite{Dodelson:1993je,Shi:1998km} and dark sector particles. In contrast to ULDM, these candidates may be either thermal or non-thermal in origin, depending on their interactions and cosmological history. The keV scale is of particular significance, as it marks the lower bound for \emph{warm dark matter} inferred from structure formation~\cite{Bode:2000gq,Bond:1983hb} and Lyman-\(\alpha\) forest~\cite{PhysRevD.71.063534} constraints. Below this scale, free-streaming effects ~\cite{Bond:1982uy,Dayal:2023nwi} suppress the formation of small-scale structure.

\paragraph{Weakly Interacting Massive Particles (WIMPs):} At masses around the Electroweak scale (GeV-TeV), one encounters the traditional paradigm of \emph{weakly interacting massive particles} (WIMPs). These candidates are typically thermal relics whose present-day abundance is set by freeze-out~\cite{Gondolo:1990dk} in the early Universe. The WIMP paradigm historically motivated a broad experimental programme including direct, indirect, and collider searches~\cite{Bergstrom:2012fi,Feng:2010gw,Bertone:2004pz}.

An important theoretical upper bound arises from partial-wave unitarity, which limits the mass of a thermal relic to roughly $m_{\rm DM} \lesssim 100\ \mathrm{TeV}$~\cite{PhysRevLett.64.615}. Above this scale, DM cannot be produced as a standard thermal relic without violating unitarity.

\paragraph{Composite dark matter:} Beyond the WIMP scale, DM may take the form of \emph{composite objects}, such as
Q-balls~\cite{Kusenko:2001vu}, nuggets~\cite{Ge:2019voa}, or other bound states arising from new strong dynamics. These candidates are typically macroscopic, with masses far exceeding those of elementary particles, and their phenomenology is governed by their extended structure rather than point-like interactions~\cite{Deliyergiyev:2019vti}.

\paragraph{Primordial black holes:} At the extreme end of the mass spectrum lie \emph{primordial black holes}~\cite{Hawking:1971ei,Chapline:1975ojl}, which may have formed from enhanced density fluctuations in the early Universe rather than from particle physics interactions. Their masses can range from microscopic values up to astrophysical scales, including masses of order $M \sim 10\,M_\odot$, comparable to stellar-mass black holes. In this case, DM is not a particle at all, but a population of compact gravitational objects. Such candidates are probed through gravitational wave (GW) observations of black hole mergers~\cite{Bird:2016dcv,Sasaki:2016jop}, as well as microlensing~\cite{Niikura:2017zjd} and other astrophysical and cosmological constraints.

This mass hierarchy emphasises that DM encompasses a wide range of possible physical realisations, each associated with distinct theoretical frameworks, production mechanisms, and experimental probes. The ultralight regime, in particular, motivates novel searches using precision and quantum technologies, which will be the focus of the following discussion.

\subsection{A Particle Physicist's Problem}

Historically, the dominant paradigm for particle DM has been that of \emph{thermal weakly interacting massive particles} (WIMPs). In this framework, DM is assumed to be a massive, stable particle that was once in thermal equilibrium with the SM plasma in the early Universe. Its present-day abundance is determined by thermal freeze-out, leading to the well-known ``WIMP miracle''~\cite{PhysRevLett.39.165}, whereby weak-scale masses and interaction strengths naturally reproduce the observed DM density. Experimental searches for thermal WIMPs are traditionally divided into three complementary strategies: direct detection, collider searches, and indirect detection.

\paragraph{Direct detection: } Direct detection experiments aim to observe the elastic scattering of a WIMP off an atomic nucleus in a terrestrial detector. A DM particle from the Galactic halo enters the detector and transfers a small amount of kinetic energy to a detector nucleus, producing a detectable recoil signal~\cite{PhysRevD.31.3059}. The expected event rate depends on the local DM density, the DM velocity distribution, the WIMP-nucleus scattering cross section, and the detector material.

These experiments are optimised to detect extremely rare and low-energy nuclear recoils, requiring ultra-low backgrounds and deep underground operation. The main observable is the DM-nucleon cross section as a function of the DM mass~\cite{Akerib:2022ort}.

\paragraph{Collider searches:} Collider searches probe DM production in high-energy particle collisions, such as those at the LHC. In this approach, SM particles collide and produce DM particles in the final state. Since DM does not interact electromagnetically, it escapes the detector, leading to missing transverse energy (MET) signatures.

The typical collider signal involves visible SM particles emitted with missing energy, for example, in mono-jet, mono-photon, or mono-\(Z\) channels~\cite{Abercrombie:2015wmb,Boveia:2022syt}. Collider searches are particularly sensitive to electroweak-scale DM candidates with sizable couplings to SM particles and probe the underlying particle nature and interaction structure rather than the astrophysical properties of DM.

\paragraph{Indirect detection:} Indirect detection experiments search for the products of DM annihilation or decay in astrophysical environments. In regions of high DM density, such as the Galactic centre or dwarf spheroidal galaxies, WIMPs may annihilate into SM particles. These processes subsequently produce observable gamma rays, as well as cosmic rays, X-rays, and radio emission arising from electrons, positrons, neutrinos, protons, and antiprotons~~\cite{Cirelli:2024ssz,Ando:2022kzd}.

The characteristic indirect detection signature is an excess in one or more of these channels above known astrophysical backgrounds, with an energy spectrum and spatial distribution consistent with DM annihilation or decay.

Despite decades of experimental effort and substantial improvements in sensitivity, no unambiguous signal of thermal WIMPs has yet been observed. The absence of a discovery, together with the approach of direct detection experiments to the neutrino background limit~\cite{PhysRevLett.134.111802}, has motivated growing interest in alternative DM candidates and search strategies. In particular, these developments have driven a shift toward exploring light MeV-scale DM, non-thermal production scenarios, and ultralight dark matter using precision and quantum sensing techniques, which probe observables fundamentally different from those targeted by traditional WIMP searches.

\section{Wave-like Dark Matter}
\label{sec:uldm}

A particularly important class of DM candidates arises when DM consists of \emph{spin-0 bosonic fields} with extremely small masses. In this case, DM exhibits intrinsically wave-like behaviour on astrophysical scales.

\subsection{Features}

Wave-like or ultralight DM~\cite{Eberhardt:2025caq} corresponds to spin-0 particles in the approximate mass range
\[
10^{-22}\ \mathrm{eV} \;\lesssim\; m_\phi \;\lesssim\; 1\ \mathrm{eV}.
\]
At such small masses, the de Broglie wavelength of the DM particle becomes macroscopic. In particular, for typical galactic halo velocities, the associated wavelength can be comparable to astrophysical length scales. The maximum coherence length is effectively bounded by the size of the DM halo, with a characteristic scale  $\lambda \sim \mathcal{O}(\mathrm{kpc})$, set by the halo extent.

When light bosons constitute most or all of DM, their number density in galactic halos is extremely large. As a result, the occupation number within a coherence volume satisfies
\[
n_\phi \left(\frac{\lambda_{\rm coh}}{2\pi}\right)^3 \gg 1,
\]
 where $\lambda_{\rm coh}$ and $n_\phi$ denote, respectively, the coherence length and number density of the DM field. This regime justifies a \emph{classical field description} of DM. Rather than treating DM as an ensemble of individual particles, it is more appropriate to model it as a coherently oscillating classical field whose dynamics are governed by a classical wave equation. Therefore, an ultralight DM field can be written as~\cite{Hui:2016ltb}
\begin{align}
\phi(\vec{x},t) \;\approx\;
\phi_0 \cos\bigl(m_\phi (t + \vec{\beta}\cdot\vec{x})\bigr)
\label{eq:ULDM}
\end{align}
where
\begin{enumerate}
    \item The DM amplitude is determined by requiring that the field reproduces the observed local DM density $\rho_{\phi}$, which for a coherently oscillating scalar field is given by
\[
\rho_\phi \;=\; \frac{1}{2} m_\phi^2 \phi_0^2,
\]
which implies
\[
\phi_0 \;=\; \frac{\sqrt{2\rho_\phi}}{m_\phi}
\]
Thus, lighter ULDM fields correspond to larger field amplitudes.
\item The temporal oscillation of the field is primarily determined by the rest mass of the DM particle. To leading order, the angular frequency of the oscillation is
\[
\omega \simeq m_\phi.
\]
This reflects the fact that the field oscillates at its Compton frequency.
\item  The vector $\vec{\beta} \equiv \frac{\vec{v}_\phi}{c}$ in \cref{eq:ULDM} encodes the DM velocity in units of the speed of light. For virialised DM in the Milky Way halo, $|\vec{\beta}| \sim 10^{-3}$. The spatial dependence of the field, \(\vec{\beta}\cdot\vec{x}\), therefore introduces a position-dependent phase. Since the DM velocity distribution is not monochromatic, this spatial term effectively appears as a \emph{random phase} over laboratory length scales.

\end{enumerate}

DM particles in the Galactic halo are not exactly at rest, but possess a small kinetic energy associated with their virialised motion. This induces a small spread in the oscillation frequency of the field. The relative size of
this effect is
\[
\frac{\Delta \omega}{\omega}
\;\sim\;
\frac{\langle v_\phi^2\rangle}{c^2}
\;\sim\;
10^{-6}
\]
where \(\langle v_\phi^2\rangle\) denotes the velocity dispersion of DM in the halo. These corrections are small but play an important role in setting the coherence properties of the field.

Lastly, the finite frequency spread implies that the oscillations remain phase coherent only over a finite time interval. The coherence time is set by the inverse of the frequency spread~\cite{PhysRevD.104.055037},
\[
\tau_{\mathrm{coh}} \sim \frac{2\pi}{\Delta \omega}.
\]
Using the estimate above for \(\Delta \omega\), this can be written as
\[
\tau_{\mathrm{coh}} \sim 10^{6}\, T_{\mathrm{osc}},
\]
where \(T_{\mathrm{osc}} = 2\pi/\omega\) is the oscillation period of the field. Thus, the ULDM field oscillates coherently for roughly a million oscillation cycles.

The combination of a well-defined oscillation frequency, a large field amplitude, and a long coherence time is a defining feature of ultralight dark matter. These properties make ULDM particularly amenable to searches using precision measurements and quantum technologies, which are sensitive to coherent, time-dependent signals.

\subsection{Production in the Early Universe}

ULDM candidates can be produced in the early Universe through the \emph{misalignment mechanism}~\cite{PhysRevD.28.1243}, a non-thermal production process that is generic for ultralight bosonic fields and does not require thermal equilibrium with the Standard Model plasma.

In an expanding Friedmann-Robertson-Walker (FRW) Universe, a homogeneous classical scalar field \(\phi(t)\) evolves according to the Klein--Gordon equation with Hubble friction,
\[
\ddot{\phi} + 3H(t)\dot{\phi} + m_\phi^2 \phi = 0,
\]
where \(H(t)\) is the Hubble expansion rate and \(m_\phi\) is the mass of the ultralight field. The second term acts as a friction term induced by cosmic expansion.

At very early times, the expansion rate of the Universe is large. When $3H(t) \gg m_\phi$, the equation of motion is overdamped. In this regime, the friction term dominates over the mass term, and the ultralight field cannot roll efficiently. As a result, the field remains approximately constant in time, effectively ``frozen'' at its initial value. This initial displacement from the minimum of the potential is commonly referred to as the \emph{misalignment angle}.

As the Universe expands, the Hubble rate decreases. Oscillations of the field begin when the Hubble friction becomes comparable to the mass term $3H(t_{\mathrm{osc}}) \simeq m_\phi$. At this time, the field starts to roll towards the minimum of its potential and undergoes coherent oscillations about that minimum. This transition marks the onset of dynamical evolution of the ULDM field.

At later times, when $H(t) \ll m_\phi$, the expansion-induced friction becomes negligible over an oscillation period. The field executes rapid oscillations, and its dynamics are well approximated by those of a harmonic oscillator in an expanding background.

In this regime, the energy density stored in the oscillating bosonic field averages to $\rho_\phi \propto a^{-3}$, where \(a(t)\) is the scale factor of the Universe. This scaling is identical to that of pressureless cold dark matter, implying that ULDM produced via the misalignment mechanism behaves as cold dark matter at late times.

The present-day relic abundance of ULDM produced by misalignment depends on two key parameters---(i) the mass of the DM particle \(m_\phi\), which sets the time at which oscillations begin, and (ii) the initial field displacement (misalignment angle), which determines the initial energy stored in the field. Different choices of these parameters lead to different DM abundances, allowing ULDM to constitute either a fraction or the entirety of the observed DM density.

\subsection{Axions and Axion-Like Particles as ULDM Candidates}

\subsubsection{Axions}

Axions arise as a well-motivated extension of the Standard Model and provide a compelling solution to the \emph{strong CP problem}~\cite{Preskill:1982cy,Abbott:1982af, Dine:1982ah} of QCD. In addition, axions are excellent candidates for cold dark matter and naturally fit within the framework of ULDM discussed above.

\paragraph{Axions and the strong CP problem:} The QCD Lagrangian admits a CP-violating term of the form
\[
\mathcal{L}_{\theta} \;=\; \frac{\alpha_s}{8\pi}\,\theta_{\rm QCD}\,
G_{\mu\nu}^a \tilde{G}^{a\,\mu\nu},
\]
where \(G_{\mu\nu}^a\) is the gluon field-strength tensor and \(\tilde{G}^{a\,\mu\nu}\) its dual. Experimental bounds on the neutron electric dipole moment require \(|\theta_{\rm QCD}| \lesssim 10^{-10}\)~\cite{Abel:2020pzs}, posing the strong CP problem.

The Peccei--Quinn (PQ) mechanism promotes \(\theta_{\rm QCD}\) to a dynamical field by introducing a global \(U(1)_{\rm PQ}\) symmetry that is spontaneously broken at an energy scale \(f_a\). The associated pseudo Nambu--Goldstone boson is the axion \(a\). In this framework,
\[
\theta_{\rm QCD} \;\propto\; \frac{a}{f_a},
\]
and the effective Lagrangian contains
\[
\mathcal{L} \supset \frac{\alpha_s}{8\pi}
\left(\theta + \frac{a}{f_a}\right)
G_{\mu\nu}^a \tilde{G}^{a\,\mu\nu}.
\]
Minimisation of the axion potential dynamically drives the vacuum expectation value to
\[
\langle a \rangle = - f_a \theta,
\]
thereby cancelling the CP-violating term and solving the strong CP problem.

\paragraph{Axion mass and interactions:} Non-perturbative QCD effects generate a potential for the axion, giving it a small but non-zero mass. Parametrically, the axion mass is related to the PQ scale as~\cite{Preskill:1982cy,Abbott:1982af, Dine:1982ah}
\begin{align}
m_a \;\propto\; \frac{\Lambda_{\rm QCD}^2}{f_a}
\label{eq:axion}
\end{align}
indicating that lighter axions correspond to higher PQ-breaking scales.

A more precise relation is obtained from chiral perturbation theory~\cite{GrillidiCortona:2015jxo},
\[
m_a f_a \simeq m_\pi f_\pi
\sqrt{\frac{m_u m_d}{(m_u + m_d)^2}},
\]
which leads numerically to
\[
m_a \simeq 5.7~\mu{\rm eV}
\left(\frac{10^{12}~{\rm GeV}}{f_a}\right).
\]
Thus, the axion mass is not a free parameter but is directly tied to its interactions with Standard Model particles. 

The same underlying symmetry structure that fixes the axion mass also strongly constrains the form of its interactions. The axion is a pseudo Nambu-Goldstone boson associated with the spontaneous breaking of the PQ symmetry. As a result, its interactions with fermions are predominantly \emph{derivative} couplings of the form
\[
\mathcal{L} \supset
\frac{\partial_\mu a}{f_a}\,
\bar{\psi}\gamma^\mu\gamma_5 \psi,
\]
where \(\psi\) denotes a SM fermion. These derivative couplings reflect the approximate shift symmetry of the axion field. For practical low-energy calculations, it is often convenient to parametrise their effects in terms of an equivalent pseudoscalar axion–fermion coupling, which scales as~\cite{Bauer:2021mvw,DiLuzio:2020wdo}
\[
g_{af} \sim C_f \frac{m_f}{f_a},
\]
where \(C_f\) is a model-dependent coefficient and \(m_f\) is the fermion mass.

Axions also couple to photons through the electromagnetic anomaly,
\[
\mathcal{L} \supset
- \frac{1}{4} g_{a\gamma}\, a F_{\mu\nu}\tilde{F}^{\mu\nu},
\]
with the coupling given by
\[
g_{a\gamma} =
\frac{\alpha}{2\pi f_a}
\left(\frac{E}{N} - 1.92\right).
\]
where \(E/N\) denotes the ratio of electromagnetic to colour anomalies, which is determined by the UV completion of the theory. In particular, KSVZ~\cite{Shifman:1979if,Kim:1979if} and DFSZ~\cite{Zhitnitsky:1980tq,Dine:1981rt} axion models predict different values of \(E/N\).

Axions produced via the misalignment mechanism naturally behave as cold dark matter~\cite{Chadha-Day:2021szb}. Depending on the values of \(f_a\) and the initial misalignment angle, axions can constitute either a fraction or the entirety of the observed DM abundance. Their small mass, weak couplings, and long lifetime make axions an exceptionally well-motivated and theoretically robust DM candidate.

\subsection{Axion-Like Particles (ALPs)}

Axion-like particles (ALPs) are a broad class of hypothetical light scalar or pseudo-scalar bosons that share some properties with the QCD axion but are not tied to the solution of the strong CP problem. They arise naturally in many extensions of the Standard Model, including string theory and generic high-energy completions with spontaneously broken global symmetries~\cite{Svrcek:2006yi,Choi:2009jt,Arvanitaki:2009fg}.

The defining feature of the QCD axion is its role in solving the strong CP problem by dynamically relaxing the CP-violating QCD parameter \(\theta_{\rm QCD}\). In contrast, ALPs \emph{do not} play this role. Specifically, ALPs are not required to couple to the QCD topological term \(\,G\tilde{G}\) and consequently, they do not dynamically cancel \(\theta_{\rm QCD}\) and do not solve the strong CP problem.

Thus, while the QCD axion is a specific, highly constrained particle, ALPs form a much more general class of light bosons.

For the QCD axion, the mass is fixed by QCD dynamics and is related to the Peccei--Quinn symmetry-breaking scale \(f_a\) through \cref{eq:axion}. This relation does not apply to ALPs. The ALP mass is not generated by QCD effects, and therefore mass is a \emph{free parameter}, independent of the overall coupling scale. This flexibility of parameter space allows ALPs a much broader phenomenology than the QCD axion. This motivates a wide range of experimental searches~\cite{Adams:2022pbo,Irastorza:2018dyq}, including laboratory experiments, astrophysical probes, and precision measurements sensitive to ultralight dark matter fields.

\subsubsection{ALP Effective Interactions}

At low energies, ALP interactions with Standard Model fields are described by dimension--5 operators such as the axion--photon, axion--electron, and axion--nucleon couplings. These constitute the effective Lagrangian~\cite{Bauer:2021mvw}
\begin{align}
\mathcal{L} \supset 
- \frac{g_{a\gamma}}{4} \, a F_{\mu\nu}\tilde F^{\mu\nu}
- g_{ae} \, a \bar e \gamma_5 e
- g_{aN} \, a \bar N \gamma_5 N
\end{align}
where the precise values of these couplings depend on the ultraviolet completion of the theory and on how the approximate Peccei–Quinn-like symmetry is realised.

These effective interactions define the observable signatures through which ALPs can be probed experimentally, as discussed in the following section.

\section{Probing Low-Energy Dimension--5 ALP Interactions}
\label{sec:probe-dim5}

The leading dimension--5 interactions of ALPs enable their conversion into photons in the presence of electromagnetic fields. This property underpins a wide class of experimental searches, which exploit controlled laboratory environments as well as astrophysical sources to induce and detect axion–photon conversion. In particular, magnetic field-based experiments such as haloscopes and helioscopes provide complementary and well-established strategies for probing axions and ALPs across a broad range of masses and couplings.

\subsection{Resonant Cavity Haloscopes}

Cavity haloscopes are experiments designed to detect axion or ALP dark matter through its resonant conversion into photons~\cite{PhysRevD.32.2988,Sikivie:1983ip}. They exploit the fact that galactic DM axions form a coherently oscillating classical field with a well-defined Compton frequency set primarily by the ALP mass,
\begin{align}
\nu_a \simeq \frac{m_a}{2\pi} \left( 1 + \mathcal{O}(v^2) \right),
\end{align}
where \(v \sim 10^{-3}\) is the typical virial velocity of DM in the Milky Way. As a result, ALP-induced signals are extremely narrow in frequency, with a fractional
linewidth \(\Delta\nu/\nu \sim 10^{-6}\). The detection principle relies on the ALP--photon interaction,
\begin{align}
\mathcal{L}_{a\gamma\gamma}
= \frac{g_{a\gamma\gamma}}{4} \, a \, F_{\mu\nu} \tilde{F}^{\mu\nu}
= g_{a\gamma\gamma} \, a \, \vec{E}\cdot\vec{B},
\label{eq:cavityEM}
\end{align}
which allows ALPs to convert into photons in the presence of a static magnetic field.

In a cavity haloscope, a high-\(Q\) microwave resonator is immersed in a strong external magnetic field \(\vec{B}_0\).
The ALP field couples to the virtual photons associated with this magnetic field and acts as a source term for an oscillating electromagnetic field inside the cavity. The conversion power is resonantly enhanced when the axion Compton frequency matches the frequency of one of the cavity’s electromagnetic modes.

\begin{figure}[!ht]
    \centering
    \includegraphics[width=0.5\linewidth]{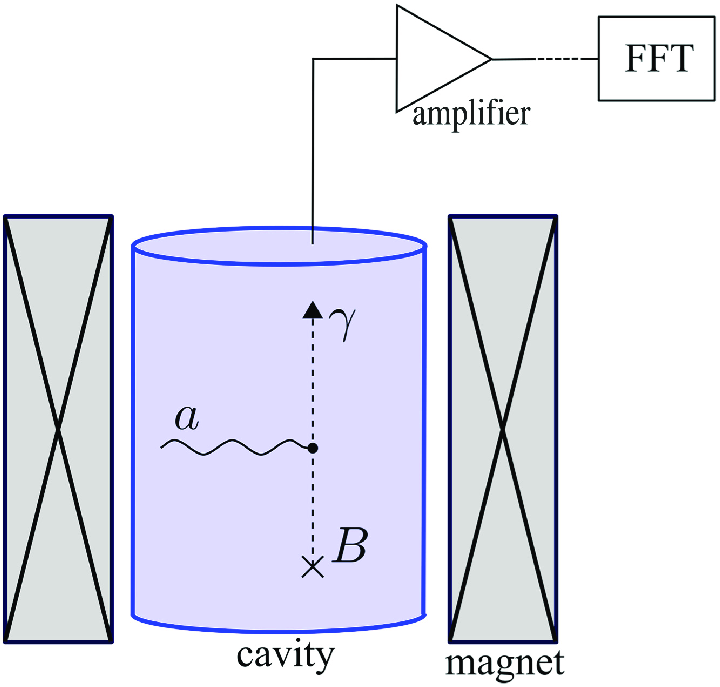}
    \caption{Schematic illustration of a haloscope experiment for axion dark matter detection.}
\end{figure}


This highlights the essential ingredients of a cavity haloscope: strong magnetic fields, large cavity volumes, high quality factors, and low-noise microwave readout.

Haloscopes are optimally suited to the detection of cold dark matter axions and ALPs in the \(\mu\text{eV}\) mass range, corresponding to microwave frequencies. The original proposal of this technique was put forward by Sikivie~\cite{Sikivie:1983ip}, and it forms the basis of a wide experimental programme including ADMX and its successors (see review in Ref.~\cite{Adams:2022pbo}).

\subsubsection{Signal Formation and Readout in Cavity Haloscopes}

In a haloscope, the axion-induced electromagnetic signal appears as a narrowband excess of power at a frequency set by the axion/ALP mass. Resonant conversion occurs when the axion Compton frequency lies within the bandwidth of a microwave cavity mode,
\begin{align}
|\nu_a - \nu_{\rm cav}| \lesssim \frac{\nu_{\rm cav}}{Q_L},
\end{align}
where \(Q_L\) is the loaded quality factor of the cavity and $\nu_{\rm cav}$ is the cavity resonance frequency.

Because the ALP mass is unknown, the cavity frequency must be tunable. Experiments therefore, scan over frequency by mechanically adjusting the cavity geometry (e.g.\ with tuning rods), stepping through narrow frequency intervals and integrating at each setting. The narrow bandwidth of the ALP signal implies that sensitivity improves only slowly with integration time, while the total scan rate is constrained by the achievable noise temperature and cavity bandwidth.

Axion-to-photon conversion inside the cavity generates excess electromagnetic power in the resonant mode. The signal is extracted via a weakly coupled antenna and passed through a low-noise amplification chain. Typically, the microwave signal is first amplified using near-quantum-limited amplifiers, then mixed with a local oscillator to shift it to an intermediate frequency, and finally digitised. 

The presence of an axion signal is searched for in the frequency domain by performing a Fourier transform of the time-domain voltage data, where it appears as a narrow spectral line above the noise background.

The expected signal power on resonance can be written schematically as~\cite{ADMX:2018gho}
\begin{equation}
P_{a \to \gamma}
= g_{a\gamma\gamma}^2
\frac{\rho_{\rm DM}}{m_a}
B_0^2 \, V \, C \,
\min(Q_L, Q_a),
\label{eq:haloscope}
\end{equation}
where \(\rho_{\rm DM}\) is the local dark matter density, \(B_0\) the external magnetic field strength, \(V\) the cavity volume, and \(C\) a dimensionless form factor encoding the overlap between the cavity mode electric field and the external magnetic field. The factor \(\min(Q_L, Q_a)\) reflects the fact that the enhancement is limited either by the cavity bandwidth or by the intrinsic coherence time of the axion field, characterised by an effective axion quality factor \(Q_a \sim 10^6\).

This expression highlights several key features of cavity haloscopes. First, sensitivity improves for larger magnetic fields and cavity volumes. Second, high-\(Q\) cavities are essential, but gains saturate once the cavity bandwidth becomes narrower than the axion signal itself. Finally, because the signal power is extremely small, achieving low system noise temperatures and long integration times is critical for resolving the axion-induced spectral excess.

\subsubsection{Key Scaling Relations in Cavity Haloscopes}

The signal power in the haloscope as described in \cref{eq:haloscope} highlights several key scaling behaviours that guide the design of haloscope experiments. First, the signal power scales quadratically with the magnetic field strength. As a result, stronger magnets directly translate into larger conversion rates, making
high-field superconducting magnets a central component of haloscope sensitivity.

Second, the signal power scales linearly with the cavity volume. Since the resonant frequency of a cavity mode is inversely related to its physical size, larger cavities naturally probe lower axion masses. This creates a strong correlation between accessible mass range and achievable volume, with low-mass searches favouring large-scale cavities and higher-mass searches requiring progressively smaller structures.

Third, the quality factor \(Q\) controls the resonant enhancement of the signal. A higher \(Q\) corresponds to a sharper resonance and a larger stored electromagnetic energy for a given input power. In practice, \(Q\) is limited by material losses, surface resistance, and coupling to the readout chain. Moreover, once \(Q\) exceeds the effective ALP quality factor \(Q_a \sim 10^6\), further increases do not enhance the signal power, as the axion field itself loses coherence on shorter timescales.

These principles are exemplified by leading haloscope experiments such as Axion Dark Matter eXperiment (ADMX)~\cite{ADMX:2018gho,ADMX:2019uok,PhysRevD.64.092003,ADMX:2021nhd} and HAYSTAC~\cite{HAYSTAC:2020kwv,HAYSTAC:2023cam}, which employ high-field superconducting solenoids and microwave cavities with characteristic dimensions of $\mathcal{O}$\,(0.1-1)~m. Operating at GHz frequencies, corresponding to axion masses in the $\mu$eV range, these experiments achieve resonant enhancement through high-$Q$ cavities coupled to ultra–low noise amplification chains. ADMX has demonstrated sensitivity to QCD axion models in the classic mass window, while HAYSTAC has pioneered the use of quantum-limited amplifiers to extend sensitivity at higher frequencies.

More generally, the scalings in \cref{eq:haloscope} imply an inherent trade-off between cavity volume and operating frequency: large-volume cavities maximise signal power but are limited to low frequencies, whereas searches at higher axion masses require progressively smaller cavities with reduced volume. This tension has motivated the development of alternative strategies, such as cavity arrays, multi-mode readout, and novel resonator concepts, aimed at preserving sensitivity while extending the accessible mass range.

\subsubsection{Probing Higher Masses: Dielectric Haloscopes}

A promising next-generation approach to axion DM detection is the \emph{dielectric haloscope}, exemplified by the MADMAX experiment~\cite{refId0,Li:2021oV}. Unlike cavity haloscopes, which rely on resonant enhancement in a closed microwave cavity, dielectric haloscopes exploit coherent electromagnetic emission from multiple dielectric interfaces placed in a strong external magnetic field.

The basic idea is to stack a series of dielectric disks with different dielectric constants \(\varepsilon\), separated by vacuum gaps, in front of a reflecting mirror and immersed in a static magnetic field \( \vec{B}_e \)~\cite{c749-419q,PhysRevLett.118.091801}.

\begin{figure}[!ht]
    \centering
    \includegraphics[width=1.0\linewidth]{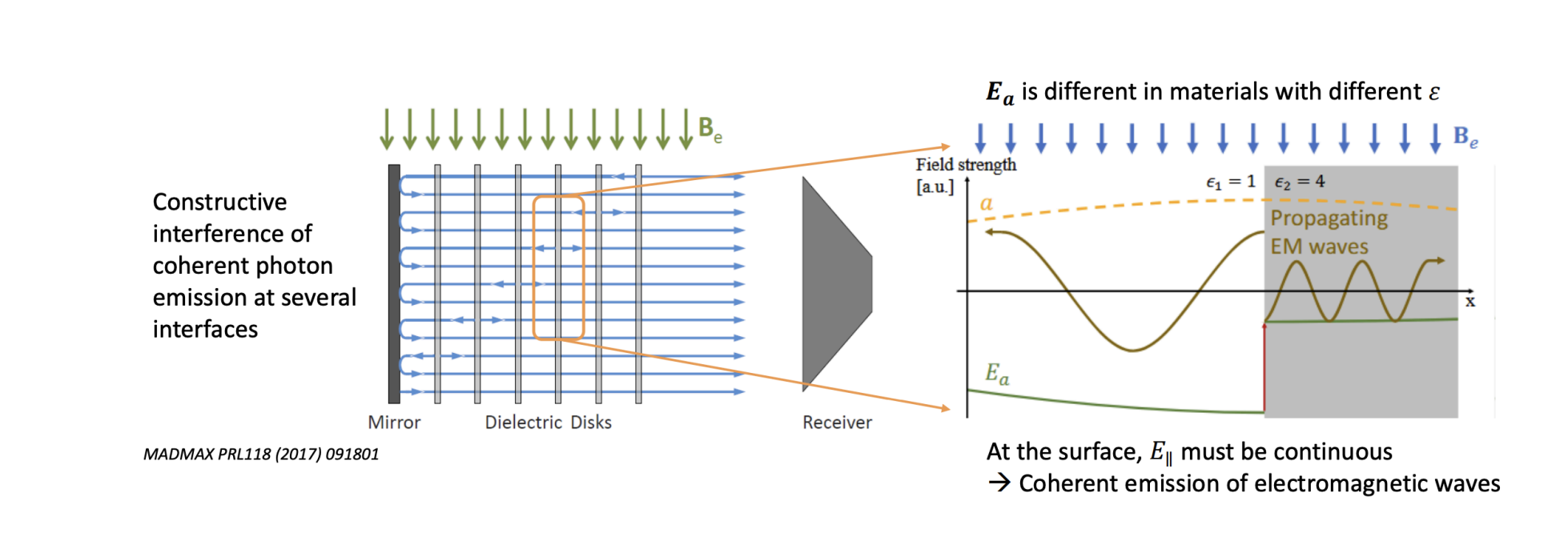}
    \caption{Schematic of the MADMAX dielectric haloscope, illustrating axion-photon conversion and coherent emission from multiple dielectric interfaces in a strong magnetic field. Figure from Ref.~\cite{PhysRevLett.118.091801}.}
\end{figure}

As given in \cref{eq:cavityEM}, the axion field induces an effective oscillating electric field proportional to the external magnetic field. When the axion encounters a boundary between materials with different dielectric constants, this induced field changes discontinuously.

At each dielectric interface, the parallel component of the electric field must remain continuous. As a consequence, the axion-induced electromagnetic response generates outgoing electromagnetic waves at every interface. By carefully choosing the thicknesses of the disks and the separations between them, these emitted waves can be arranged to interfere constructively in a given direction. The result is a coherent enhancement of the emitted signal, analogous to a phased antenna array.

A key advantage of MADMAX  is that the enhancement does not rely on a single narrow cavity resonance. Instead, the response can be engineered over a broader frequency range by adjusting the disk positions, allowing sensitivity to axion masses well above the \(\mu\text{eV}\) scale. In particular, dielectric haloscopes are designed to probe axion masses up to a few orders of magnitude higher than those accessible to conventional large-volume microwave cavities.

From a scaling perspective, the signal power still benefits from a strong magnetic field and large transverse area, but the relevant length scale is no longer set by a resonant cavity mode. Rather, the enhancement arises from the number of dielectric interfaces and their coherent arrangement. This makes dielectric haloscopes especially attractive for higher-frequency (axion mass) searches, where cavity volumes necessarily become small and traditional haloscope sensitivity degrades.

Overall, MADMAX represents a conceptually distinct strategy for axion DM detection, trading resonant quality factor for coherent multi-interface emission. It provides a clear path toward extending haloscope searches into previously inaccessible mass ranges while retaining strong control over signal coherence and background rejection.

\subsubsection{Towards Lower Frequencies: Superconducting Radio-Frequency Cavities}

An alternative next-generation strategy for axion and ALP DM searches is based on \emph{superconducting radio-frequency (SRF) cavities}~\cite{Berlin:2019ahk,Giaccone:2022pke,PhysRevLett.123.021801}. 

The key motivation is that superconducting cavities can achieve extremely high quality factors, typically \(Q \sim 10^{9}\) or higher, far exceeding those of normal conducting microwave cavities. This opens up new detection regimes where sensitivity is driven primarily by ultra-narrow resonances rather than large physical volumes.

The basic principle remains axion-photon conversion via the inverse Primakoff interaction in the presence of an external magnetic field. However, SRF-based proposals exploit resonant frequency conversion between distinct cavity modes. An axion field oscillating at frequency \(\omega_a \simeq m_a\) can mediate energy transfer between two electromagnetic modes whose frequency separation matches the axion mass.

In this sense, the axion acts as a weak parametric drive that couples otherwise orthogonal cavity modes. Two limiting regimes are commonly discussed. In the first, the signal frequency is directly set by the axion mass, \(\omega_{\rm sig} \sim m_a\), and the signal bandwidth is controlled by the axion coherence time, corresponding to a fractional linewidth \(\Delta \omega / \omega \sim 10^{-6}\). In this case, the effective signal enhancement is governed by the minimum of the axion quality factor \(Q_a\) and the cavity quality factor \(Q\).

In the second regime, the axion induces transitions between two cavity modes with frequencies \(\omega_0\) and \(\omega_1\), such that \(\omega_1 - \omega_0 \simeq m_a\). Here, the signal appears at a much higher frequency than the axion Compton frequency itself, while still retaining sensitivity to the axion mass through the mode spacing. This frequency up-conversion allows access to extremely small axion masses using cavities operating at radio or microwave frequencies.

\begin{figure}[!ht]
    \centering
    \includegraphics[width=0.9\linewidth]{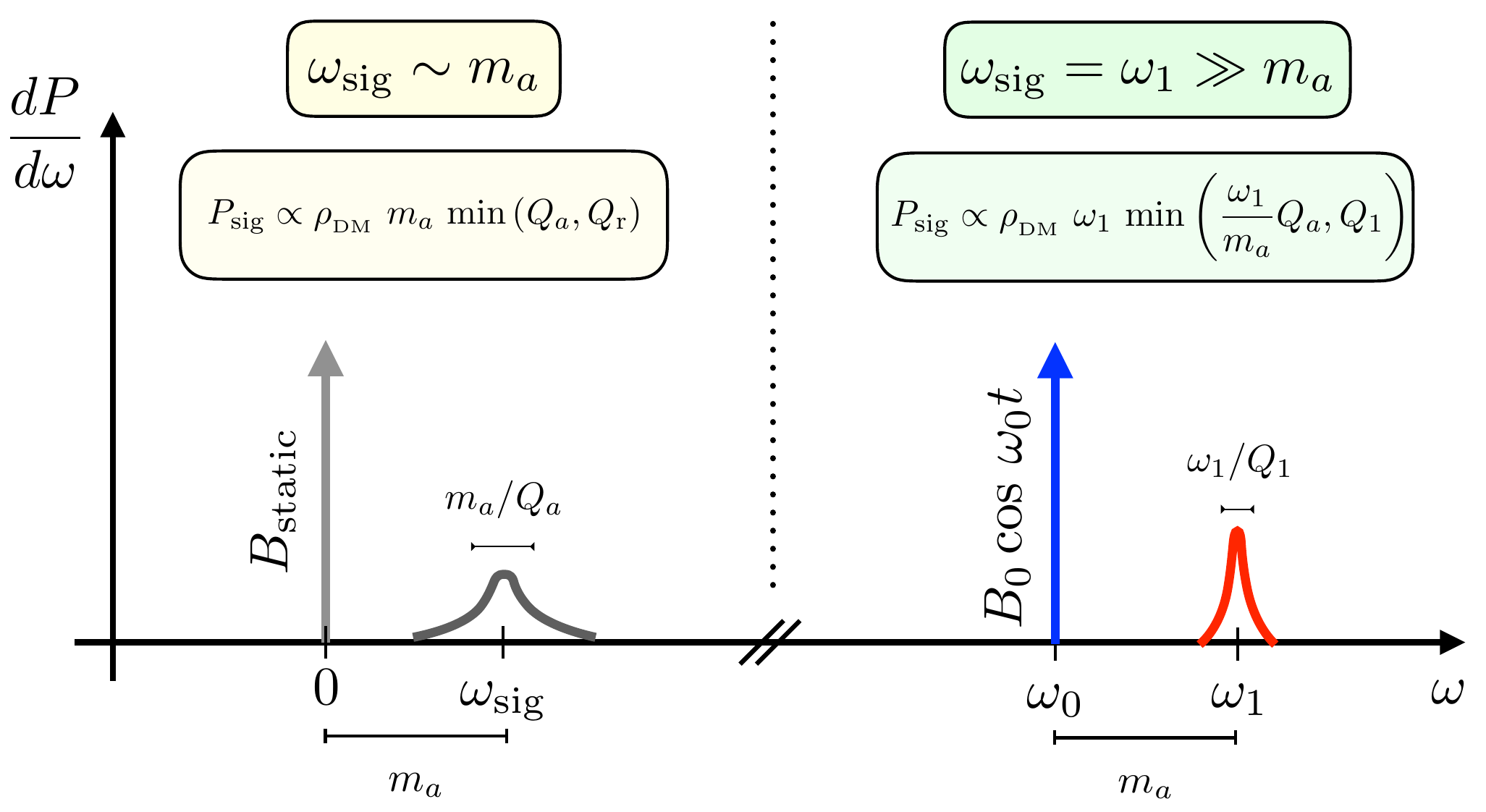}
    \caption{Comparison of static-field and driven SRF cavity schemes for axion dark matter detection, illustrating the resulting signal frequencies and power scaling. Figure from  Ref.~\cite{Berlin:2019ahk}.}
\end{figure}

The defining advantage of SRF cavities is their extraordinarily high quality factor, which enables very long signal integration times and strong resonant enhancement. This makes SRF-based experiments particularly powerful at low axion masses, where conventional haloscope volumes become prohibitively large. 

However, maintaining superconductivity in the presence of strong magnetic fields is experimentally challenging, and typically requires careful magnetic shielding or innovative cavity geometries that spatially separate the RF mode from the external field.

From a sensitivity perspective, SRF cavity proposals can probe axion--photon couplings well below existing limits in the ultralight mass range, complementing both microwave haloscopes and dielectric haloscopes. Rather than competing directly on volume or magnetic field strength, they leverage coherence and ultra-high \(Q\) to explore a qualitatively different region of parameter space.

Overall, superconducting RF cavities represent a conceptually distinct direction in axion detection, emphasising resonant frequency conversion and precision electromagnetic control. They highlight how advances in accelerator and superconducting cavity technology can be repurposed to address fundamental questions in DM searches.

\subsection{Helioscopes}

Helioscopes are experiments designed to search for axions and ALPs produced in the Sun and subsequently converted into detectable photons in a laboratory magnetic field~\cite{vanBibber:1988ge}. Their working principle relies on three essential steps:

\begin{enumerate}
\item Production of axions in the solar interior.
\item Free propagation of axions from the Sun to Earth.
\item Conversion of axions into photons in a strong laboratory magnetic field via the inverse Primakoff effect.
\end{enumerate}

Because the expected signal rate is extremely small, the sensitivity of a helioscope depends critically on both theoretical input and experimental design choices such as magnetic field strength, magnet length, aperture area, exposure time, and coherence conditions.

Since the axion signal is directional, helioscopes must track the Sun in order to maximise exposure. The sensitivity therefore depends not only on the magnetic field strength and geometry, but also on the total tracking time, detector performance, and background suppression.

Helioscopes provide a laboratory-based and relatively model-independent probe of axion couplings, complementary to astrophysical constraints. They are particularly powerful in the sub-eV mass range, where coherence can be maintained and solar axion production is efficient. This approach has motivated successive generations of helioscope experiments, culminating in the design of next-generation facilities aimed at improving sensitivity by several orders of magnitude.

In the following subsections, we will explain how axion/ALP models translate into a solar axion flux, how this flux is converted into detectable photons, and how experimental parameters shape the sensitivity of helioscope experiments such as CAST and IAXO.

\subsubsection{Axion Production inside the Sun}

The Sun provides an intense and continuous source of axions and ALPs, produced in its hot and dense interior. Temperatures of order \(T \sim \mathcal{O}(\text{keV})\) and large densities of charged particles make the solar plasma an efficient environment for axion emission through weakly coupled processes.

\paragraph{Primakoff production:} The dominant axion production mechanism in the Sun for models with an axion--photon coupling is the Primakoff process~\cite{vanBibber:1988ge}. In the electromagnetic fields of charged particles (ions and electrons), thermal photons can convert into axions via
\begin{equation}
\gamma + \gamma^\ast \;\to\; a ,
\end{equation}
where \(\gamma^\ast\) denotes a virtual photon provided by the Coulomb field of the plasma. This process directly probes the axion--photon interaction
\(
\mathcal{L} \supset -\tfrac{1}{4} g_{a\gamma} a F_{\mu\nu}\tilde{F}^{\mu\nu}
\),
and produces a continuous axion spectrum with typical energies of a few keV, set by the solar core temperature.

\paragraph{Electron-induced processes:} If the axion couples to electrons, additional production channels become
relevant. These are collectively referred to as \emph{ABC processes}~\cite{Redondo:2013wwa} and include:

\begin{itemize}
\item Compton-like scattering:
\(
\gamma + e^- \to e^- + a
\),
\item electron--ion bremsstrahlung:
\(
e^- + I \to e^- + I + a
\),
\item electron--electron bremsstrahlung:
\(
e^- + e^- \to e^- + e^- + a
\),
\item axiorecombination and axio-deexcitation involving bound atomic states.
\end{itemize}

These channels depend on the axion-electron coupling \(g_{ae}\) and dominate the solar axion flux in models where this coupling is sizable.

Compared to Primakoff axions, the resulting energy spectrum can exhibit distinctive features associated with atomic physics in the solar plasma.

\paragraph{Nuclear transitions:} Axions can also be produced in nuclear processes in the Sun. A particularly important example is the magnetic dipole (M1) transition of \({}^{57}\mathrm{Fe}\)~\cite{CAST:2009jdc,DiLuzio:2021qct}, which emits monoenergetic axions with energy \(E_a = 14.4\,\mathrm{keV}\). This channel probes axion-nucleon couplings and leads to a line-like axion signal rather than a continuous spectrum.

\paragraph{Escape from the Sun:} Once produced, axions interact so weakly with solar matter that they escape the Sun without further scattering. As a result, the solar axion flux arriving at Earth preserves both the energy spectrum and the directional information, pointing back to the solar core. This property is crucial for helioscope experiments, which exploit the directional nature of the signal to suppress backgrounds.

\subsubsection{Solar Axion Fluxes on the Earth}
\label{Sec:helioFlux}

Axions produced in the solar interior stream freely out of the Sun and reach the Earth as an intense, continuous flux. The differential flux depends on the underlying production mechanism and, crucially, on the axion couplings involved. In helioscope searches, the observable signal rate is obtained by folding these fluxes with the axion-photon conversion probability in the laboratory magnetic field.

\paragraph{Primakoff flux:} For axions coupled to photons, the dominant contribution arises from the Primakoff process in the solar plasma. A convenient parametrisation of the resulting differential flux at Earth is~\cite{RevModPhys.54.767,Vogel:2023rfa}
\begin{equation}
\left.\frac{\mathrm{d}\Phi_a}{\mathrm{d}\omega}\right|_{\mathrm{P}}
\simeq
2.0 \times 10^{18}
\left( \frac{g_{a\gamma}}{10^{-12}\,\mathrm{GeV}^{-1}} \right)^2
\omega^{2.45}\,
\mathrm{e}^{-0.829\,\omega}
\quad
\mathrm{m^{-2}\,yr^{-1}\,keV^{-1}},
\end{equation}
where \(\omega\) is the axion energy in keV. The spectrum peaks at a few keV, reflecting the temperature of the solar core, and scales quadratically with the axion--photon coupling \(g_{a\gamma}\).

\paragraph{Electron-induced fluxes (ABC processes):} As described in the previous subsection, for dominant ALP-electron couplings, ABC processes contribute. Representative parametrisations are~\cite{Barth:2013sma}
\begin{align}
\left.\frac{\mathrm{d}\Phi_a}{\mathrm{d}\omega}\right|_{\mathrm{C}}
&\simeq
4.2 \times 10^{18}
\left( \frac{g_{ae}}{10^{-13}} \right)^2
\omega^{2.987}\,
\mathrm{e}^{-0.776\,\omega},
\\[0.5em]
\left.\frac{\mathrm{d}\Phi_a}{\mathrm{d}\omega}\right|_{\mathrm{B}}
&\simeq
8.3 \times 10^{20}
\left( \frac{g_{ae}}{10^{-13}} \right)^2
\frac{\omega}{1 + 0.667\,\omega^{1.278}}\,
\mathrm{e}^{-0.77\,\omega},
\end{align}
corresponding to Compton-like and bremsstrahlung processes, respectively. These fluxes depend on the axion-electron coupling \(g_{ae}\) and typically populate the low-energy (\(\lesssim\) few keV) part of the spectrum more strongly
than the Primakoff contribution.

\begin{figure}[!ht]
    \centering
    \includegraphics[width=0.9\linewidth]{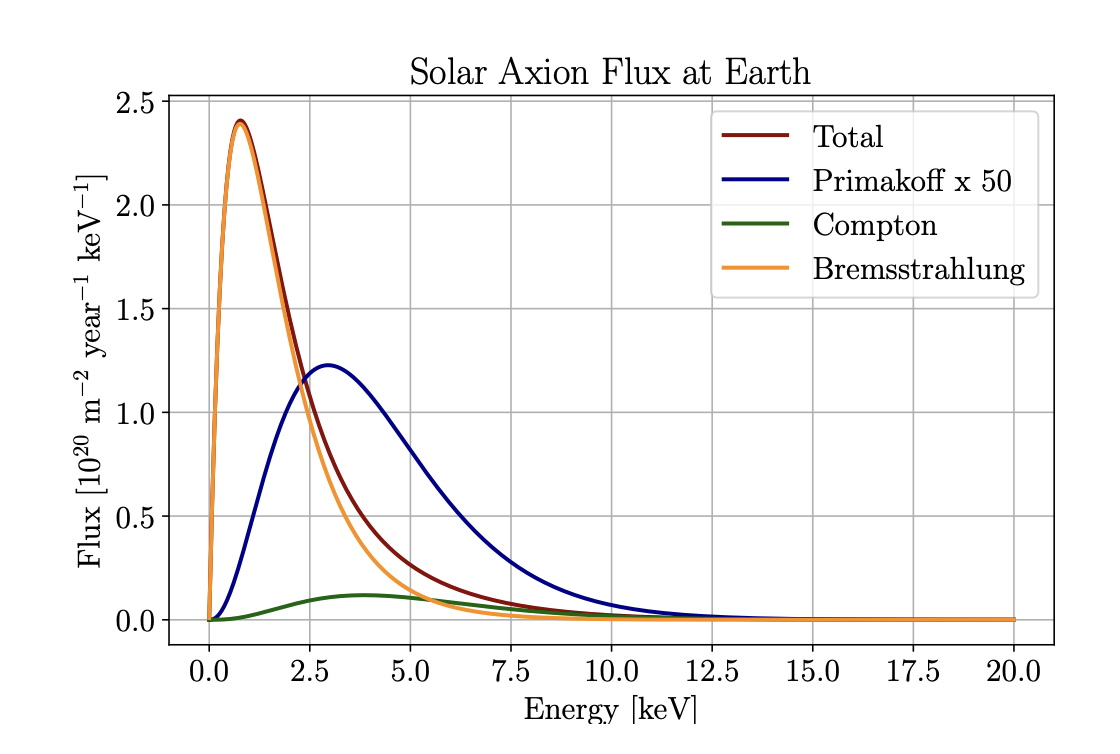}
    \caption{Differential solar axion flux at Earth, showing the contributions from Primakoff, Compton, and bremsstrahlung production processes.}
     \label{fig:helioFlux}
\end{figure}

\paragraph{Total solar axion spectrum:} The total solar axion flux at Earth is obtained by summing all relevant contributions,
\begin{equation}
\frac{\mathrm{d}\Phi_a}{\mathrm{d}\omega}
=
\left.\frac{\mathrm{d}\Phi_a}{\mathrm{d}\omega}\right|_{\mathrm{P}}
+
\left.\frac{\mathrm{d}\Phi_a}{\mathrm{d}\omega}\right|_{\mathrm{C}}
+
\left.\frac{\mathrm{d}\Phi_a}{\mathrm{d}\omega}\right|_{\mathrm{B}}
+ \cdots .
\end{equation}
Depending on the underlying axion model, different terms dominate. Photon-coupled axions lead to a Primakoff dominated spectrum, while models with sizeable electron couplings can produce significantly enhanced low-energy
fluxes.

\paragraph{Implications for helioscopes:} The broad, keV-scale energy distribution of solar axions motivates the use of low-threshold X-ray detectors in helioscope experiments. Moreover, the distinct spectral shapes associated with different production mechanisms provide an opportunity to disentangle axion couplings by combining spectral information with precision background control.

\subsubsection{Solar Axion Propagation and Signal Characteristics}

\paragraph{Free streaming of solar axions:} Photons produced in the solar core interact frequently with charged particles in the plasma. Their mean free path is extremely short compared to the solar radius, \(\ell_\gamma \ll R_\odot\), and energy transport therefore proceeds via diffusion.

Axions and ALPs, by contrast, interact only through their extremely weak couplings to photons, electrons, or nucleons. As a result, once produced, they propagate almost freely out of the Sun. This justifies treating the Sun as a point-like source of axions and motivates helioscope searches on Earth.

For axions coupled dominantly to photons, the inverse Primakoff process controls their absorption rate in the solar plasma. The corresponding mean free path can be estimated parametrically as~\cite{Vogel:2023rfa}
\begin{equation}
\ell_a \;\simeq\;
6 \times 10^{24}\,
g_{10}^{-2}\ \mathrm{cm}
\;\simeq\;
8 \times 10^{13}\, g_{10}^{-2}\, R_\odot ,
\end{equation}
where
\[
g_{10} \equiv \frac{g_{a\gamma}}{10^{-10}\,\mathrm{GeV}^{-1}} .
\]
For a representative coupling $g_{a\gamma} =10^{-10}\,\mathrm{GeV}^{-1}$, one finds $\ell_a \;\sim\; 10^{14}\, R_\odot$ , which is vastly larger than the solar radius. Axions, therefore, free-stream out of the Sun without significant reabsorption.

To reduce the axion mean free path to the size of the solar radius, the axion-photon coupling would need to be increased by many orders of magnitude,
\begin{equation}
g_{a\gamma} \;\sim\; 10^{-3}\,\mathrm{GeV}^{-1},
\end{equation}
i.e., roughly seven orders of magnitude larger than the values probed by helioscopes and constrained by stellar evolution. Only in this extreme regime would axions become trapped and behave analogously to photons inside the Sun.

Because solar axions are free-streaming, the axion flux at Earth is directly determined by the production rate in the solar interior. Moreover, transport effects inside the Sun can be neglected, and helioscope experiments can treat the incoming axion beam as an undistorted, weakly interacting flux. These properties underline the robustness of solar axion searches and their direct connection to axion couplings.

\paragraph{Clean solar axion signals on the Earth:} A central advantage of helioscope experiments is that solar axions provide one of the cleanest and most theoretically controlled signals available in astroparticle physics. This is a direct consequence of how axions are produced and how they propagate from the Sun to the Earth.

For most conventional astrophysical signals, what we observe at Earth is not the same as what was produced at the source. Photons and gamma rays undergo scattering, absorption, and energy loss as they travel through dense environments, interstellar media, and radiation backgrounds. Charged cosmic rays are further complicated by diffusion and deflection in magnetic fields, losing all directional information in the process.

As a result, interpreting astrophysical photon or cosmic-ray data typically requires detailed modeling of transport effects, including interaction cross sections with intervening matter, magnetic field configurations, diffusion and energy-loss mechanisms, and secondary particle production. These effects introduce significant theoretical uncertainties and can obscure the connection between the observed signal and the underlying particle physics.

Solar axions are produced deep inside the Sun, primarily in the core, through processes such as the Primakoff effect, Compton scattering, and bremsstrahlung. Despite originating in an extremely dense and hot environment, axions interact so weakly with matter that, once produced, they escape the Sun essentially without further interaction.

Quantitatively, the axion mean free path in the Sun is many orders of magnitude larger than the solar radius for all couplings of interest. As a result, solar axions leave the Sun immediately after production and do not thermalise, scatter, or get reabsorbed.

After leaving the Sun, axions propagate from the solar surface to the Earth over a distance of approximately one astronomical unit. Over this distance, axions continue to free-stream. There are no relevant interactions with interplanetary matter, radiation fields, or magnetic fields that would modify their energy or direction.

Crucially, this means that : (i) the axion energy spectrum at Earth is identical to the spectrum at production, (ii) no propagation-induced distortions occur, and (iii) the axions retain precise directional information pointing back to the Sun.

Because axions propagate freely, the solar axion flux measured at Earth is related to the production rate in the Sun by nothing more than geometric dilution~\cite{Irastorza:2011gs},
\begin{equation}
\Phi_a(E)\big|_{\oplus}
\;=\;
\frac{1}{4\pi d^2}
\frac{\mathrm{d}N_a}{\mathrm{d}E}\bigg|_{\text{Sun}} .
\end{equation}
The only difference between the spectrum at the source and at Earth is the \(1/(4\pi d^2)\) suppression due to the spreading of particles over a sphere of radius \(d\). This is a remarkably simple relation compared to most astrophysical messengers.

The free-streaming nature of solar axions has several powerful implications for helioscope searches because (i) the expected signal is theoretically clean and robustly predictable from solar physics and axion couplings. (ii) There are no uncertainties associated with propagation or reprocessing. (iii) The signal is highly directional, allowing experiments to suppress backgrounds by tracking the Sun. (iv) Any excess observed when the magnet is aligned with the Sun can be directly interpreted as axion-photon conversion.

For these reasons, helioscope experiments such as CAST and the proposed IAXO probe axion physics in an exceptionally clean and controlled experimental environment, making solar axions one of the most compelling targets in the search for new light particles.

\subsubsection{Inside a Helioscope}

A helioscope is designed to detect axions produced in the Sun by converting them back into photons/X-rays in a controlled laboratory magnetic field on Earth. The detection principle exploits the axion–photon coupling underlying Primakoff production in stellar plasmas, enabling the coherent conversion of solar axions into photons in a laboratory magnetic field.

\paragraph{From the Sun to the laboratory:} Axions are produced in the solar core with typical energies of a few keV, set by the temperature of the Sun. Once produced, they escape the Sun freely and propagate to the Earth without scattering or energy loss. The time of flight from the Sun to the Earth is approximately $t \simeq 500~\text{s}$, which plays no dynamical role in the detection but illustrates the macroscopic scale over which axions free-stream.

\paragraph{Axion-photon conversion in the magnet:} When solar axions enter the helioscope magnet, they traverse a region with a strong static magnetic field oriented transverse to their direction of motion. In this field, axions can convert into photons via the inverse Primakoff effect, mediated by the axion-photon coupling \( g_{a\gamma} \).

Physically, the external magnetic field provides a virtual photon, allowing the axion field to oscillate into a real electromagnetic wave. The conversion probability grows with the square of the magnetic field strength and with the length of the magnet, making long, high-field dipole magnets essential for helioscope experiments.

\paragraph{Energy of the converted photons:} The conversion process is coherent and conserves energy. As a result, the photon produced in the magnet carries essentially the same energy as the incoming axion. Since solar axions are produced thermally, the expected photon spectrum peaks at energies of a few keV, corresponding to soft X-rays. This spectral property strongly guides the choice of detector technologies.

\paragraph{Detection of the X-rays:} At the end of the magnet bore, X-ray detectors are placed to register photons produced by axion conversion. Several detector technologies are commonly used~\cite{Irastorza:2011gs,Abbon:2007ug,Kuster:2007ue,Autiero:2007uf}:

\begin{itemize}
\item \textbf{CCDs}, which provide pixel-level imaging and good sensitivity in the keV range,
\item \textbf{Micromegas and time projection chambers (TPCs)}, which are gaseous detectors with excellent background rejection capabilities,
\item \textbf{Cryogenic calorimeters}, which offer very high energy resolution and extremely low detection thresholds.
\end{itemize}

These detectors are optimised specifically for low-energy X-rays and are operated in ultra-low-background conditions.

\paragraph{Background suppression:} Because the expected axion signal rate is extremely small, controlling backgrounds is critical. Helioscope experiments employ multiple layers of background mitigation, including passive shielding, active anticoincidence systems, careful material selection, and low-noise electronics. In addition, the directional nature of the signal allows data to be taken both while tracking the Sun and while pointing away from it, providing an in-situ background measurement.

\subsubsection{Coherence in Axion-Photon Conversion}
\label{sec:coh}

The efficiency of axion-photon conversion in a helioscope is controlled not only by the strength of the magnetic field or the magnet length, but also by the phase coherence between the axion field and the electromagnetic wave produced during conversion.

\paragraph{Conversion probability and momentum mismatch:} In a homogeneous transverse magnetic field of length \(L\), the axion-photon conversion probability in vacuum is given by~\cite{Vogel:2023rfa,Irastorza:2011gs}
\begin{equation}
P_{a\to\gamma}
=
\left( \frac{g_{a\gamma} B L}{2} \right)^2
\left(
\frac{\sin(qL/2)}{qL/2}
\right)^2 
\label{eq:coherence}
\end{equation}
where the quantity
\begin{equation}
q \;\equiv\; k_a - k_\gamma \;\simeq\; \frac{|m_a^2 - m_\gamma^2|}{2E}
\end{equation}
represents the momentum mismatch between the axion and the photon. Physically, \(q\) measures how quickly the axion and photon waves drift out of phase as they propagate through the magnetic field.

\paragraph{Coherence and use of buffer gas :}Efficient conversion requires that the axion and photon remain approximately in
phase over the full length of the magnet. This translates into the coherence condition
\begin{equation}
\frac{qL}{2} \;\lesssim\; \pi .
\end{equation}
When this condition is satisfied, the conversion amplitudes generated at different positions along the magnet add constructively. If the phase mismatch becomes too large, the oscillatory factor \(\sin(qL/2)/(qL/2)\) suppresses the signal due to destructive interference.

In vacuum, the photon is effectively massless (\(m_\gamma \simeq 0\)), while the
axion has a finite mass \(m_a\). As a result, \(q \neq 0\) for sufficiently large \(m_a\), and the axion and photon waves gradually drift out of phase.

\begin{figure}[!ht]
    \centering
    \includegraphics[width=1.0\linewidth]{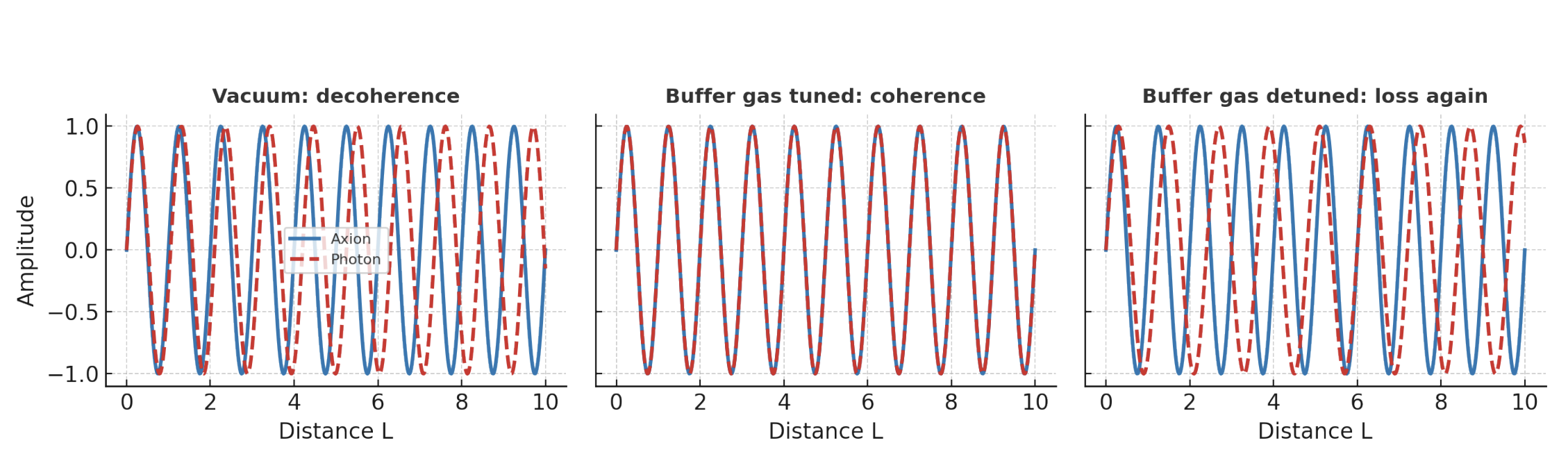}
    \caption{Illustration of axion-photon coherence in a helioscope, showing decoherence in vacuum, restored coherence with tuned buffer gas, and loss of coherence when detuned.}
    \label{fig:helioCoh}
\end{figure}

This behaviour is illustrated in the left panel of \cref{fig:helioCoh}, where the axion (solid blue) and photon (dashed red) waves progressively lose alignment. Beyond a certain magnet length, additional propagation no longer increases the conversion probability.

To recover coherence at higher axion masses, helioscopes introduce a low-pressure buffer gas into the magnet bore. In a medium, photons acquire an effective mass \(m_\gamma\) determined by the plasma frequency of the gas~\cite{Irastorza:2011gs}
\begin{equation}
m_\gamma^2 = \omega_p^2 \propto n_e ,
\end{equation}
where $n_e$ is the electron density of the gas. By tuning the gas pressure, one can adjust $m_\gamma$ such that $m_\gamma \simeq m_a$, thereby reducing the momentum mismatch \(q\)  to nearly zero. This restores coherence between axion and photon states and recovers the full conversion probability over the length of the magnet.

This situation is shown in the central panel of \cref{fig:helioCoh}, where the axion and photon waves remain locked in phase over the full length of the magnet, leading to maximal conversion.

The corresponding axion mass window covered at a fixed gas pressure is approximately
\begin{equation}
\Delta m_\gamma \lesssim \frac{2\pi E}{L m_a} .
\end{equation}
As a result, each pressure setting probes only a narrow interval in axion mass. To scan a broad mass range, the experiment must therefore operate in many discrete pressure steps. At each step, the buffer gas density is adjusted to shift the effective photon mass, and data are taken for a fixed exposure time. The total mass coverage is obtained by combining hundreds of such pressure settings, each corresponding to a small axion mass interval.

In practice, different gases are used to access different mass ranges. In CAST, He-4 has been employed to probe axion masses up to approximately $m_a \sim 0.4~\mathrm{eV}$, while He-3, which allows higher achievable gas densities at cryogenic temperatures, extends the sensitivity up to $m_a \sim 1.2~\mathrm{eV}$~\cite{CAST:2011rjr,CAST:2013bqn}. The pressure resolution required to scan adjacent mass intervals is determined by
\begin{equation}
\frac{\Delta P}{P} = \frac{\Delta m_\gamma}{m_\gamma} ,
\end{equation}
highlighting the need for precise pressure control and stability.

The buffer gas technique thus enables helioscope experiments to maintain coherence at higher axion masses, at the cost of increased experimental complexity and scanning time. This approach was successfully implemented in CAST and forms a key component of the strategy for next-generation helioscopes such as IAXO.

If the gas density is not tuned correctly, the effective photon mass overshoots or undershoots the axion mass. The momentum mismatch reappears, coherence is again lost, and the conversion probability decreases. This loss of coherence is illustrated in the right panel of \cref{fig:helioCoh}, where the axion and photon waves drift out of phase despite the presence of the buffer gas.

\subsubsection{Expected Photon Yield in a Helioscope}

The observable quantity in a helioscope experiment is the number of photons generated by the conversion of solar axions inside the magnetic field region of the detector. The expected number of signal photons can be written as~\cite{Vogel:2023rfa}
\begin{equation}
N_\gamma
=
\int dE_a \;
\frac{d\Phi_a(E_a, g_{a\gamma}^2)}{dE_a}
\,
P_{a\to\gamma}(E_a, m_a, g_{a\gamma}^2)
\,
\varepsilon(E_a)
\,
\Delta t \,
A ,
\label{eq:N_gamma}
\end{equation}
where each factor has a clear physical interpretation.

The quantity $\frac{d\Phi_a}{dE_a}$ is the differential solar axion flux at Earth, which depends on the axion production mechanisms inside the Sun and scales quadratically with the axion–photon coupling. The conversion probability $P_{a\to\gamma}$ describes axion–photon conversion inside the helioscope magnet via the inverse Primakoff effect and depends on the axion energy, mass, coupling, magnetic field strength, and magnet length. The detector efficiency $\varepsilon(E_a)$ accounts for energy-dependent acceptance and detection thresholds, while $\Delta t$ and $A$ denote the total exposure time and the effective cross-sectional area of the magnet bore, respectively.

For relativistic axions propagating through a homogeneous transverse magnetic field $B$ over a length $L$, the conversion probability $P_{a\to\gamma}$ in vacuum is given by \cref{eq:coherence}. In the coherent regime, defined by $qL \ll 1$, the oscillatory factor approaches unity and the probability scales quadratically with the magnet length, $P_{a\to\gamma} \propto L^2$.

This scaling highlights a key design principle of helioscope experiments: extending the magnetic field length is one of the most effective ways to increase sensitivity, provided coherence is maintained. Consequently, next-generation helioscopes aim to maximise the product $B L$ while preserving low background levels and high detector efficiency.

\subsubsection{Detector Parameters and Sensitivity Scaling}

The sensitivity of a helioscope experiment is determined not only by the axion--photon conversion probability inside the magnet, but also by the performance of the detector system used to register the converted photons. It is therefore useful to separate the experimental parameters into those controlling the signal yield and those governing the background level.

Following \cref{eq:N_gamma}, the expected number of signal photons detected during an exposure time $t$ can be written schematically as~\cite{Vogel:2023rfa}
\begin{equation}
N_\gamma \propto B^2 L^2 A \, \epsilon_d \, \epsilon_0 \, \epsilon_t \, g_{a\gamma}^4 \, t ,
\end{equation}
where $B$ is the magnetic field strength, $L$ the magnet length, and $A$ the cross-sectional area of the magnet bore. The factors $\epsilon_d$, $\epsilon_0$, and $\epsilon_t$ denote the detector efficiency, optics throughput, and tracking efficiency, respectively. The strong dependence on the axion--photon coupling, $N_\gamma \propto g_{a\gamma}^4$, reflects the fact that the coupling enters both the solar axion production rate and the conversion probability in the helioscope magnet.

The dominant background contribution arises from detector noise and environmental radiation. The expected number of background counts accumulated during the same exposure time is
\begin{equation}
N_b = b \, a \, \epsilon_t \, t ,
\end{equation}
where $b$ is the background rate per unit area and energy, and $a$ is the effective spot size on the detector defined by the X-ray optics. Reducing the spot size is therefore crucial, as it suppresses background counts without affecting the signal rate.

The statistical significance of a potential signal is commonly quantified by the signal-to-noise ratio,
\begin{equation}
\mathrm{SNR} = \frac{N_\gamma}{\sqrt{N_b}} .
\end{equation}
For background-dominated searches, this scaling highlights two complementary strategies for improving discovery potential: increasing the signal yield (through larger $B$, $L$, $A$, or longer exposure) and reducing the background (through improved shielding, focusing optics, and low-noise detectors).

From the SNR scaling, one can infer how the sensitivity to the axion--photon coupling improves with experimental upgrades. Requiring a fixed discovery threshold $N^\ast$ leads to an approximate coupling reach
\begin{equation}
g_{a\gamma} \sim \left( \frac{N^\ast}{\sqrt{N_b}} \right)^{-1/4} .
\end{equation}
This weak power-law dependence illustrates a key feature of helioscope experiments: large gains in magnet strength, aperture, or background suppression translate into relatively modest but steady improvements in coupling sensitivity. As a result, next-generation helioscopes aim to optimise all components of the detection toolkit simultaneously, rather than relying on a single dominant parameter.

\subsubsection{Figure of Merit for Helioscope Experiments}

The sensitivity of a helioscope experiment can be conveniently characterised through a single figure of merit that captures how efficiently a given setup converts solar axions into a statistically significant photon signal. Since helioscope searches are background-limited, the relevant quantity is the signal significance rather than the raw signal rate.

A useful definition of the figure of merit is~\cite{,Irastorza:2011gs}
\begin{equation}
f \;=\; \frac{N^*}{\sqrt{N_b}} \, ,
\end{equation}
where $N^*$ denotes the expected number of signal photons for a reference axion--photon coupling, and $N_b$ is the number of background counts accumulated during the same exposure. This quantity directly controls the signal-to-noise ratio and therefore the discovery potential of the experiment.

Crucially, this figure of merit can be factorised into three independent contributions,
\begin{equation}
f \;=\; f_M \, f_{DO} \, f_T \, ,
\end{equation}
each corresponding to a different experimental subsystem.

The first factor,
\begin{equation}
f_M = B^2 L^2 A \, ,
\end{equation}
encodes the performance of the magnet. It reflects the fact that axion--photon conversion in a helioscope scales with the square of the transverse magnetic field strength $B^2$, the square of the magnetic length $L^2$, and linearly with the magnet aperture $A$. This term highlights that magnet design is the dominant lever arm for improving helioscope sensitivity, motivating the development of large-aperture, long, high-field magnets in next-generation experiments.

The second factor,
\begin{equation}
f_{DO} = \frac{\varepsilon_d \, \varepsilon_0}{\sqrt{b\, a}} \, ,
\end{equation}
captures the combined performance of the detector and X-ray optics. Here $\varepsilon_d$ is the detector efficiency, $\varepsilon_0$ is the optics throughput, $b$ is the background rate per unit area, and $a$ is the effective focal spot size. This term shows explicitly that focusing optics play a dual role: they increase signal acceptance while simultaneously suppressing background by concentrating the signal onto a small detector area.

The third factor,
\begin{equation}
f_T = \sqrt{\varepsilon_t \, t} \, ,
\end{equation}
accounts for tracking efficiency $\varepsilon_t$ and total exposure time $t$. Since background fluctuations grow as $\sqrt{t}$, longer exposure improves sensitivity only as the square root of observing time, making efficient Sun tracking essential.

This factorised form of the figure of merit is particularly powerful because it cleanly separates the impact of magnet design, detector performance, and operational strategy. It allows different helioscope concepts to be compared on equal footing and makes transparent where technological improvements yield the greatest gains. In practice, substantial advances in sensitivity require simultaneous optimisation of all three components, with the magnet term providing the largest potential improvement.

\paragraph{Changing aperture and optics:}Combining all three factors, the compact expression for the figure of merit becomes
\begin{equation}
f \;\sim\; \left( B^2 L^2 A \right)\;
\frac{\varepsilon_d \varepsilon_0}{\sqrt{b\, a}}\;
\sqrt{\varepsilon_t t} \, 
\label{Eq:fom}
\end{equation}
where the explicit dependence on the magnet aperture $A$ appears in the magnet factor. At first sight, this suggests that increasing the aperture always improves sensitivity. However, the situation is more subtle once background considerations are included.

A larger magnet aperture directly increases the number of signal photons, since the axion flux intercepted by the magnet scales linearly with $A$. In this sense,
\begin{equation}
A \uparrow \quad \Rightarrow \quad N_\gamma \uparrow \, .
\end{equation}
At the same time, however, a larger aperture also admits more background photons if the signal is recorded directly on a detector placed behind the magnet. In the absence of focusing optics, the background rate scales with the detector area, which itself must grow with the magnet aperture. As a result, increasing $A$ without additional mitigation strategies yields diminishing returns, since both signal and background grow together.

This trade-off is resolved through the use of X-ray focusing optics. While the signal rate still scales with the full magnet aperture $A$, the background rate is instead determined by the size of the focal spot $a$ on the detector:
\begin{equation}
N_b \;\propto\; b\, a\, t \, ,
\end{equation}
where $a \ll A$ for a well-designed optical system. In this configuration, the optics decouple signal acceptance from background collection: signal photons produced anywhere across the magnet aperture are focused onto a small detector region, while diffuse backgrounds are not.

Consequently, with X-ray optics, the figure of merit benefits linearly from increasing the magnet aperture, while the background contribution remains controlled:
\begin{equation}
\text{signal} \;\propto\; A,
\qquad
\text{background} \;\propto\; a \, .
\end{equation}
This separation is one of the key conceptual advances in modern helioscope design and explains why large-aperture magnets combined with focusing optics lead to dramatic improvements in sensitivity.

In practical terms, this insight underlies the design philosophy of next-generation helioscopes such as IAXO, where a large magnet aperture is paired with dedicated X-ray optics for each bore, maximising signal collection while keeping backgrounds low.

\paragraph{Changing magnet length:} It is obvious from \cref{Eq:fom} that the figure of merit contains a strong dependence on the magnet length, reflecting the fact that axion-photon conversion occurs coherently along the magnetic field region. In the idealised limit of perfect coherence, the conversion probability
scales as
\begin{equation}
P_{a \to \gamma} \;\propto\; (B L)^2 \, ,
\end{equation}
so increasing the magnet length directly enhances the signal.

However, this quadratic scaling only holds as long as the axion and photon fields remain phase matched throughout the magnet. In vacuum, the axion and photon have different dispersion relations, leading to a momentum mismatch as described in \cref{sec:coh}. As a result, destructive interference sets in and the conversion probability no longer grows as $L^2$. Instead, the signal saturates and can even decrease with further increases in length. This loss of coherence is the fundamental limitation of vacuum helioscopes at higher axion masses.

The coherence can be restored by introducing a low-density buffer gas into the magnet bores, as discussed previously. In a medium, photons acquire an effective mass $m_\gamma$, allowing the momentum mismatch to be tuned. By adjusting the gas pressure, one can match $m_\gamma$ to a target axion mass, thereby recovering coherent conversion over the full magnet length. In this regime, the $L^2$ scaling is effectively restored.

In summary, increasing the magnet length is highly beneficial for helioscope sensitivity, but only insofar as coherence is maintained. Buffer gas operation is essential for exploiting long magnets at higher axion masses, allowing experiments to fully realise the potential $L^2$ enhancement encoded in the figure of merit.

\paragraph{Changing magnetic field strength:} Following \cref{Eq:fom}, the figure of merit has a strong dependence on the magnetic field strength. The quadratic scaling with $B$ reflects the underlying axion--photon conversion process, which proceeds via the inverse Primakoff effect in an external magnetic field.

At the amplitude level, the axion-photon mixing term is proportional to $g_{a\gamma\gamma} B$. Since the conversion probability is obtained by squaring the amplitude and integrating coherently over the magnet length, the resulting signal scales as
\begin{equation}
P_{a \to \gamma} \;\propto\; (g_{a\gamma\gamma} B L)^2 \, .
\end{equation}
Consequently, increasing the magnetic field strength directly enhances the expected number of signal photons, making $B$ one of the most powerful single levers for improving helioscope sensitivity.

This strong dependence explains why modern helioscopes rely on high-field superconducting magnets, such as repurposed accelerator dipoles. In practice, however, increasing $B$ is far from trivial. High magnetic fields require advanced superconducting technology, robust cryogenic infrastructure, and substantial mechanical support to withstand large Lorentz forces. As a result, very high-field magnets are expensive, technically demanding, and difficult to scale to large apertures.

Moreover, magnet design involves unavoidable trade-offs. Achieving higher fields often comes at the expense of bore size or total length, both of which also enter the figure of merit. For helioscopes, where sensitivity depends simultaneously on $B^2$, $L^2$, and $A$, optimisation typically favours magnets that balance moderately high fields with long lengths and large apertures rather than maximising $B$ alone.

\paragraph{Varying integration time and statistical scaling:}The figure of merit in \cref{Eq:fom} encodes the dependence on the total integration time \(t\) and the tracking efficiency \(\varepsilon_t\). This factor reflects the statistical nature of signal extraction in the
presence of background.

In the standard background-limited regime, the number of signal photons scales linearly with time,
\begin{equation}
N_\gamma \propto t ,
\end{equation}
while the number of background counts also grows linearly,
\begin{equation}
N_b \propto t .
\end{equation}
As a result, the signal-to-noise ratio behaves as
\begin{equation}
\mathrm{SNR} \;=\; \frac{N_\gamma}{\sqrt{N_b}} \;\propto\; \sqrt{t} .
\end{equation}
This square-root scaling implies that increasing the integration time leads to diminishing returns: doubling the exposure improves the sensitivity only by a factor of \(\sqrt{2}\). Consequently, very long data-taking campaigns alone cannot compensate for the limited magnet performance or high background levels.

A qualitatively different situation arises in the background-free regime, where background counts are negligible over the relevant exposure. In this case,
\begin{equation}
\mathrm{SNR} \;\propto\; N_\gamma \;\propto\; t ,
\end{equation}
and the sensitivity improves linearly with integration time. Achieving this regime requires extremely low background rates, typically through a combination of strong shielding, efficient background rejection, and tight focusing of the signal onto a small detector area.

For helioscopes, most operating conditions fall within the background-limited regime, making the \(\sqrt{t}\) scaling the relevant one for sensitivity estimates. This highlights the importance of background suppression: reducing the background rate \(b\) is often more effective than simply increasing the exposure time. Only when backgrounds are sufficiently controlled does a longer integration time become a powerful lever for improving sensitivity.

\subsubsection{CAST: The CERN Axion Solar Telescope}

The CERN Axion Solar Telescope (CAST)~\cite{CAST:2004gzq,CAST:2007jps,CAST:2008ixs,CAST:2011rjr,CAST:2013bqn,CAST:2015qbl,CAST:2017uph} was the first large-scale helioscope experiment designed to search for axions produced in the solar core. CAST operated at CERN from 2003 to 2017 and established the helioscope technique as a mature and competitive approach to axion detection.

The core of the CAST experiment was a decommissioned LHC prototype dipole magnet~\cite{Zioutas:1998cc}, providing a strong transverse magnetic field of $B \simeq 9~\mathrm{T}$ over a length of $L \simeq 9.3~\mathrm{m}$. The magnet featured two parallel bores, each with a cross-sectional area of approximately $14.5~\mathrm{cm}^2$, allowing axions streaming from the Sun to convert into X-ray photons via the inverse Primakoff effect while traversing the magnetic field.

CAST was mounted on a movable platform that tracked the Sun for approximately 1.5 hours during sunrise and sunset each day. During these tracking periods, axions emitted from the solar core could convert into X-rays inside the magnet bores and be detected at the magnet ends. The expected signal was a small excess of X-ray photons with energies in the keV range, matching the solar axion production spectrum.

\begin{figure}[htb!]
    \centering
    \includegraphics[width=0.75\linewidth]{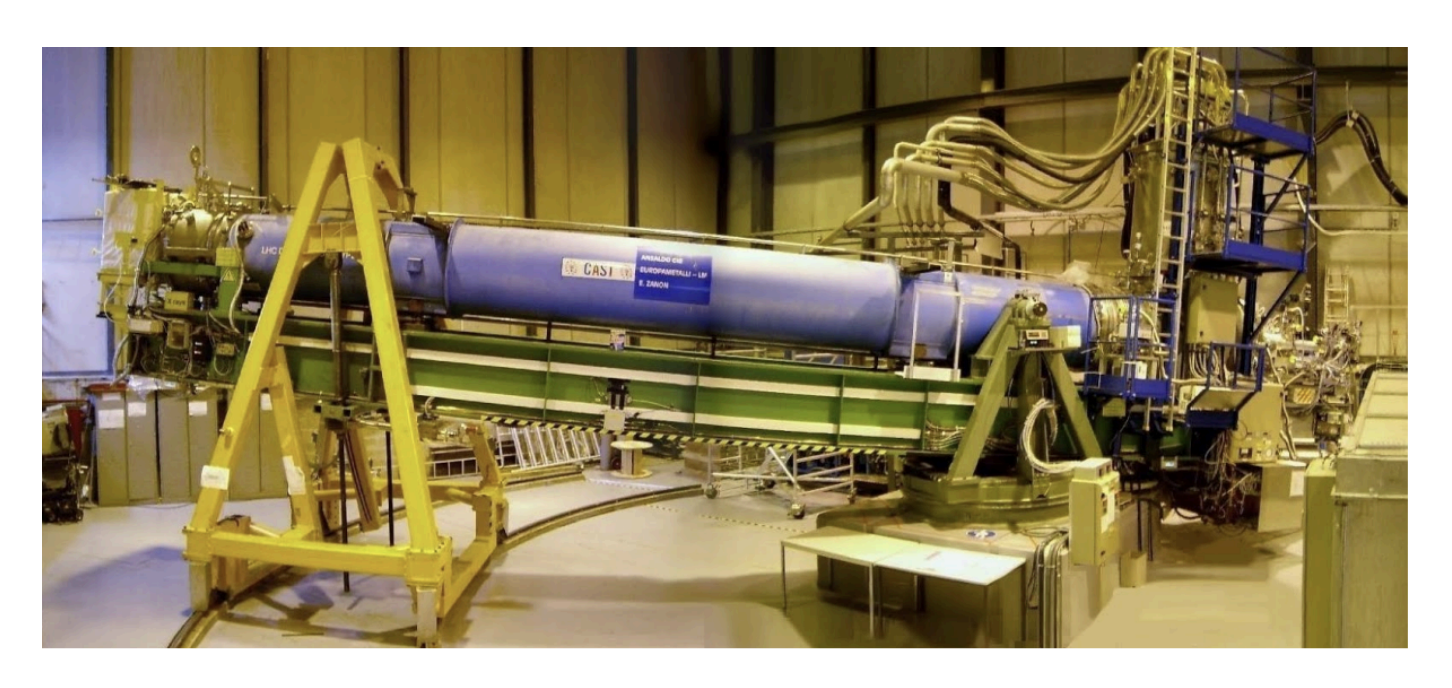}
    \caption{CAST, a helioscope experiment searching for solar axions via axion-photon conversion in a strong magnetic field. Figure from Ref.~\cite{Vogel:2023rfa}. }
\end{figure}

The experiment operated in several distinct phases. In its initial phase~\cite{CAST:2004gzq,CAST:2007jps,CAST:2008ixs}, CAST ran in vacuum, which ensured full coherence for very light axions. To extend sensitivity to higher axion masses, CAST subsequently introduced buffer gas into the magnet bores, first using $^4$He and later $^3$He~\cite{CAST:2011rjr,CAST:2013bqn,CAST:2015qbl}. The presence of buffer gas endowed the photon with an effective mass, restoring coherence for axion masses up to $\sim \mathcal{O}(1~\mathrm{eV})$. By stepping through many discrete gas pressure settings, CAST was able to scan a wide range of axion masses in a controlled and systematic way.

At the detector end, CAST employed a variety of low-background X-ray detectors, including Micromegas detectors, time projection chambers (TPCs), and CCDs. One of the bores was equipped with focusing X-ray optics repurposed from a spare ESA telescope, which concentrated the converted photons onto a small detector area. This significantly reduced background by limiting the effective spot size on the detector.

The dominant backgrounds in CAST arose from detector noise and cosmic rays, rather than from astrophysical X-ray sources. Extensive shielding, careful detector design, and off-Sun background measurements allowed these backgrounds to be characterised and subtracted with high precision.

\begin{figure}[htb!]
    \centering
    \includegraphics[width=0.7\linewidth]{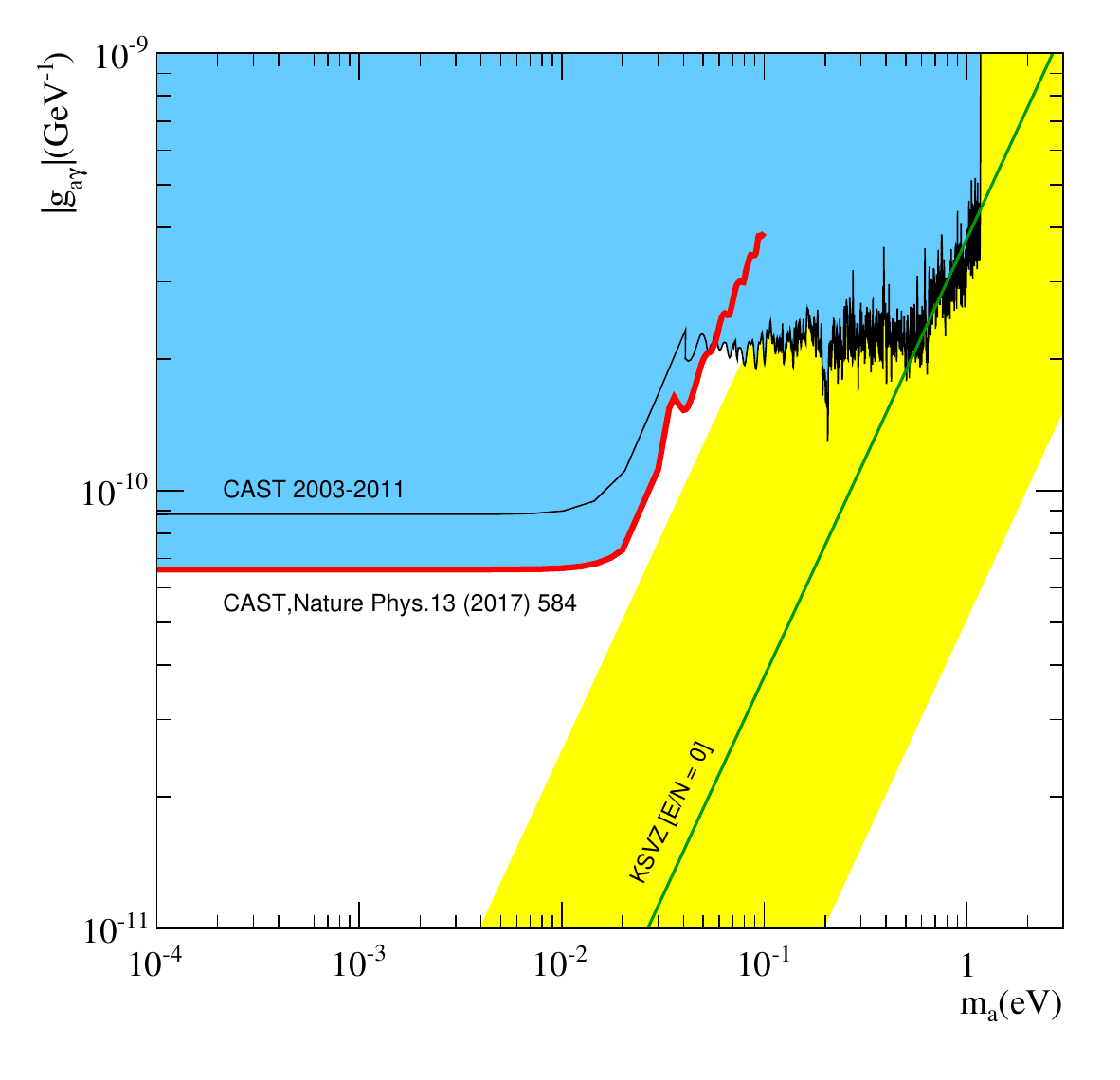}
    \caption{CAST constraints on the axion-photon coupling as a function of axion mass, including cumulative results from both vacuum and buffer gas phases. Figure adapted from Ref.~\cite{CAST:2017uph}.}
\end{figure}

CAST achieved several landmark results. Most notably, it was the first helioscope experiment to probe axion-photon couplings below the KSVZ benchmark in a broad axion mass range. CAST set the strongest laboratory bounds on the axion-photon coupling for axion masses up to approximately $1~\mathrm{eV}$ and demonstrated the scalability of the helioscope concept.

Beyond its direct physics results, CAST established the experimental techniques, background control strategies, and analysis frameworks that now underpin next-generation helioscopes such as IAXO. In this sense, CAST represents both a pioneering experiment and the foundation upon which the future of solar axion searches is built.

\subsubsection{IAXO: The International Axion Observatory}

The International Axion Observatory (IAXO)~\cite{Giannotti:2016drd,Armengaud:2014gea,IAXO:2019mpb} represents the next-generation helioscope concept, designed to improve the sensitivity to solar axions by several orders of magnitude relative
to CAST. Rather than relying on incremental upgrades, IAXO is a purpose-built experiment whose design directly targets all dominant factors in the helioscope figure of merit.

The central element of IAXO is a large toroidal superconducting magnet, inspired by accelerator technology but optimised for axion searches. The baseline design consists of an eight-coil toroidal magnet with a length of approximately \(L \simeq 20\)--25~m, providing eight independent bores with diameters of order \(60~\mathrm{cm}\). This geometry dramatically increases the effective aperture \(A\) while maintaining a strong transverse magnetic field over a long conversion region.

Each bore is equipped with dedicated X-ray optics and an independent low-background detector. The use of focusing X-ray telescopes is a key innovation: while the axion-induced signal scales with the full magnet aperture \(A\), the background scales with the much smaller focal spot
area \(a\). This separation allows IAXO to exploit the large-aperture advantage without paying a corresponding background penalty, significantly enhancing the signal-to-noise ratio.

\begin{figure}[htb!]
    \centering
    \includegraphics[width=0.65\linewidth]{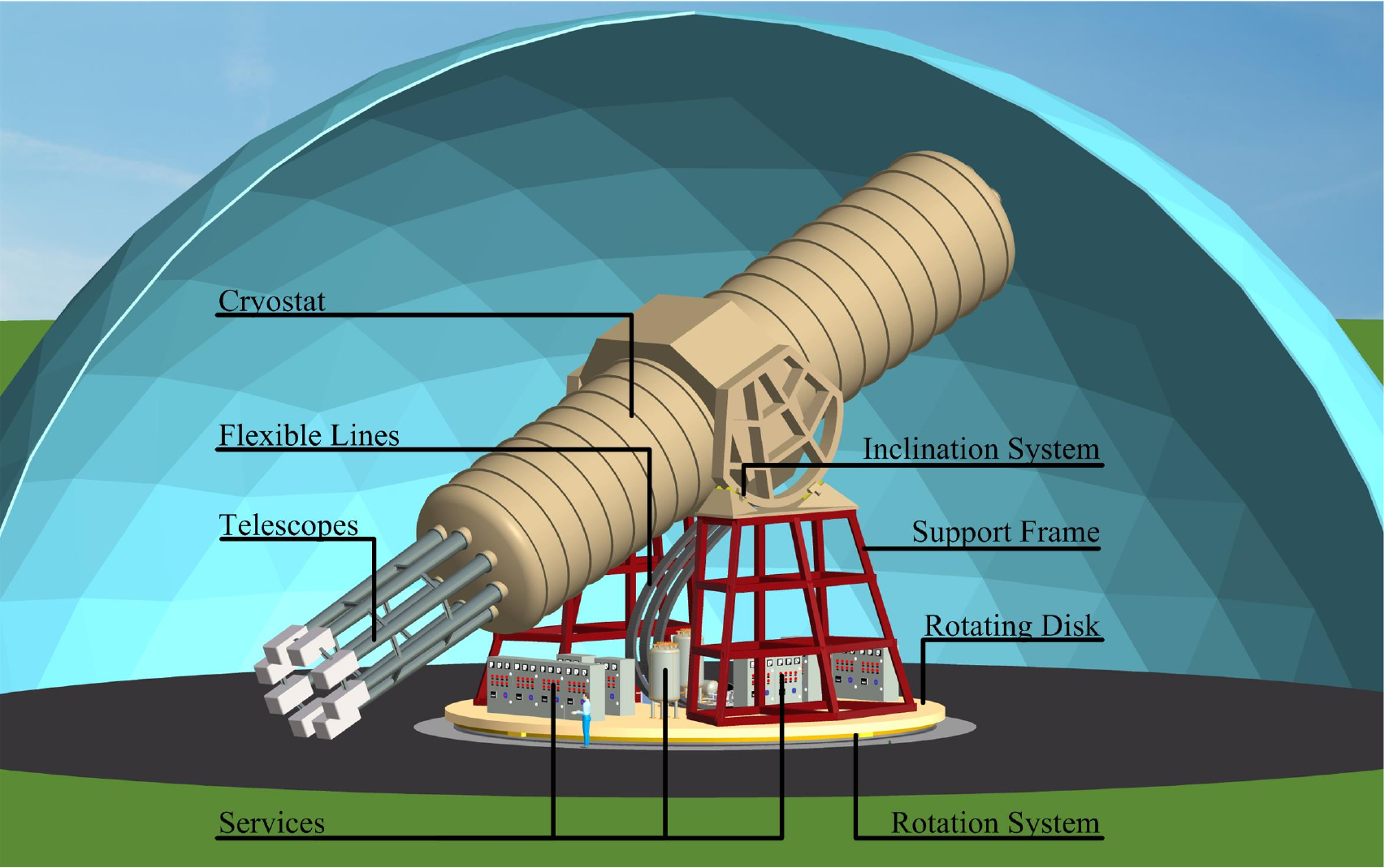}
    \caption{Schematic of the International Axion Observatory (IAXO) helioscope. Figure from Ref.~\cite{Armengaud:2014gea}.}
    \label{fig:iaxo}
\end{figure}

The expected performance gain can be understood directly from the helioscope figure of merit in \cref{Eq:fom}. Relative to CAST, IAXO improves:
\begin{itemize}
  \item the magnetic term \(B^2 L^2 A\) through a longer magnet and a vastly larger aperture,
  \item the detector--optics term through dedicated X-ray focusing and ultra-low background detectors,
  \item the exposure term via continuous solar tracking on a fully rotating platform.
\end{itemize}

When combined, these improvements lead to an expected gain in signal-to-noise ratio of approximately four to five orders of magnitude~\cite{IAXO:2025ltd} compared to CAST. This enhancement translates directly into sensitivity to axion--photon couplings well below current helioscope limits. In particular, IAXO is expected to probe deep into the parameter space predicted by QCD axion models such as KSVZ and DFSZ over a broad axion mass range, extending from the coherent vacuum regime to higher masses using buffer gas operation.

IAXO therefore represents not merely a scaled-up CAST, but a qualitatively new experimental regime for solar axion searches. Its design is explicitly driven by the figure of merit analysis, making it an instructive example of how theoretical scaling arguments translate into concrete experimental architecture.

\subsubsection{Solar Tracking and Exposure Time}

An essential practical ingredient of helioscope sensitivity is the ability to track the Sun. Since solar axions arrive from a fixed direction on the sky, the magnet must be aligned with the Sun in order to maintain sensitivity. The total tracking time therefore directly controls the effective exposure and enters the signal-to-noise ratio through the integration time \(t\).

In CAST, the magnet was mounted on a limited mechanical structure that allowed only partial solar tracking. As a result, the experiment could follow the Sun for approximately \(2\)–\(3\) hours per day, typically split between sunrise and sunset. For the remainder of the day, the magnet was misaligned and data-taking was necessarily off-source, contributing only to background measurements.

In contrast, IAXO is designed from the outset with full solar tracking capability. The entire magnet assembly is mounted on a rotating platform with independent azimuthal and elevation drives, enabling continuous alignment with the Sun over essentially the full diurnal cycle. This allows tracking times of order \(10\)–\(12\) hours per day, limited primarily by horizon constraints and operational considerations rather than by mechanical design.

The impact of solar tracking can be understood directly from the figure of merit scaling. In the background-dominated regime relevant for helioscopes, the signal-to-noise ratio scales as $\mathrm{SNR}\propto \sqrt{t}$, so increasing the tracking time leads to a parametric sensitivity improvement proportional to \(\sqrt{t}\). While this scaling is weaker than the quadratic dependence on magnetic field or magnet length, it represents a robust and cumulative gain that compounds with other improvements.

Equally important, extended solar tracking improves experimental robustness. Longer on-source periods reduce sensitivity to short-term fluctuations, allow better control of systematics, and enable more effective background subtraction using interleaved off-Sun data.

In this sense, the full azimuthal and elevation tracking of IAXO is not merely a technical upgrade but a structural enhancement of the experiment. By maximising exposure time, it ensures that the gains achieved through magnet geometry, optics, and detector performance are fully realised in the final sensitivity to the axion-photon coupling.

\subsubsection{Probing ALP--Electron Couplings with Helioscopes}

So far, the discussion has focused on solar axions produced dominantly through the axion-photon coupling and detected via inverse Primakoff conversion in a laboratory magnetic field. However, axions may also couple directly to electrons~\cite{Redondo:2013wwa}, opening additional production channels in the Sun and modifying both the expected flux and spectral shape.

The axion--electron interaction is described by the derivative coupling
\begin{equation}
\mathcal{L}_{aee} = g_{ae}\, a\, \bar{\psi}_e \gamma^\mu \gamma^5 \psi_e ,
\end{equation}
which enables axion production through electron-related processes in the solar plasma. The dominant contributions are atomic recombination and de-excitation, bremsstrahlung, and Compton-like scattering, commonly grouped together as the ABC processes, as described in \cref{Sec:helioFlux}. These mechanisms are efficient in the dense solar interior and generate an axion flux that depends quadratically on the electron coupling \(g_{ae}\).

In addition to ABC production, axion-electron couplings can indirectly contribute to plasmon-axion conversion in large-scale solar magnetic fields, although this effect is subdominant compared to direct photon-coupling-induced Primakoff production for typical parameter choices.

\begin{figure}[htb!]
    \centering
    \includegraphics[width=0.65\linewidth]{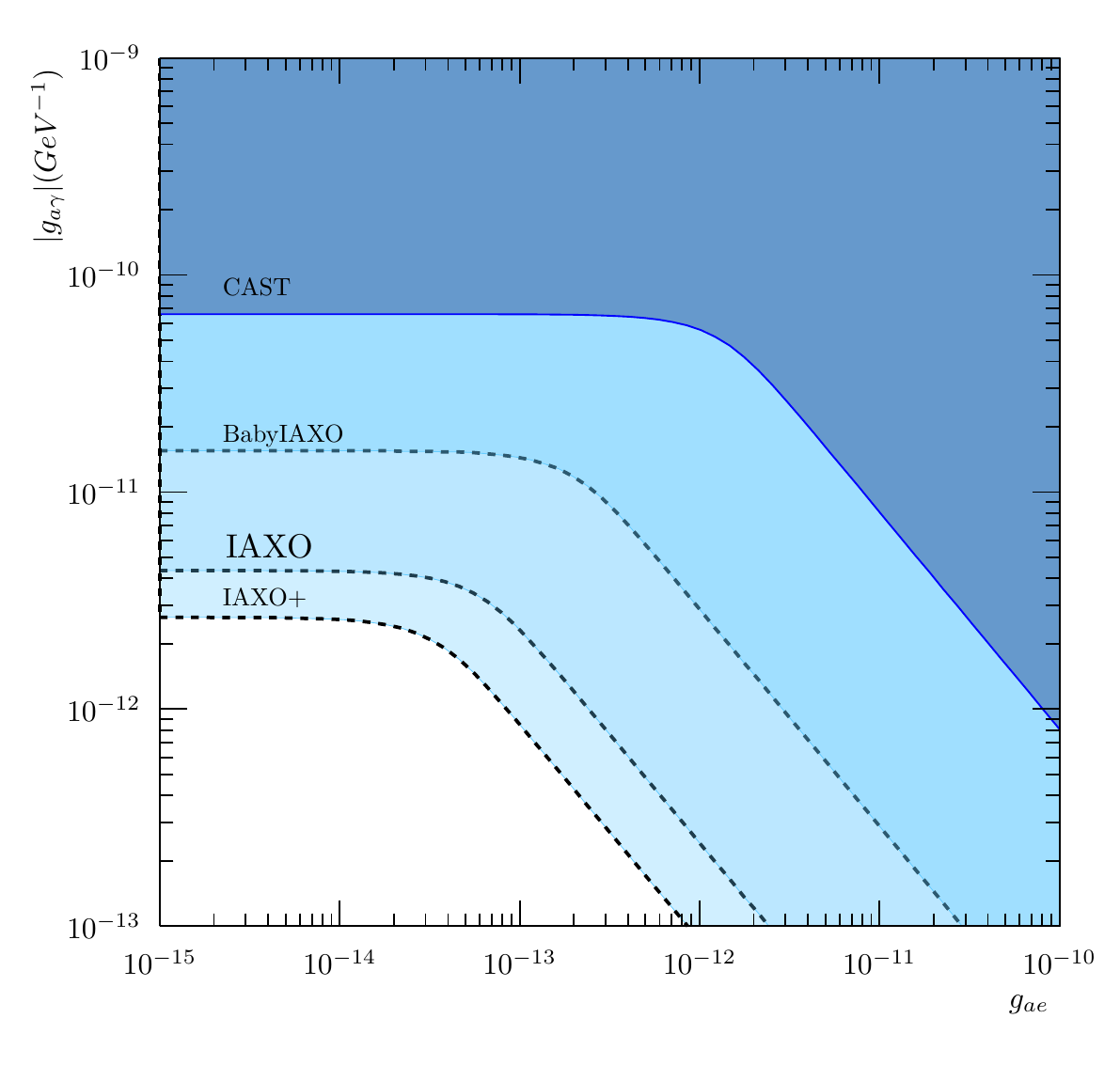}
    \caption{CAST limits and projected IAXO sensitivities in the $\left(g_{a\gamma},\, g_{ae}\right)$ plane for solar axions produced via ABC processes. Figure from Ref.~\cite{IAXO:2019mpb}.}
    \label{fig:alp-e}
\end{figure}

A key qualitative difference between axions produced via \(g_{ae}\) and those produced via \(g_{a\gamma}\) lies in their energy spectra. Axions originating from electron processes are generally softer, with a spectrum that peaks at lower energies and exhibits distinct atomic features reflecting the underlying solar plasma physics, as shown in \cref{fig:helioFlux}. In contrast, Primakoff-produced axions yield a smoother spectrum peaking at a few keV. This spectral distinction provides a powerful handle for disentangling different coupling
structures experimentally.

On the detection side, helioscopes remain sensitive to axion--electron couplings~\cite{Barth:2013sma,IAXO:2019mpb} because the axions are still converted into X-rays via the axion--photon coupling inside the magnetic field. However, the requirements on detector performance become more stringent. Since the electron-induced axion spectrum extends to lower energies, low detector thresholds down to the 10–100 eV range are essential to fully exploit sensitivity to \(g_{ae}\).

Large magnet apertures play a particularly important role, as they increase the geometric acceptance for the incoming axion flux. At the same time, good energy resolution is needed to resolve spectral features and to discriminate axion-induced signals from residual backgrounds. This motivates the use of focusing X-ray optics combined with low-background detectors such as Micromegas chambers or cryogenic calorimeters, which offer both excellent energy resolution and strong background rejection.

As seen in \cref{fig:alp-e}, resulting sensitivity reach in the \((m_a, g_{ae})\) plane reflects the interplay between production and detection. Current bounds from CAST already probe axion-electron couplings in the \(g_{ae} \sim 10^{-13}\)–\(10^{-12}\) range, while next-generation helioscopes such as IAXO are expected to improve this sensitivity by more than an order of magnitude. This places helioscopes in a unique position to test well-motivated axion models and to explore regions of parameter space relevant for stellar cooling anomalies and other astrophysical hints.

Overall, axion-electron couplings provide a complementary discovery channel to the photon coupling, enriching both the phenomenology of solar axions and the physics reach of helioscope experiments.

\subsubsection{Probing ALP-Nucleon Couplings with Helioscopes}

In addition to couplings to photons and electrons, axions generically interact with nucleons. In fact, axion-nucleon couplings are unavoidable in QCD axion models, arising from the axion coupling to quarks and its mixing with pseudoscalar mesons.

At the hadronic level, the interaction can be written as
\begin{equation}
\mathcal{L}_{aN} = - i\, a\, \bar{N}\gamma_5
\left( g_{aN}^0 + g_{aN}^3 \tau^3 \right) N ,
\end{equation}
where \(g_{aN}^0\) and \(g_{aN}^3\) denote the isoscalar and isovector axion--nucleon couplings, respectively.

A particularly important consequence of this interaction is the production of monoenergetic solar axions from nuclear transitions. The most prominent example is the magnetic dipole (M1) transition in \({}^{57}\mathrm{Fe}\)~\cite{CAST:2009jdc,DiLuzio:2021qct}, which occurs at an energy of 14.4~keV. Thermally excited \({}^{57}\mathrm{Fe}\) nuclei in the solar core can de-excite by emitting either a photon or an axion, leading to a narrow axion line at this fixed energy. Unlike Primakoff or ABC axions, which produce continuous spectra, this mechanism yields a sharply defined spectral feature.

Recent updates to the relevant nuclear matrix elements and transition rates indicate that the expected axion emission from the \({}^{57}\mathrm{Fe}\) line is approximately 30\% stronger than previously estimated. This enhances the discovery potential of helioscopes for axion-nucleon couplings and motivates renewed experimental interest in this channel.

\begin{figure}[htb!]
    \centering
    \includegraphics[width=0.7\linewidth]{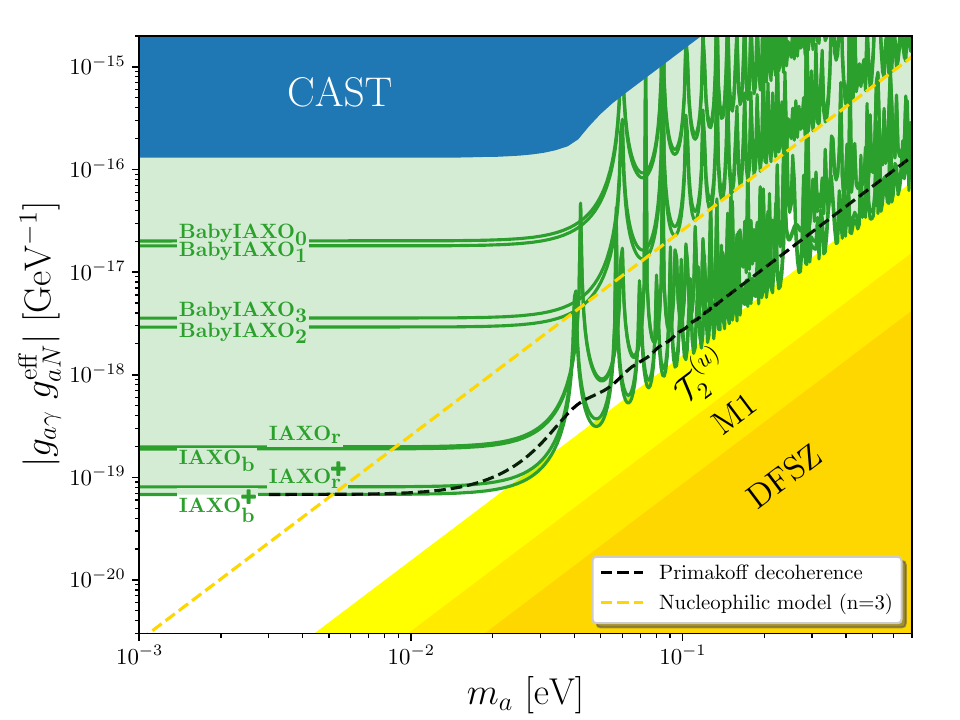}
    \caption{Helioscope sensitivity to nucleophilic ALPs, including contributions from solar axion production via nucleon processes and the \({}^{57}\mathrm{Fe}\ {\rm M}1\) transition. Figure from Ref.~\cite{DiLuzio:2021qct}.}
    \label{fig:alp-nuc}
\end{figure}

Once produced, these 14.4~keV axions propagate freely to Earth and can be detected in a helioscope via conversion into X-rays in a laboratory magnetic field. The conversion probability retains its oscillatory form $P_{a\to\gamma} \propto \left( \sin(qL/2) \right)^2$ , where \(q\) is the axion-photon momentum mismatch and \(L\) is the magnet length. As a result, the signal strength exhibits characteristic oscillations as a function of axion mass, which are visible as ripples in the sensitivity curves shown in \cref{fig:alp-nuc}.

From an instrumental perspective, axion-nucleon searches pose distinct challenges. Standard X-ray optics used in helioscopes are typically optimized for energies below 10~keV, and their efficiency drops significantly at 14.4~keV. Moreover, resolving the narrow axion line requires detectors with good energy resolution to distinguish it from the Primakoff axion continuum, which acts as an irreducible background at similar energies.

Consequently, detector choice becomes critical. Micromegas detectors offer very low background levels and good stability, while CZT and cryogenic calorimeters provide superior energy resolution and lower thresholds, making them particularly well suited for isolating narrow spectral features. The combination of improved detector technology and large-aperture, long-baseline magnets envisioned for next-generation helioscopes such as IAXO significantly extends sensitivity to axion-nucleon couplings.

Overall, axion-nucleon interactions open a complementary discovery channel that is spectrally distinct, theoretically well motivated, and experimentally accessible. Their study strengthens the physics case for helioscopes as multi-coupling axion observatories rather than single-channel experiments.

\subsubsection{Helioscope Optimisation in Generic ALP Models}

A defining feature of ALPs, in contrast to QCD axions, is the
absence of fixed coupling relations. In a generic ALP framework, the particle may couple independently to photons, electrons, nucleons, or other Standard Model fields, with strengths that are not constrained by a single underlying symmetry. As a result, there is no unique production or detection channel that dominates across the full parameter space.

This model flexibility poses a fundamental optimisation problem for helioscope experiments: If the ALP can couple to ``anything'', how should an experiment be designed? Should one prioritise larger magnetic fields and longer conversion regions, or instead focus on detector performance such as energy resolution, thresholds, and background rejection? The answer is that no single optimisation strategy is universally optimal; rather, the optimal design depends on which coupling dominates the solar ALP flux.

Let us consider the dependence of sensitivity on magnet length \(L\) and aperture area \(A\). For Primakoff-dominated scenarios, where ALPs couple primarily to photons, the solar flux is smooth and continuous, and the signal scales strongly with magnet geometry. In this regime, increasing the magnetic length and aperture directly enhances the conversion probability and signal rate, but also increases the collected background. Consequently, efficient background rejection becomes crucial, favouring designs with strong X-ray focusing optics that concentrate signal photons onto a small detector area.

By contrast, in ABC-dominated scenarios where ALPs are produced via couplings to electrons, the solar spectrum contains richer structure and extends to lower energies. Here, sensitivity is driven less by sheer conversion volume and more by detector performance. Low energy thresholds and good energy resolution are essential to access soft axions and to disentangle spectral features from background. In this regime, improvements in detector technology can outweigh gains from increasing magnet size.

The continuous transition between these regimes implies a trade-off surface rather than a single optimum. Large-aperture, long magnets push experiments toward Primakoff sensitivity, while high-resolution, low-threshold detectors favour electron-coupling searches. The most powerful helioscope designs therefore aim to operate in a balanced region of parameter space, combining large magnetic volumes with advanced optics and detectors capable of suppressing background and resolving spectral information.

This perspective motivates the philosophy behind next-generation experiments such as IAXO: rather than optimising for a single coupling or production mechanism, the goal is to build a flexible, multi-channel observatory. Such an approach maximises discovery potential across a wide range of ALP models and ensures that sensitivity is retained even if Nature chooses a coupling structure that differs from canonical expectations.

\subsection{Global Landscape of Axion-Photon Constraints}

\begin{figure}[htb!]
    \centering
    \includegraphics[width=0.8\linewidth]{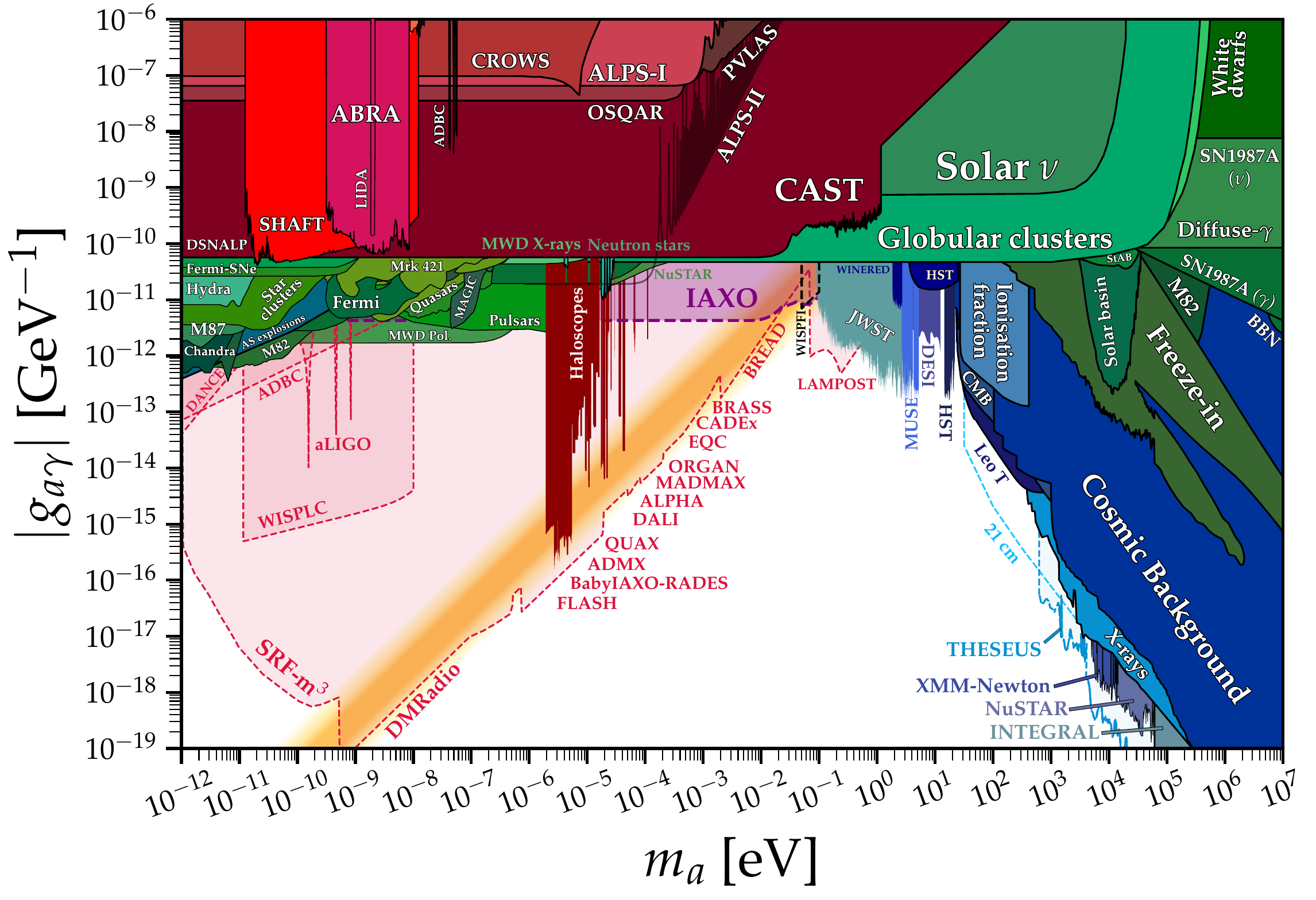}
    \caption{Summary of existing constraints and projected sensitivities on the axion-photon coupling as a function of axion mass from laboratory, astrophysical, and cosmological probes. Figure from Ref.~\cite{axionlimits}.}
     \label{fig:axions}
\end{figure}

\cref{fig:axions} summarises the current experimental, astrophysical, and cosmological constraints on the dimension-5 axion--photon coupling \(g_{a\gamma}\) as a function of the axion mass \(m_a\). This plot provides a unified view of how very different search strategies populate complementary regions of parameter space, reflecting the wide range of physical environments in which axions and ALPs can be produced and detected.

Laboratory experiments dominate the low-to-intermediate mass region, typically below the eV scale. Purely laboratory-based searches such as light-shining-through-walls (LSW) experiments probe relatively large couplings at low masses, limited primarily by achievable magnetic fields, optical power, and detector sensitivity.

Helioscopes extend this reach by exploiting the Sun as an intense axion source, allowing sensitivity to much smaller couplings through solar axion production and magnetic-field-induced reconversion. The projected sensitivity of IAXO illustrates how instrumental improvements translate directly into large gains in coupling reach.

At even lower masses, resonant cavity haloscope searches target axions constituting the local dark matter halo. These experiments are characterised by extreme sensitivity at the price of narrow mass coverage, scanning the parameter space sequentially. Projections from dielectric haloscope  such as MADMAX substantially improve sensitivity at higher masses, while the SRF cavities dominate the low-mass coverage.

Astrophysical observations impose some of the most stringent constraints over a wide mass range. Stellar cooling limits, globular cluster observations, and solar neutrino measurements restrict axion couplings independently of terrestrial instrumentation. These bounds are powerful but intrinsically model-dependent, relying on assumptions about axion production rates, stellar environments, and plasma effects. Nevertheless, they carve out large excluded regions that guide the design of laboratory experiments.

At higher masses, cosmological and high-energy astrophysical probes become dominant. Constraints from the CMB, BBN, and diffuse photon backgrounds limit axion couplings over many orders of magnitude in mass. X-ray and gamma-ray observations further constrain heavier ALPs through photon--axion conversion in cosmic magnetic fields. These bounds are sensitive to large-scale magnetic field models and cosmic axion populations, and therefore complement direct laboratory searches rather than replacing them.

Taken together, this global picture highlights several key features. First, no single experimental strategy can cover the entire axion parameter space. Second, laboratory experiments are essential for providing model-independent, controlled tests of axion/ALP interactions. Finally, the most compelling future progress lies in the coordinated development of complementary approaches, combining resonant searches, broadband detection, helioscopes, and astrophysical observations to map the axion landscape as completely as possible.

This synthesis emphasises that axion sensitivity is not set by a single figure of merit, but by the combined interplay of experimental design, source physics, and signal coherence.

\section{Probing Low-Energy Dimension-6 ALP Interactions}
\label{sec:probe-dim6}

\subsection{ALP Interactions at Different Scales}

ALPs interact with Standard Model fields in a way that depends crucially on the energy scale under consideration. A consistent description therefore, requires following ALP interactions from the ultraviolet (UV) scale, where they are defined in terms of quarks and gluons, down to the infrared (IR), where hadronic degrees of freedom become relevant. This connection is achieved through renormalisation 
group (RG) running, threshold matching, and chiral EFT.

At energy scales above the QCD confinement scale, \(\mu > \Lambda_{\rm QCD}\), ALP interactions are conveniently described by an effective Lagrangian written in terms of quark and gluon fields. Restricting to operators of dimension five, the effective Lagrangian takes the schematic form~\cite{Bauer:2020jbp,Bauer:2024hfv}
\begin{align}
\mathcal{L}^{D\leq 5}_{\rm eff}(\mu > \Lambda_{\rm QCD})
\;\supset\;
\frac{\partial^\mu a}{2f}
\left(
c_{uu}\,\bar{u}\gamma_\mu\gamma_5 u
+
c_{dd}\,\bar{d}\gamma_\mu\gamma_5 d
\right)
+
c_{GG}\,\frac{\alpha_s}{4\pi}\,
\frac{a}{f}\,
G_{\mu\nu}^a \tilde{G}^{a\,\mu\nu}
+\dots
\label{eq:RGE}
\end{align}
Here \(a\) denotes the ALP field, \(f\) is the ALP decay constant, and the coefficients \(c_{uu}\), \(c_{dd}\), and \(c_{GG}\) encode the model-dependent couplings of the ALP to quarks and gluons. The ellipsis indicates additional possible couplings to heavier quarks, leptons, or electroweak gauge bosons.

As the energy scale is lowered from the UV toward the QCD scale, the ALP couplings evolve due to RGE running~\cite{Chala:2020wvs}. Loop effects induce mixing between different operators, modifying the effective couplings at lower scales.

When crossing mass thresholds of heavy particles, such as heavy quarks, the EFT must be matched onto a theory with fewer active degrees of freedom. This \emph{threshold matching} ensures continuity of physical observables and generates additional effective ALP couplings in the low-energy theory.

Together, RGE running and threshold matching determine how the original UV ALP interactions map onto the effective interactions relevant at hadronic scales. For a detailed RGE analysis of the effective Lagrangian in \cref{eq:RGE}, see the appendix of Ref.~\cite{Bauer:2024hfv}.

\paragraph{A chiral lagrangian description:}Below the QCD confinement scale, the appropriate degrees of freedom are hadrons rather than quarks and gluons. In this regime, ALP interactions are encoded in a chiral effective Lagrangian. The leading-order chiral Lagrangian involving the ALP field can be written as~\cite{GrillidiCortona:2015jxo}
\[
\mathcal{L}_{\chi{\rm PT}}
=
\frac{f_\pi^2}{4}\,
\mathrm{Tr}
\left[
\Sigma\, m_q(a)^\dagger
+
m_q(a)\,\Sigma^\dagger
\right],
\]
where \(f_\pi\) is the pion decay constant and \(\Sigma\) is the nonlinear Goldstone boson field,
\[
\Sigma = \exp\!\left( i\sqrt{2}\,\Pi / f_\pi \right),
\]
with \(\Pi\) denoting the pseudoscalar meson matrix.

A crucial feature of the low-energy theory is that the quark mass matrix becomes dependent on the ALP field through chiral rotation. Explicitly, the quark mass matrix takes the form~\cite{Bauer:2024hfv}
\begin{align}
m_q(a)
=
e^{- i \kappa_q \frac{a}{f} c_{GG}}\,
m_q\,
e^{- i \kappa_q \frac{a}{f} c_{GG}}
\label{eq:ALPquark}
\end{align}
where \(m_q\) is the usual quark mass matrix and \(\kappa_q\) encodes the appropriate chiral charges. This ALP dependence reflects the fact that ALP couplings to gluons can be traded, via chiral rotations, for phases in the quark mass matrix. As a consequence, the ALP induces mixings with pseudoscalar mesons and modifies hadronic interactions in a systematic and calculable way.

\subsubsection{Origin of ALP Quadratic Interactions:} Beyond the leading linear (derivative) couplings, ALPs generically induce \emph{quadratic interactions} at higher order in the EFT expansion. For ALPs interacting with gluons at the UV, quadratic interactions arise naturally once the ALP dependence of the quark mass matrix is taken into account.

The quark mass matrix in the presence of an ALP background
takes the form as described in \cref{eq:ALPquark}. Expanding this expression to second order in the ALP field, one finds that hadronic observables acquire corrections proportional to \(a^2/f^2\).

In particular, the pion mass squared depends on the ALP field through
\[
m_\pi^2(a)
\;\propto\;
\mathrm{Tr}\!\left[m_q(a)\right]
\simeq
\mathrm{Tr}[m_q]
-
\frac{a^2}{2f^2}\,
\mathrm{Tr}\!\left[\{\kappa_q^2, m_q\}\right],
\]
where \(\{\cdot,\cdot\}\) denotes the anticommutator.

Ref.~\cite{Kim:2023pvt,Bauer:2024hfv} calculated the ALP dependent shift in the pion mass, originating from the EFT Lagrangian described in \cref{eq:RGE}. This shift can be parametrised as
\begin{align}
m_{\pi,\mathrm{eff}}^2(a)
=
m_\pi^2 \left( 1 + \delta_\pi(a) \right).
\end{align}
At leading order in the ALP field, the fractional correction is given by~\cite{Bauer:2024hfv}
\begin{align}
\delta_\pi(a)
=
- \frac{c_{GG}^2}{2}\,
\frac{a^2}{f^2}
\left(
1 - \frac{\Delta_m^2}{\hat{m}^2}
\right)
+ \mathcal{O}(\tau_a^2),
\end{align}
where
\[
\hat{m} = \frac{m_u + m_d}{2}, 
\qquad
\Delta_m = \frac{m_u - m_d}{2},
\qquad
\tau_a = \frac{m_a^2}{m_\pi^2}.
\]
This expression shows explicitly that quadratic ALP interactions induce scalar-like modifications of hadronic masses.

\paragraph{Nucleon mass corrections:} The modification of the pion sector feeds into the nucleon sector through chiral perturbation theory. At second order, the relevant chiral Lagrangian term takes
the form~\cite{Hoferichter:2015tha}
\[
\mathcal{L}_{\chi\mathrm{PT}}^{(2)}
=
c_1 \, \mathrm{Tr}[\chi_+] \, \bar{N}N + \dots ,
\]
where \(N\) denotes the nucleon field and \(\chi_+\) encodes the quark mass dependence. Substituting the ALP-dependent pion mass into this expression yields an
effective scalar interaction between the ALP field and nucleons,
\[
\mathcal{L}
\supset
4 c_1 m_\pi^2 \, \delta_\pi(a) \, \bar{N} N + \dots .
\]
where the low-energy constant is denoted by $c_1 \simeq -1.26~\mathrm{GeV}^{-2}$, as determined from phenomenological analyses.

\begin{figure}[htb!]
    \centering
    \includegraphics[width=0.75\linewidth]{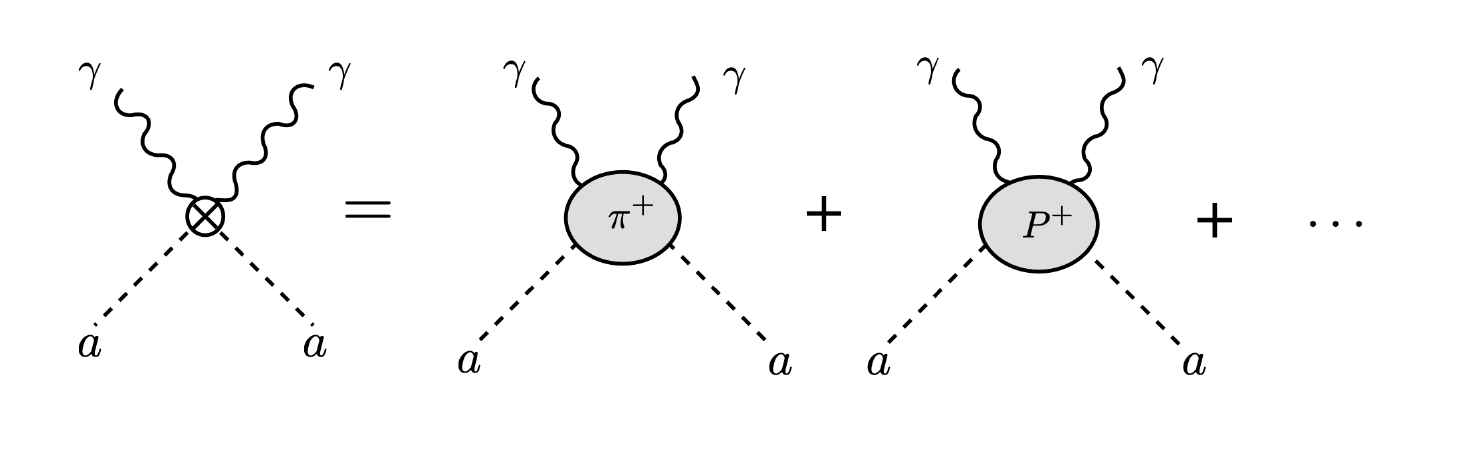}
    \caption{Representative loop-level contributions to the quadratic ALP–photon coupling~\cite{Bauer:2024hfv}.}
    \label{fig:ALPphotondim6}
\end{figure}

\paragraph{ALP-photon quadratic interactions:} Through charged pion exchange, hadronic loops, and additional higher-order effects captured at low energies within the EFT (see diagrams in \cref{fig:ALPphotondim6}), ALPs induce an effective quadratic coupling to photons of the form~\cite{Kim:2022ype,Kim:2023pvt,Beadle:2023flm,Bauer:2024hfv}
\[
\mathcal{L} \supset
C_\gamma(\mu)\,\frac{a^2}{4f^2}\,
F_{\mu\nu}F^{\mu\nu}.
\]
The corresponding Wilson coefficients specific to \cref{eq:RGE} can be written as~\cite{Bauer:2024hfv}
\[
C_\gamma(\mu)
=
\frac{\alpha}{24\pi}\,c_{GG}^2
\left(
-1
+
\frac{32\,c_1\,m_\pi^2}{M_N}
\right)
\left(
1 - \frac{\Delta_m^2}{\hat{m}^2}
\right),
\]
where \(m_\pi\) is the pion mass, \(M_N\) is the nucleon mass, \(\hat{m} = (m_u+m_d)/2\), and \(\Delta_m = (m_u-m_d)/2\).

\begin{figure}[htb]
    \centering
    \includegraphics[trim={0 0 7.25cm 0},clip, width=0.75\linewidth]{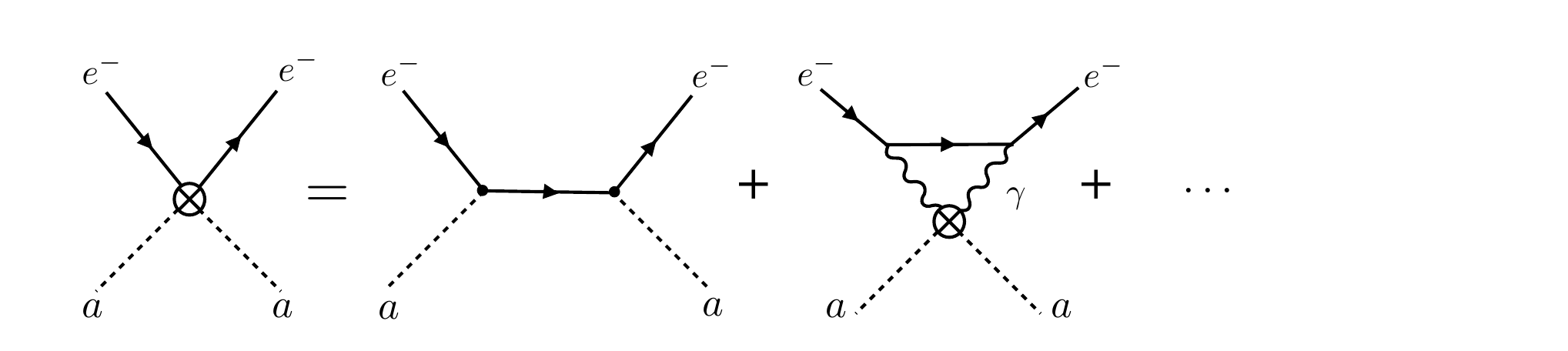}
    \caption{Representative loop-level contributions to the quadratic ALP–electron coupling~\cite{Bauer:2024hfv}.}
    \label{fig:ALPelectrondim6}
\end{figure}

\paragraph{ALP-electron quadratic interactions:} Within the EFT under consideration, quadratic ALP interactions also include an effective dimension--6 coupling to electrons,
\[
\mathcal{L} \supset
C_E\,\frac{a^2}{f^2}\,\bar{e}e.
\]
This operator is generated radiatively at the two-loop level through diagrams involving the quadratic ALP-photon interaction (see diagram in \cref{fig:ALPelectrondim6}). The dominant contribution can be expressed as~\cite{Bauer:2024hfv,Kim:2023pvt}
\[
C_E
=
- m_e \frac{3\alpha}{4\pi}\,
C_\gamma\,
\ln\!\left(\frac{m_\pi^2}{m_e^2}\right),
\]
where \(m_e\) is the electron mass.

Collecting the leading quadratic ALP effects at energies below the QCD scale, the resulting dimension--6 effective Lagrangian can be written as
\begin{align}
\mathcal{L}_{\mathrm{eff}}^{D=6}(\mu \lesssim \Lambda_{\rm QCD})
=
\bar{N}
\left[
C_N(\mu)\,\mathbf{1}
+
C_\delta(\mu)\,\tau
\right]
N\,\frac{a^2}{f^2}
+
C_e(\mu)\,\frac{a^2}{f^2}\,\bar{e}e
+
C_\gamma(\mu)\,\frac{a^2}{4f^2}\,
F_{\mu\nu}F^{\mu\nu}
\label{eq:LowEnergyDim6}
\end{align}
where the Wilson coefficients \(C_N\), \(C_\delta\), \(C_e\), and \(C_\gamma\) as defined above encode the effects of QCD dynamics, RG running, and threshold matching, as described above.

Quadratic ALP interactions are \emph{scalar} in nature and therefore differ qualitatively from the derivative couplings characteristic of linear ALP interactions. In the presence of an ultralight ALP dark matter background, \(a(t) \sim a_0 \cos(m_a t)\), these operators generate oscillatory, time-dependent shifts in: (i) the electromagnetic sector through \(F_{\mu\nu}F^{\mu\nu}\), (ii) fermion masses via \(\bar{e}e\), and \(\bar{N}N\) more generally, (iii) atomic and molecular energy levels. These effects provide a powerful handle for precision experiments, such as atomic clocks, resonant-mass detectors, and optical cavities, which are sensitive to coherent time oscillations.

\subsubsection{Shifts in Fundamental Constants}

When the ALP constitutes ultralight dark matter, the field oscillates coherently as
\[
a(t) \simeq a_0 \cos(m_a t),
\]
so that \(a^2(t)\) contains a constant component and an oscillatory component at frequency \(2m_a\)~\cite{Kim:2022ype}.

The effective Lagrangian in \cref{eq:LowEnergyDim6} implies the low-energy SM-ALP interactions scale as $a^2$, thereby generating a coherent periodic component oscillating at $2\,m_a$. This leads to the fundamental constants of nature becoming dynamical in time as follows~\cite{Bauer:2024hfv}

\begin{enumerate}
    \item The fine-structure constant becomes ALP-field dependent,
\begin{align}
\alpha^{\mathrm{eff}}(a)
=
\bigl(1 + \delta_\alpha(a)\bigr)\,\alpha,
\end{align}
with the fractional shift given by
\begin{align}
\delta_\alpha(a)
=
\frac{1}{12\pi}
\left(
1 - \frac{32 c_1 m_\pi^2}{M_N}
\right)
\delta_\pi(a)
\end{align}
\item The effective electron mass becomes
\begin{align}
m_e^{\mathrm{eff}}(a)
=
m_e\bigl(1 + \delta_e(a)\bigr),
\end{align}
where
\begin{align}
\delta_e(a)
=
\frac{3\alpha}{4\pi}\,
C_\gamma\,
\frac{a^2}{f^2}
\ln\!\left(\frac{m_\pi^2}{m_e^2}\right).
\end{align}
\item The effective nucleon mass becomes
\begin{align}
M_N(a)
=
M_N\bigl(1 + \delta_N(a)\bigr),
\end{align}
with
\begin{align}
\delta_N(a)
=
- 4 c_1 \frac{m_\pi^2}{M_N}\,\delta_\pi(a).
\end{align}
\end{enumerate}

The exact form of the ALP-induced shifts is model dependent. Nevertheless, a key implication for precision sensing is that atomic structure, molecular spectra, and nuclear energy levels depend sensitively on fundamental constants. Oscillations of these constants therefore imply that the \emph{building blocks of atoms oscillate in time}. This leads to a powerful and distinctive experimental signature of ultralight ALP dark matter, accessible to precision measurements such as atomic clocks, high-resolution spectroscopy, resonant-mass detectors, and other quantum sensors.

\subsection{Searches for Scalar-type Effects from Ultralight ALPs}

Time-dependent oscillations of fundamental constants can be probed using a wide range of precision experiments, including atomic clocks and optical cavities, together with gravitational wave detectors that can be repurposed both as laser interferometers and as mechanical resonators.

\subsubsection{Atomic Clocks}

Atomic clocks are among the most precise measurement devices ever built and provide a powerful probe of tiny variations in fundamental constants. Their extreme sensitivity makes them particularly well suited for searches for ULDM that induces coherent, time-dependent shifts in atomic energy levels~\cite{Arvanitaki:2014faa,Flambaum:2004tm,Karshenboim:2004tx,Wynands_2005,ludlow2015opticalatomicclocks}.

An atomic clock operates by stabilising the frequency of a laser or microwave oscillator to a well-defined atomic transition. The basic components of an atomic clock are:

\begin{itemize}
\item a probing laser or microwave oscillator,
\item a well-defined atomic transition frequency,
\item and a feedback loop that locks the oscillator frequency to the atomic transition.
\end{itemize}

The oscillator frequency is used to interrogate a transition between a ground state and an excited state. Through a feedback loop, the oscillator is continuously adjusted and locked such that its frequency remains resonant with the atomic transition. In practice, the resonance condition is determined by scanning the radiation frequency and locking to the peak of the spectral line. The finite linewidth \(\Delta\nu\) of the transition sets the ultimate precision of the clock~\cite{ludlow2015opticalatomicclocks}. 

\begin{figure}[htb!]
    \centering
    \includegraphics[width=0.65\linewidth]{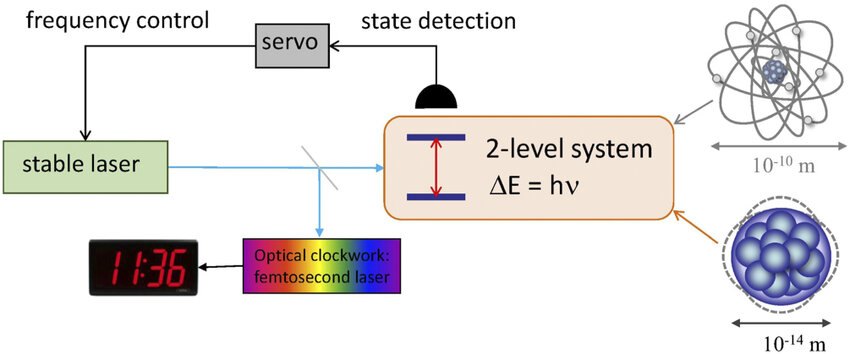}
    \caption{Conceptual layout of an atomic clock based on laser interrogation of an atomic transition. Figure from Ref.~\cite{Peik:2020cwm}.}
    \label{fig:clockschematic}
\end{figure}

Atomic clocks can be exploited as quantum sensors by comparing the frequency ratios of different atomic, vibrational, or nuclear transitions. Owing to their exceptional precision, even extremely small variations in the underlying spectral lines lead to measurable shifts in clock frequencies. In the presence of ultralight dark matter, which induces oscillatory variations in fundamental constants, these frequencies acquire a small, time-dependent modulation. Such signals can be searched for by continuously monitoring clock frequency ratios over time.

\paragraph{Clock sampling and stability:} To detect tiny, time-dependent signals such as those induced by ULDM, atomic clocks must be both \emph{ultra-stable} and \emph{coherent} over long periods of time. The statistical properties of clock noise and stability are therefore central to understanding how clock data are sampled and analysed.

Clock performance is typically characterised in terms of the \emph{fractional frequency deviation}
\[
y(t) \equiv \frac{\nu(t) - \nu_0}{\nu_0},
\]
where \(\nu(t)\) is the instantaneous clock frequency and \(\nu_0\) is its nominal reference value. In practice, the clock signal is sampled over finite time intervals of duration \(\tau\), yielding a sequence of averaged fractional frequencies
\[
y_i \equiv \frac{1}{\tau}\int\limits_{t_i}^{t_i+\tau} y(t)\,dt.
\]

\paragraph{Allan deviation:} The standard measure of clock stability is the \emph{Allan deviation}~\cite{1966IEEEP..54..221A},
\[
\sigma_y(\tau)
=
\sqrt{\frac{1}{2}\,
\left\langle
\left(y_{i+1} - y_i\right)^2
\right\rangle},
\]
where the angle brackets denote an average over all adjacent pairs of samples. The Allan deviation quantifies how the fractional frequency fluctuations depend on the averaging time \(\tau\).

Physically, \(\sigma_y(\tau)\) characterises how much the clock frequency drifts between successive measurement intervals of duration \(\tau\).

\subparagraph{Noise types and scaling laws:} The performance of an atomic clock is fundamentally limited by noise. Different physical noise sources imprint characteristic fluctuations on the clock frequency, which manifest as distinct scaling behaviours of the Allan deviation \(\sigma_y(\tau)\) as a function of the averaging time \(\tau\). Understanding these scalings is essential both for interpreting experimental data and for optimising clock operation in precision measurements.

Recall that the Allan deviation quantifies the fractional frequency stability of a clock over a given averaging time \(\tau\). Its dependence on \(\tau\) reveals the dominant noise processes affecting the clock.

\subparagraph{White (thermal) frequency noise:} White frequency noise corresponds to uncorrelated, random fluctuations of the clock frequency from one measurement interval to the next. Physically, this type of noise often arises from thermal fluctuations, photon shot noise, or other stochastic processes in the detection and interrogation system.

For white noise, the Allan deviation scales as~\cite{1966IEEEP..54..221A}
\[
\sigma_y(\tau) \propto \frac{1}{\sqrt{\tau}}.
\]
This inverse square-root scaling reflects the fact that uncorrelated noise averages down with longer integration time. Each additional measurement provides independent information, so increasing the averaging time \(\tau\) improves the frequency stability of the clock.

This regime typically dominates at short averaging times, where statistical noise sources outweigh long-term correlations.

\paragraph{Flicker noise:} Flicker noise, often referred to as \(1/f\) noise, is characterised by correlations that persist across timescales. Common sources include technical noise in electronics, laser frequency noise, and imperfections in the reference oscillator.

In the presence of flicker noise, the Allan deviation becomes approximately independent of the averaging time~\cite{5570702},
\[
\sigma_y(\tau) \simeq \text{const.}
\]
In this regime, increasing \(\tau\) no longer improves the clock stability. Instead, the Allan deviation reaches a plateau that represents a fundamental noise floor for the system. This behaviour reflects the fact that correlated fluctuations do not average out with longer integration times. Flicker noise therefore, sets a practical limit on achievable stability in many high-performance clocks.

\subparagraph{Drift (random-walk frequency noise):} Drift or random-walk frequency noise arises from slow, cumulative changes in the clock frequency over time. These can originate from environmental effects such as temperature variations, mechanical stress, ageing of components, or slow changes in the atomic reference itself.

For random-walk frequency noise, the Allan deviation increases with averaging time~\cite{804271},
\[
\sigma_y(\tau) \propto \sqrt{\tau}.
\]
In this case, longer integration times worsen the apparent clock stability, because low-frequency drifts dominate the frequency fluctuations. Unlike white noise, these variations are strongly correlated and accumulate over time.

This regime typically becomes relevant at long averaging times, where slow systematic effects can no longer be neglected.

In a typical clock, white noise dominates at short times, flicker noise sets a stability floor at intermediate times, and drift effects become important at long times. The transition between these regimes reflects the underlying physical noise sources in the system.

The optimal averaging time for a given experiment is therefore determined by the minimum of the Allan deviation curve~\cite{Kessler:2013nez}. At this point, the clock achieves its best fractional frequency stability, balancing statistical noise against long-term correlated fluctuations. This optimal operating point is especially important in precision measurements and searches for time-dependent signals, such as those induced by ULDM.

\paragraph{Single-clock measurements -- drift versus signal:} When using a single atomic clock, slow variations in the measured frequency cannot be unambiguously attributed to either instrumental systematics or a physical signal. This ambiguity is particularly important in the context of ULDM searches, where the signal itself is oscillatory but may appear as a slow drift when sampled sparsely.

\begin{figure}[htb!]
    \centering
    \includegraphics[width=0.75\linewidth]{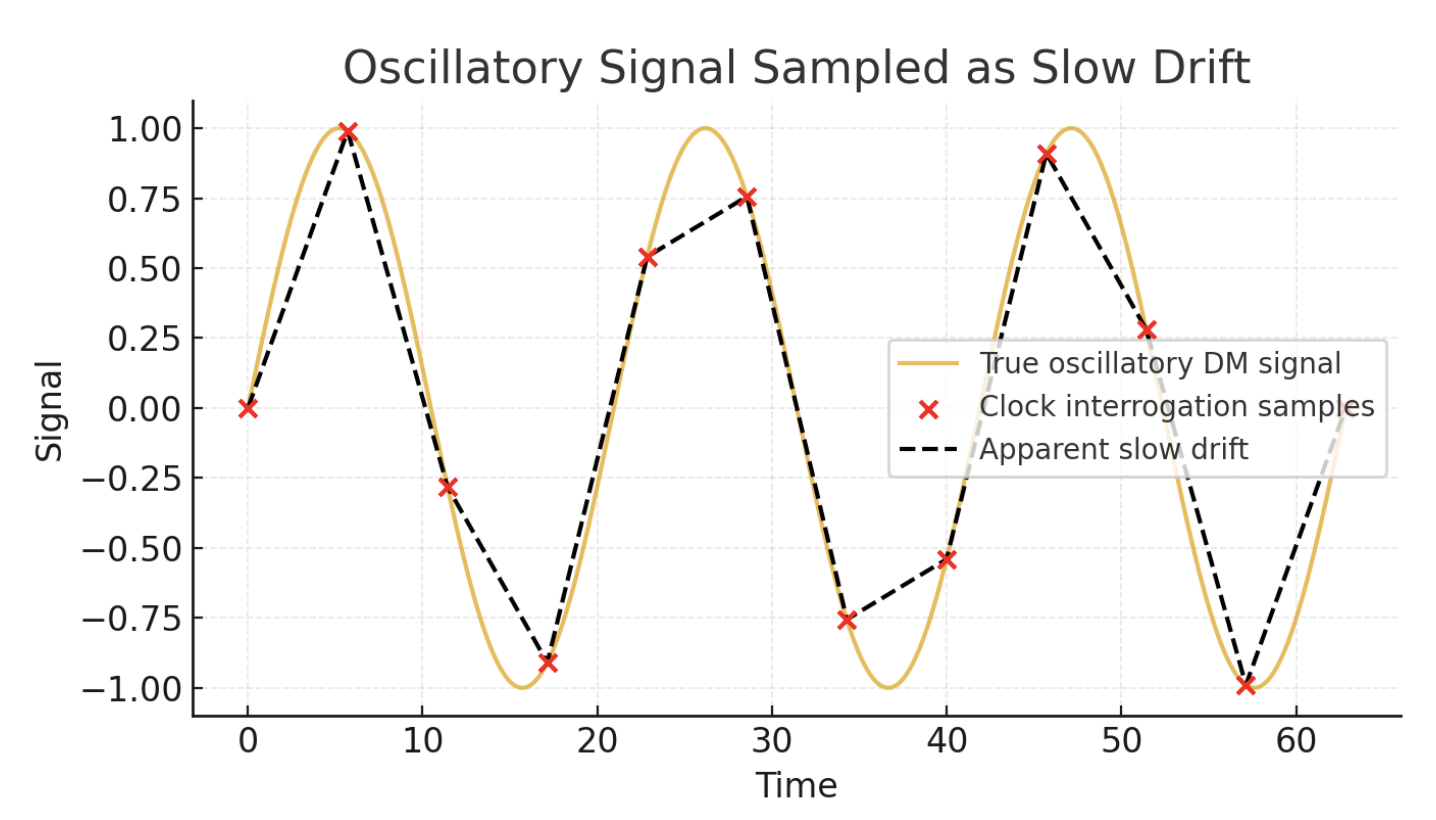}
    \caption{Discrete sampling of an oscillatory dark matter signal leading to an apparent slow drift.}
    \label{fig:singleclock}
\end{figure}

Consider an underlying oscillatory dark matter signal that induces a time-dependent fractional frequency shift, with a long oscillation period. In practice, an atomic clock does not measure this signal continuously, but rather samples it at discrete interrogation times, as shown in \cref{fig:singleclock}. Each interrogation yields a single averaged frequency point.

If the sampling cadence is much slower than the oscillation frequency, or if only a few points are collected per oscillation period, the true oscillatory signal can be misinterpreted as a monotonic or slowly varying drift. In this case, connecting successive sampled points produces an apparent slow trend, even though the underlying signal is purely oscillatory.

An oscillatory DM signal sampled at discrete times can therefore appear as a linear drift, a quadratic trend, or a general low-frequency systematic variation. From a single clock alone, it is impossible to distinguish whether such a drift originates from systematics or from a genuine physical signal.

For typical atomic clocks, the interrogation time is very short compared to the oscillation period of interest for ULDM. For example, in Rb and Cs fountain clocks, the interrogation time is $\sim 0.5-1\, {\rm s}$~\cite{Arvanitaki:2014faa}. By contrast, for ULDM in the mass range probed by clocks, the typical oscillation period can be extremely long, of the order of months. It means that only a small fraction of an oscillation is sampled over realistic measurement campaigns, enhancing the risk of confusing an oscillatory signal with slow drift.

The key limitation of a single-clock measurement is therefore the lack of a reference against which to compare. Without an independent clock, one cannot reliably separate common-mode systematics from genuine signals, and any observed drift remains ambiguous in origin. As a result, conclusions drawn from a single clock alone are inherently uncertain. This motivates the use of clock comparisons and networks of clocks, where correlated signals can be identified and systematics rejected.

\paragraph{Two co-located clocks:} A powerful strategy for improving sensitivity to variations of fundamental constants, while simultaneously mitigating systematic effects, is the use of two atomic clocks operated at the same physical location. Co-located clock comparisons form the baseline configuration for many precision metrology experiments and provide a controlled environment in which genuine physical signals can be cleanly separated from technical and environmental noise.

When two clocks are co-located, their frequencies can be compared directly and continuously in real time. The primary observable is typically the ratio of the two clock frequencies~\cite{Arvanitaki:2014faa},
\[
r(t) = \frac{\nu_1(t)}{\nu_2(t)} ,
\]
which is a dimensionless quantity and is highly sensitive to variations in fundamental constants. Many physical effects enter multiplicatively in the individual clock frequencies, so taking a ratio enhances sensitivity to genuine physical variations while suppressing absolute frequency drifts. At the same time, noise sources that affect both clocks in a similar way are strongly reduced in the ratio. For these reasons, direct frequency comparison between co-located clocks is among the most effective methods for probing temporal variations of fundamental constants.

Because both clocks experience the same environmental fluctuations, relative dephasing between them is strongly suppressed. This suppression of differential noise allows the clocks to remain phase coherent over long timescales, enabling long interrogation times and highly stable frequency comparisons. In practical terms, this translates into improved statistical sensitivity and a reduced Allan deviation at long averaging times, which is crucial for resolving weak, slowly varying signals such as those expected from ULDM-induced oscillations of fundamental constants.

Despite the shared environment, not all systematic effects cancel perfectly in practice. Residual systematics remain due to intrinsic differences between the two clocks, including species-specific sensitivities arising from different atomic or nuclear structures, as well as setup-dependent effects such as differences in trapping geometries, laser configurations, or interrogation schemes. Consequently, co-located clock comparisons significantly reduce but do not eliminate systematic uncertainties. Careful characterisation and control of these residual effects are therefore essential in high-precision measurements.

\paragraph{Clock comparison:} The frequency ratio of atomic transitions in two different atomic clocks, labeled $A$ and $B$, can be parameterised in terms of fundamental constants as
\[
\nu_{A/B} \;\propto\; \alpha^{k_\alpha}
\left( \frac{m_e}{m_p} \right)^{k_e}
\left( \frac{m_q}{\Lambda_{\mathrm{QCD}}} \right)^{k_q},
\]
where $\alpha$ is the fine-structure constant, $m_e$ and $m_p$ are the electron and proton masses, $m_q$ denotes the light quark mass, $\Lambda_{\mathrm{QCD}}$ is the QCD scale. The dimensionless coefficients $k_\alpha$, $k_e$, and $k_q$ are sensitivity coefficients that depend on the specific atomic transition under consideration~\cite{Flambaum:2006ip}. These coefficients quantify how strongly a given transition frequency responds to variations in the corresponding fundamental constants and are typically calculated using atomic and nuclear structure theory.

The observable fractional variation of the frequency ratio is therefore given by
\[
\frac{\delta (\nu_{A/B})}{\nu_{A/B}}
= \Delta k_\alpha \, \frac{\delta \alpha}{\alpha}
+ \Delta k_e \, \frac{\delta (m_e/m_p)}{(m_e/m_p)}
+ \Delta k_q \, \frac{\delta (m_q/\Lambda_{\mathrm{QCD}})}{(m_q/\Lambda_{\mathrm{QCD}})},
\]
where $\Delta k_i = k_i^{(A)} - k_i^{(B)}$ denotes the difference in sensitivity coefficients between clocks $A$ and $B$.

A non-zero signal in a clock comparison arises only if the two clocks have different sensitivity coefficients, ie, $\Delta k_i \neq 0$. If the sensitivities were identical, variations in fundamental constants would affect both clocks in the same way and cancel in the ratio.

Clock comparisons therefore, provide a powerful and selective probe of oscillatory variations of fundamental constants. By choosing atomic species and transitions with sufficiently different sensitivity coefficients, one can isolate specific combinations of constants and maximise sensitivity to new physics, such as ULDM backgrounds.

Using the effective low-energy parametrisation of quadratic ALP-induced variations described previously, the frequency comparison ratio  can be rewritten as~\cite{Bauer:2024hfv}
\[
\frac{\delta \nu_{A/B}}{\nu_{A/B}}
= k_\alpha \, \delta_\alpha(a)
+ k_e \, \delta_e(a)
- \left( k_e + 2 k_q \right) \delta_p(a)
+ k_q \, \delta_\pi(a),
\]
where $\delta$'s denote the shifts in the fine-structure constant, electron mass, quark mass, and hadronic masses and proportional to $a^2/f^2$. In the presence of an oscillating ULDM background, the ALP field behaves as a coherently oscillating classical field. For quadratic couplings, the fractional frequency shift scales as
\[
\frac{\delta \nu_{A/B}}{\nu_{A/B}} \;\propto\; a^2
= \frac{2 \rho_{\mathrm{DM}}}{m_a^2} \cos^2(m_a t)
= \frac{\rho_{\mathrm{DM}}}{m_a^2}
\left[ 1 + \cos(2 m_a t) \right],
\]
where $\rho_{\mathrm{DM}}$ and $m_a$ are the local dark matter energy density and the ALP mass, respectively. The signal therefore contains both a constant offset and an oscillatory component at angular frequency $\omega \simeq 2 m_a$.

\paragraph{Microwave and optical atomic clocks:} Atomic clocks can be broadly classified into microwave and optical clocks, according to the characteristic frequency of the reference transition that defines the clock. This distinction is central to clock performance, as the transition frequency and linewidth directly determine the achievable fractional stability and sensitivity to new physics.

A useful heuristic for understanding clock performance is that the achievable stability scales approximately as
\[
\text{clock stability} \;\propto\; \frac{\nu_0}{\Delta \nu},
\]
where \(\nu_0\) is the clock transition frequency and \(\Delta \nu\) is the linewidth of that transition.

This relation highlights two key points. First, higher transition frequencies naturally enhance clock stability. Second, narrower linewidths, corresponding to longer-lived excited states, further improve performance. Optical clocks benefit from both effects simultaneously, as they operate at much higher frequencies and often exploit extremely narrow transitions.

Microwave clocks operate using hyperfine transitions between ground-state sublevels, with characteristic frequencies in the GHz range. Well-known examples include rubidium (Rb) and caesium (Cs) clocks, which have played a central role in timekeeping for decades.

Typical properties of microwave clocks include transition frequencies of order \(\sim\mathrm{GHz}\), relatively long interrogation and integration times, and fractional frequency uncertainties at the level of $\frac{\delta \nu}{\nu} \sim 10^{-15}$.

A canonical example is the \(^{133}\mathrm{Cs}\) hyperfine transition at $\nu_{\mathrm{Cs}} = 9.19~\mathrm{GHz}$, which defines the SI second. This transition is sensitive not only to the fine-structure constant \(\alpha\), but also to nuclear magnetic moments and hadronic parameters, making microwave clocks particularly valuable for probing variations in mass ratios such as \(m_e/m_p\).

Optical clocks are based on electronic transitions with frequencies in the THz range, typically between long-lived or metastable atomic states. These clocks can be realised using neutral atoms trapped in optical lattices, or single ions confined in electromagnetic traps.

Representative systems include neutral atoms such as Sr and Al, as well as trapped ions such as Yb\(^+\) and Hg\(^+\). Optical clocks are characterised by transition frequencies of order \(\sim\mathrm{THz}\), extremely narrow natural linewidths, and fractional frequency uncertainties reaching $\frac{\delta \nu}{\nu} \sim 10^{-17}$, or even below in state-of-the-art experiments.

Because of their high transition frequencies and narrow linewidths, optical clocks currently define the state of the art in precision frequency metrology and provide unparalleled sensitivity to tiny perturbations of fundamental constants.

\begin{figure}[htb!]
    \centering
    \includegraphics[width=0.85\linewidth]{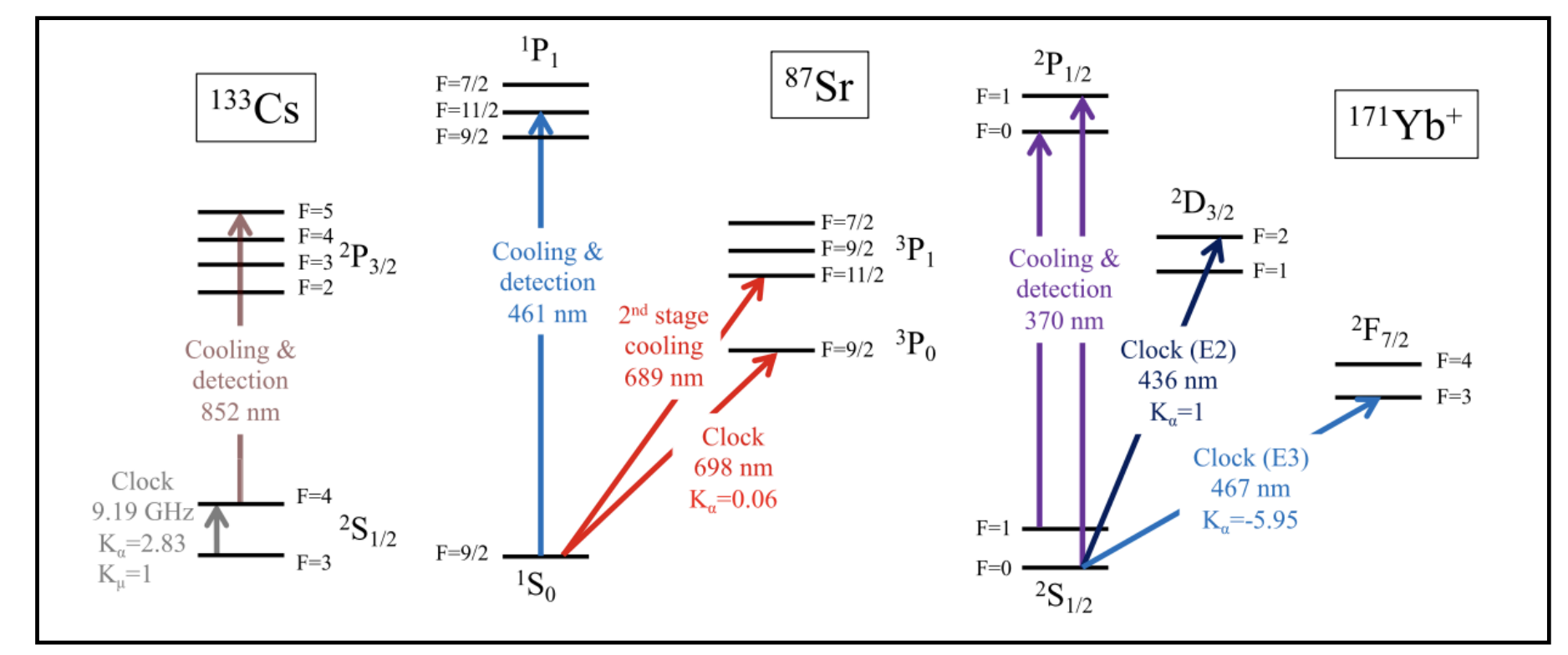}
    \caption{Atomic levels and clock transitions for representative microwave and optical atomic clock systems. Figure from Ref.~\cite{Barontini:2021mvu}.}
\end{figure}

\paragraph{Examples of clock transitions:} Concrete examples of commonly used clock systems illustrate the diversity of sensitivities available~\cite{Hees:2016gop,Sherrill:2023zah,Kobayashi_2022,Filzinger:2023zrs}:

\begin{itemize}
\item The \(^{133}\mathrm{Cs}\) microwave hyperfine transition at
\(9.19~\mathrm{GHz}\), with sensitivity coefficients
\(K_\alpha \simeq 2.83\) and \(K_\mu \simeq 1\).
\item The \(^{87}\mathrm{Sr}\) optical clock transition near \(698~\mathrm{nm}\),
which has reduced sensitivity to variations in \(\alpha\),
with \(K_\alpha \simeq 0.06\).
\item The \(^{171}\mathrm{Yb}^+\) optical electric-quadrupole (E2) transition at
\(436~\mathrm{nm}\), with \(K_\alpha \simeq 1\), and the electric-octupole (E3)
transition at \(467~\mathrm{nm}\), which exhibits enhanced sensitivity
\(K_\alpha \simeq -5.95\).
\end{itemize}

These examples demonstrate that different clock transitions respond very differently to variations in fundamental constants, even within the same atomic species.

\paragraph{Clock comparison tests impacting ULDM probe:} Different clock comparison strategies probe complementary directions in the landspace of fundamental constants. Comparisons between two microwave clocks~\cite{Hees:2016gop} are particularly powerful for detecting very low-frequency signals and phenomena with long coherence times, owing to their excellent long-term stability. Comparisons between two optical clocks~\cite{BACON:2020ubh}, by contrast, provide exceptional sensitivity to variations of the fine-structure constant, as optical transitions depend strongly on relativistic electronic effects. On the other hand, comparisons between optical and microwave clocks~\cite{Kobayashi_2022} offer broad sensitivity to multiple fundamental constants simultaneously, including both $\alpha$ and mass ratios, by combining the strengths of the two regimes.

This diversity of clock comparison strategies is essential for disentangling genuine new physics signals, from residual systematic effects. By exploiting the distinct sensitivity patterns of different clock pairs, one can both enhance discovery potential and perform consistency checks that help identify the underlying physical origin of any observed signal.

\paragraph{New developments in optical clocks:} Interactions of an oscillating DM field with QCD degrees of freedom can induce oscillations in nuclear properties. A particularly important example is a time-dependent variation of the nuclear charge radius~\cite{Banerjee:2023bjc}. This possibility opens a new and highly sensitive detection channel for ULDM using modern optical clocks, especially those based on heavy atoms and ions, where finite nuclear size effects are large and can be measured with exquisite precision.

In heavy atoms such as \(^{171}\mathrm{Yb}^+\), the total electronic energy receives a significant contribution from the \emph{field-shift energy}. This term arises because the nucleus has a finite spatial extent, so the electronic wavefunctions, especially those with substantial support near the nucleus, are sensitive to the detailed charge distribution rather than to a pointlike Coulomb potential.

To leading order, the field-shift contribution to the electronic energy can be written as
\[
E_{\mathrm{FS}} \;\simeq\; K_{\mathrm{FS}} \,\langle r_N^2 \rangle \;\propto\; A^{2/3},
\]
where \(\langle r_N^2 \rangle\) is the mean-square nuclear charge radius and \(A\) is the atomic mass number. The approximate scaling with \(A^{2/3}\) simply reflects the fact that nuclear radii grow with the size of the nucleus.

A direct consequence is that any variation of the nuclear charge radius induces a corresponding shift in atomic energy levels and therefore in atomic transition frequencies. This effect is strongly enhanced in heavy elements because the field shift is much larger than in light atoms.

\subparagraph{Effect of oscillating nuclear charge radius on  clock frequency ratios:} To understand how nuclear size variations enter clock observables, consider two atomic transitions with frequencies \(\nu_a\) and \(\nu_b\). The fractional
variation of the frequency ratio caused by a change in the nuclear charge radius
is
\[
\frac{\Delta(\nu_a/\nu_b)}{(\nu_a/\nu_b)}
\;=\;
K_{a,b}\,\frac{\Delta \langle r_N^2 \rangle}{\langle r_N^2 \rangle},
\]
where the sensitivity coefficient \(K_{a,b}\) is defined by
\[
K_{a,b}
\;\equiv\;
\frac{K_{\mathrm{FS}}^{\nu_a}\,\langle r_N^2 \rangle}{\nu_a}
\;-\;
\frac{K_{\mathrm{FS}}^{\nu_b}\,\langle r_N^2 \rangle}{\nu_b}.
\]
This form makes the key strategy transparent. Clock comparisons become highly sensitive to nuclear size variations when the two transitions have very different field-shift coefficients (or very different ratios
\(K_{\mathrm{FS}}^{\nu}/\nu\)), so that the difference defining \(K_{a,b}\) is large. In other words, one can strongly amplify sensitivity by comparing transitions whose electronic wavefunctions probe the nucleus in very different ways.

The nuclear charge radius is determined by the spatial distribution of protons inside the nucleus and by the characteristic inter-nucleon separation. Both are governed by QCD-scale physics. In particular, nuclear sizes and binding properties depend on hadronic quantities such as the pion decay constant, the pion mass, and pion-exchange forces, which set the range and strength of the nuclear interaction.

Parametrically, the fractional variation of the nuclear charge radius may be written as~\cite{Banerjee:2023bjc}
\[
\frac{\Delta \langle r_N^2 \rangle}{\langle r_N^2 \rangle}
\;\simeq\;
\alpha\,\frac{\Delta f_\pi}{f_\pi}
\;+\;
\beta\,\frac{\Delta m_\pi^2}{m_\pi^2}
\;\simeq\;
\alpha\,\frac{\Delta \Lambda_{\mathrm{QCD}}}{\Lambda_{\mathrm{QCD}}}
\;+\;
\beta\,\frac{\Delta m_\pi^2}{m_\pi^2},
\]
where \(\alpha\) and \(\beta\) are dimensionless coefficients encoding nuclear structure dependence.

An ULDM field that couples to QCD can induce time-dependent variations in quantities such as \(\Lambda_{\mathrm{QCD}}\) and \(m_\pi\), which in turn produce oscillations of \(\langle r_N^2 \rangle\). These nuclear oscillations are then imprinted as a time-dependent modulation of atomic transition frequencies measured by optical clocks. This effect has been shown to be particularly pronounced for the \(\mathrm{Yb}^+\) E\(_3\)/E\(_2\) transition, owing to its exceptional sensitivity to nuclear charge radius variations.~\cite{Banerjee:2023bjc}. 

\subparagraph{Enhanced sensitivity in \(\mathrm{Yb}^+\) ion clock transitions:} Among ionic systems, a particularly promising platform is provided by comparisons between clock transitions in the\(^{171}\mathrm{Yb}^+\) ion. In this system, the ratio of two optical transitions exhibits a markedly enhanced differential sensitivity to variations of the fine-structure constant. The relevant transitions are the electric-octupole (E\(_3\)) transition~\cite{Filzinger:2023zrs},
\[
(4f^{14}6s)^2S_{1/2}
\;\rightarrow\;
(4f^{13}6s^2)^2F_{7/2},
\]
and the electric-quadrupole (E\(_2\)) transition,
\[
(4f^{14}6s)^2S_{1/2}
\;\rightarrow\;
(4f^{14}5d)^2D_{3/2}.
\]
Because these two transitions have markedly different electronic structures, and benefit from long laser–ion coherence times with further gains expected from advanced interrogation schemes, the frequency ratio \(\mathrm{Yb}^+\) E\(_3\)/E\(_2\) is highly sensitive to oscillatory effects induced by ULDM. Ref.~\cite{Filzinger:2023zrs} has shown that the sensitivity of optical clock comparisons can be improved by several orders of magnitude relative to earlier experiments such as BACON~\cite{BACON:2020ubh}. This represents a major advance in optical clock stability and precision in modern implementations, establishing ${\rm Yb^+}$ ion clocks as the most sensitive optical systems currently available, second only to proposed nuclear clocks.

\subsubsection{Optical Cavities}

Optical cavities~\cite{PhysRevLett.123.031304,Antypas:2022asj} provide a complementary probe of oscillating dark matter through their sensitivity to variations of fundamental constants, in particular the fine-structure constant $\alpha$ and the electron mass $m_e$.

An optical cavity provides a frequency reference whose value is set by a macroscopic length scale -- the cavity length. The key point is that the length of a solid object, such as the spacer that defines an optical cavity, is not an independent parameter: it is ultimately determined by microscopic atomic length scales. Parametrically, the characteristic size of atoms is set by the Bohr radius,
\[
L_{\mathrm{solid}} \;\propto\; a_B \;=\; \frac{1}{m_e \alpha}.
\]
Therefore, if the fine-structure constant \(\alpha\) or the electron mass \(m_e\) varies in time, the equilibrium size of atoms shifts, and so does the equilibrium spacing of atoms in a solid. This induces a fractional change in the cavity length of the form
\[
\frac{\delta L_{\mathrm{solid}}}{L_{\mathrm{solid}}}
= -\frac{\delta m_e}{m_e} - \frac{\delta \alpha}{\alpha}.
\]
The negative signs simply reflect the inverse dependence of the Bohr radius on \(m_e\) and \(\alpha\): increasing either parameter shrinks the atomic length scale and hence the characteristic size of the solid.

The resonance frequencies of an optical cavity are determined by its physical length. For a given mode, one has the scaling
\[
\nu_{\mathrm{solid}} \;\propto\; \frac{1}{L_{\mathrm{solid}}}
\;\propto\; m_e \alpha.
\]
Consequently, any fractional change in the cavity length produces a fractional change in the cavity resonance frequency with the opposite sign. Using the scaling above, the induced fractional frequency shift can be written as
\[
\frac{\delta \nu_{\mathrm{solid}}}{\nu_{\mathrm{solid}}}
= \frac{\delta m_e}{m_e} + \frac{\delta \alpha}{\alpha}.
\]

\begin{figure}[htb!]
    \centering
    \includegraphics[width=0.55\linewidth]{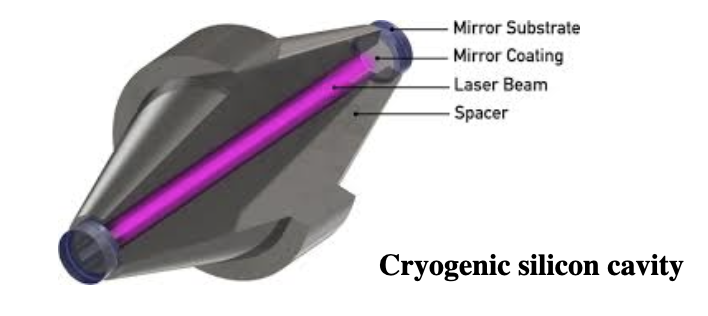}
    \caption{Schematic of a cryogenic Si optical cavity used for ultra-stable laser frequency references.}
\end{figure}

\paragraph{Comparison with atomic transitions:} It is useful to contrast cavity frequencies with purely atomic reference frequencies. Atomic electronic transition frequencies scale with the Rydberg energy,
\[
\nu_{\mathrm{atom}} \;\propto\; \mathrm{Ry} \;\propto\; m_e \alpha^2.
\]
Taking the ratio of an atomic transition frequency to a cavity resonance frequency therefore yields
\[
\frac{\nu_{\mathrm{atom}}}{\nu_{\mathrm{solid}}}
\;\propto\; \alpha.
\]
This particularly simple dependence is important: it shows that comparing an atomic transition to an optical cavity mode provides a clean probe of variations in the fine-structure constant, because the leading dependence on \(m_e\) cancels in the ratio. 

Therefore, experimentally, cavity frequencies can be compared either to atomic transition frequencies in optical clocks or to other cavity modes (in the optical or even microwave domain). Because these different references depend on fundamental constants in different ways, their comparison enables sensitive searches for oscillatory signals induced by ULDM.

\paragraph{Clock--cavity and cavity--cavity comparisons:} Clock--cavity and cavity--cavity frequency comparisons extend the reach of precision searches for ULDM by exploiting the different dependencies of atomic transitions and solid-state cavity resonances on fundamental constants. These systems probe complementary combinations of $\alpha$ and $m_e$, and access distinct DM mass ranges.

\subparagraph{Sr/Si comparison:} A prominent example is the comparison between a silicon (Si) optical cavity and an $^{87}$Sr optical lattice clock~\cite{Kennedy:2020bac}. In this case, the observable is the frequency ratio between the cavity resonance and the Sr clock transition. The Sr clock transition frequency scales approximately as
\[
\nu_{\mathrm{Sr}} \;\propto\; \alpha^{2.06} \, m_e,
\]
while the cavity reference frequency scales as $\nu_{\mathrm{cav}} \;\propto\; \alpha \, m_e$. As a result, the frequency ratio is primarily sensitive to variations of the fine-structure constant $\alpha$, with only a weak dependence on $m_e$. This comparison operates entirely in the optical domain, benefiting from high intrinsic frequency stability, narrow linewidths, and long coherent interrogation times.

Consequently, the Sr/Si system provides some of the strongest constraints on ultralight ALPs in the mass range
\[
m_a \;\approx\; 10^{-17} \;-\; 2\times10^{-16}~\mathrm{eV}.
\]

\subparagraph{H/Si comparison:} Another important configuration compares the reference frequency of a silicon optical cavity to a hydrogen maser (H maser)~\cite{Kennedy:2020bac}. The hyperfine transition frequency of the hydrogen maser scales as
\[
\nu_H \;\propto\; \alpha^4 \, m_e^2,
\]
which differs significantly from the cavity scaling $\nu_{\mathrm{cav}} \propto \alpha m_e$.

As a result, the H/Si comparison is simultaneously sensitive to variations in both $\alpha$ and $m_e$. This system operates in the microwave domain and probes
a different region of parameter space compared to purely optical setups.

\subparagraph{Cs/D$_2$ comparison:} At higher DM masses, comparisons involving electronic transitions in atoms against a laser cavity become relevant. A notable example is the Cs/D$_2$ comparison~\cite{Tretiak:2022ndx}, which compares measurements of electronic transitions between two states of $^{133}$Cs against a laser cavity reference. Sensitivity arises from variations in fundamental constants between the acoustic cut-off frequency of the cavity resonator, and the natural linewidth of the excited atomic state.

This configuration constrains ALPs in the mass range
\[
m_a \;\approx\; 4.6\times10^{-11} \;-\; 10^{-7}~\mathrm{eV}.
\]

\paragraph{Unequal time-delay cavities:} An alternative strategy for probing oscillating ULDM with optical cavities is to compare a single ultra-stable cavity to a time-delayed version of itself. Instead of comparing two independent frequency references (two clocks, or two cavities), the experiment constructs two copies of the \emph{same} optical signal separated by a fixed propagation time and then interferes them. This idea is realised experimentally in \emph{unequal time-delay interferometers}, such as DAMNED~\cite{Savalle:2020vgz}.

The motivation is straightforward. If the cavity frequency (or the effective optical path length) acquires a small oscillatory modulation due to ULDM then comparing the signal at time \(t\) to the same signal at an earlier time \(t-T_0\) naturally converts such modulations into an observable phase response. At the same time, fluctuations that are common to both copies of the signal can partially cancel, improving robustness against some sources of technical noise.

\subparagraph{Working principle:} The basic idea is to interfere the instantaneous cavity frequency \(\nu(t)\) with a copy of the same signal delayed by a fixed propagation time
\[
T_0 \;=\; \frac{L_0}{c_0},
\]
where \(L_0\) is the length of the delay line (typically an optical fibre) and \(c_0\) is the speed of light in vacuum.

Operationally, the light derived from the ultra-stable cavity output is split into two arms. One arm is measured directly and thus retains the instantaneous frequency \(\nu(t)\). The other arm propagates through the delay line and yields a time-shifted copy \(\nu(t-T_0)\). The two signals are then recombined in an unequal arm Mach--Zehnder interferometer. The measurable quantity is the phase difference accumulated between the two arms, which depends on how the cavity frequency and the effective propagation time fluctuate in time.

The phase difference between the delayed and non-delayed signals can be written as
\[
\Delta\phi(t)
= \omega_0 T_0
+ \omega_0 \int\limits_{t-T_0}^{t}
\left(
\frac{\Delta T(t')}{T_0}
+ \frac{\Delta \omega(t')}{\omega_0}
\right)\,dt'
+ \omega_0 T_0
\left(
\frac{\delta T}{T_0}
+ \frac{\delta\omega}{\omega_0}
\right)
\sin\!\left(\omega_\phi t - \frac{\omega_\phi T_0}{2}\right)
\sinc\!\left(\frac{\omega_\phi T_0}{2}\right).
\]

To interpret this expression term by term, first, \(\omega_0 T_0\) is the nominal static phase offset set by the carrier optical angular frequency \(\omega_0\) and the fixed delay \(T_0\). This is a large constant contribution that does not carry information about dark matter, but provides the baseline phase around which small fluctuations are measured.

Second, the integral
\[
\omega_0 \int\limits_{t-T_0}^{t}
\left(
\frac{\Delta T(t')}{T_0}
+ \frac{\Delta \omega(t')}{\omega_0}
\right)\,dt'
\]
accounts for \emph{slow} time-dependent variations. Here \(\Delta T(t)\) and \(\Delta\omega(t)\) represent slowly varying changes in the propagation time and in the optical angular frequency, respectively. These contributions capture long-term drifts and low-frequency technical noise sources (for example, temperature-dependent fibre length changes or slow cavity drift).

Third, the last term,
\[
\omega_0 T_0
\left(
\frac{\delta T}{T_0}
+ \frac{\delta\omega}{\omega_0}
\right)
\sin\!\left(\omega_\phi t - \frac{\omega_\phi T_0}{2}\right)
\sinc\!\left(\frac{\omega_\phi T_0}{2}\right),
\]
encodes the oscillatory contributions induced by ULDM. In this term, \(\delta T\) and \(\delta\omega\) denote oscillatory modulations of the delay time and the optical frequency, respectively. The angular frequency \(\omega_\phi\) corresponds to the oscillation frequency of the ULDM field.

A key feature of the response is the appearance of the factor \(\sinc(\omega_\phi T_0/2)\)). This factor arises because the interferometer compares the signal over a \emph{finite} delay interval \(T_0\), effectively averaging the oscillation over that time window. The result is a characteristic frequency-dependent sensitivity with suppression and nulls at specific values of \(\omega_\phi\), set by the condition that the sine wave averages to zero over the delay. These nulls are a distinctive experimental signature of the unequal
time-delay response.

\subparagraph{Sensitivity to dark matter:} In the presence of oscillating dark matter, variations of fundamental constants modify both the cavity resonance frequency (through the dependence of cavity and atomic length scales on \(\alpha\), \(m_e\), etc.) and the effective optical path length (through the propagation time of light in the delay line). In an unequal time-delay interferometer, the observable is constructed by comparing the cavity to its own past state. This comparison naturally suppresses common-mode noise associated with the absolute cavity frequency, because slow fluctuations that affect both \(\nu(t)\) and \(\nu(t-T_0)\) similarly can partially cancel.

At the same time, oscillatory signals with frequencies comparable to \(1/T_0\) are retained, because the signal at time \(t\) and the signal at time \(t-T_0\) differ significantly when the modulation period is not much longer than the delay. For this reason, unequal time-delay cavities are naturally sensitive to higher-frequency signals than conventional clock-clock or cavity-cavity comparisons that rely on long-term averaging. This extends the reach to axion (or ALP) masses well above those accessible in experiments whose sensitivity is limited by long averaging times and low-frequency noise floors. The DAMNED experiment implements this concept using an ultra-stable optical cavity, long fibre delay lines, and a high-bandwidth photonic readout chain~\cite{Savalle:2020vgz}.
. 

\subsubsection{Coverage at Low Frequencies}

\begin{figure}[htb!]
    \centering
    \includegraphics[width=0.8\linewidth]{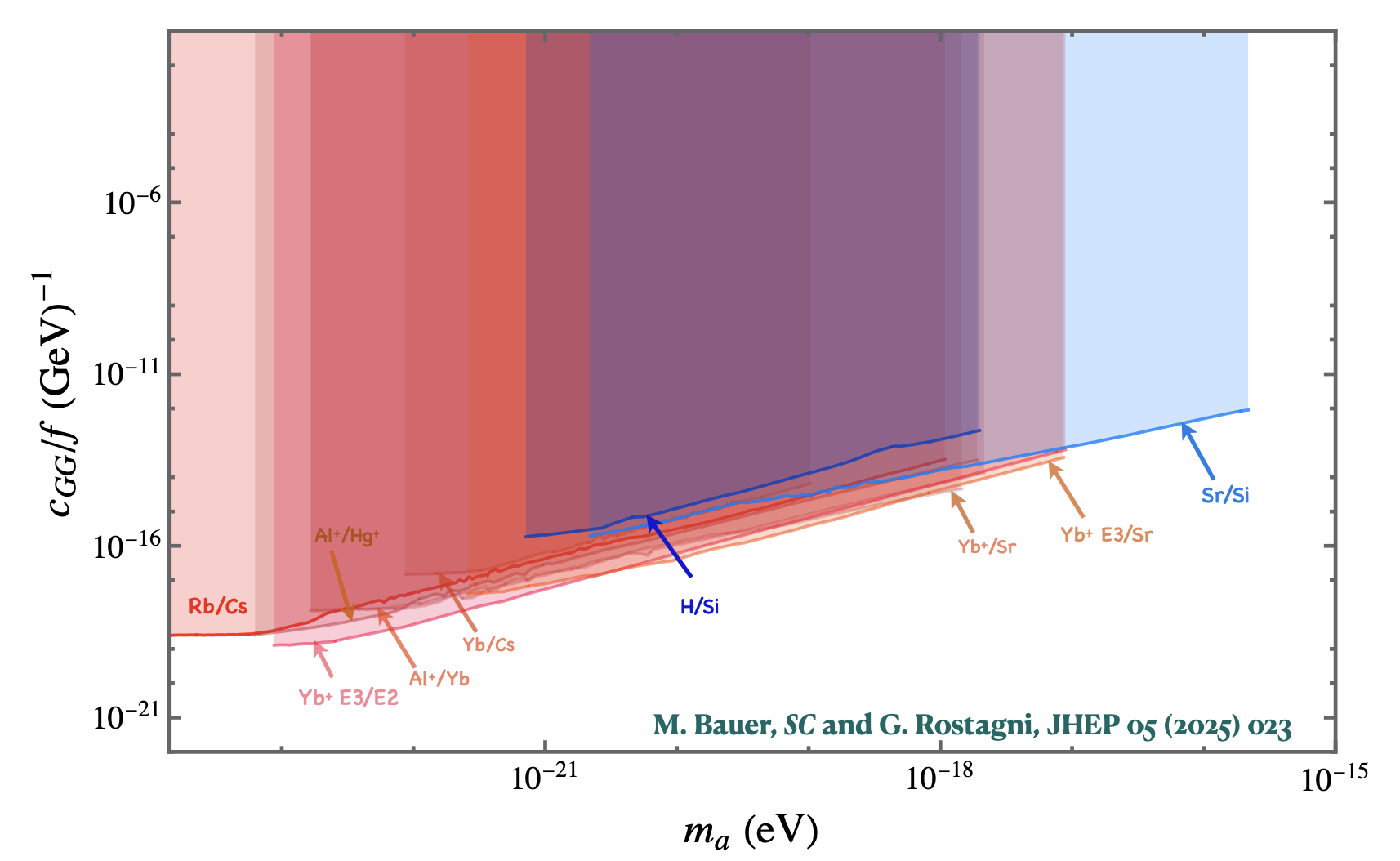}
    \caption{Sensitivity of atomic clock and clock–cavity comparisons at low frequencies to ALP–gluon couplings, accounting for quadratic interactions generated radiatively at low energies~\cite{Bauer:2024hfv}.}
    \label{fig:lowcoverage}
\end{figure}

\cref{fig:lowcoverage} summarises the sensitivity of clock--clock, clock--cavity, and cavity--cavity comparisons to ultralight ALPs in the low-frequency (small mass) regime. The horizontal axis corresponds to the axion mass $m_a$, while the vertical axis shows the effective coupling $c_{GG}/f$ inferred from the absence of a signal.

At very low masses, the axion-induced signal varies slowly compared to the interrogation cycle of atomic clocks. In this regime, the sensitivity is ultimately limited by the total duration of the experiment and by long-term frequency stability rather than by short-term noise.

Comparisons such as Rb/Cs provide excellent sensitivity at the lowest frequencies, corresponding to axion masses
\[
m_a \;\lesssim\; 10^{-20}~\mathrm{eV}.
\]
This is a direct consequence of their long operational history and extended datasets, which allow sensitivity to oscillation periods comparable to months or longer. However, their overall reach in coupling strength is limited by the intrinsic stability of microwave transitions.

Hybrid systems, such as H/Si, bridge the gap between microwave and optical experiments. Owing to their sensitivity to both $\alpha$ and $m_e$, these comparisons improve the reach at intermediate masses
\[
m_a \;\sim\; 10^{-21} \text{--} 10^{-18}~\mathrm{eV},
\]
where purely microwave clocks begin to lose sensitivity.

Optical clock comparisons (e.g.\ Yb/Sr, Yb$^+$ E$_3$/Sr, and Sr/Si cavity-clock systems) dominate the sensitivity at higher low-frequency masses,
\[
m_a \;\gtrsim\; 10^{-18}~\mathrm{eV},
\]
benefiting from narrow linewidths and superior fractional frequency stability. In particular, cavity-based comparisons such as Sr/Si provide the strongest constraints in the upper part of the low-frequency window.

Together, these techniques provide continuous coverage over several decades in $m_a$, demonstrating that precision frequency metrology is uniquely suited for probing ULDM with oscillation periods ranging from seconds to years.

\subsubsection{Optical Interferometers}

A two-arm laser interferometer is a precision instrument designed to detect very small changes in the \emph{difference} of optical path lengths between its two arms. The observable is not the absolute arm length, but the differential phase accumulated by the laser fields propagating along the two paths.

\begin{figure}[htb!]
    \centering
    \includegraphics[width=0.6\linewidth]{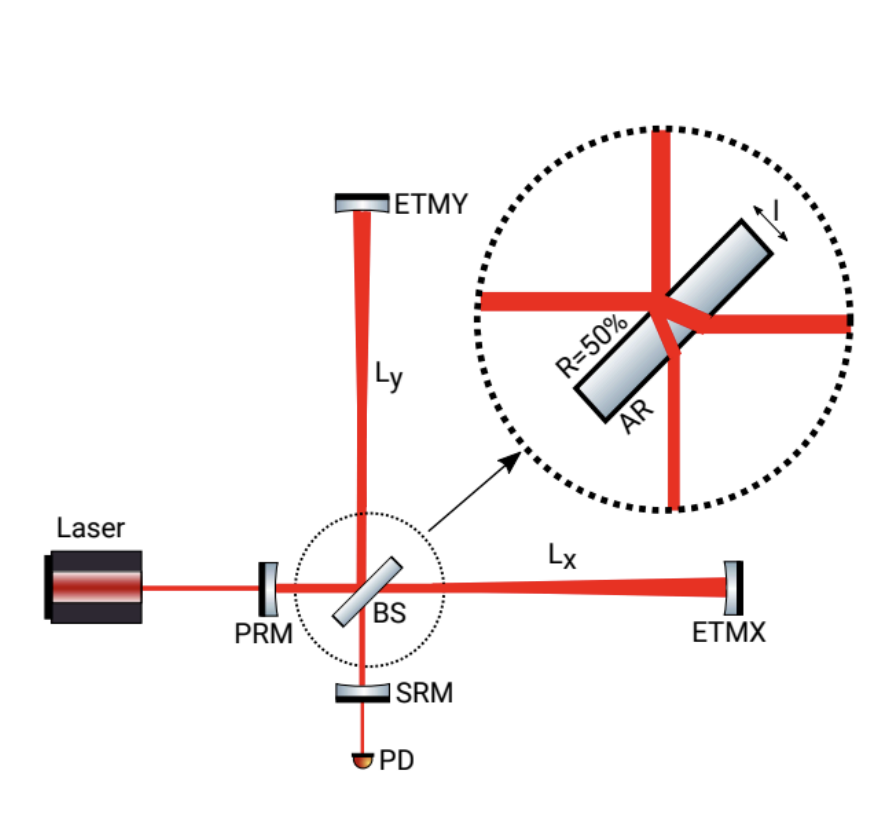}
    \caption{Interferometric layout used in precision laser interferometers such as GW detectors. Figure from Ref.~\cite{Grote:2019uvn}.}
\end{figure}

A monochromatic laser beam is split at a beam splitter into two orthogonal arms, commonly denoted $x$ and $y$, with optical path lengths $L_x$ and $L_y$, respectively. After reflection from the end mirrors, the two beams are recombined at the beam splitter and directed toward a photodetector. The detected signal depends on the phase difference
\[
\Delta\phi \;=\; \frac{2\pi}{\lambda}\,(L_x - L_y),
\]
so that any change in the differential optical path length produces a measurable change in the output power.

In practice, the two arms of large-scale interferometers are constructed to be nearly equal in length in order to suppress common-mode noise. However, the beam splitter itself introduces a geometric asymmetry: it defines the splitting and recombination point of the two arms and fixes the relative orientation of the optical paths. Even when $L_x \simeq L_y$, the interferometer is therefore sensitive to differential effects associated with the beam splitter and the arm mirrors.

In precision interferometers, such as those used for GW detection, the beam splitter and arm mirrors are freely suspended. If fundamental constants vary in time, this induces small, time-dependent changes in the physical dimensions of the optical components and in the effective refractive indices of the materials. As a result, the optical path lengths of the two arms can change differently and the beam splitter itself can undergo tiny size variations about its centre-of-mass.

These effects lead to a time-dependent displacement of the main reflecting surface that splits and recombines the laser beam.

A freely suspended beam splitter that experiences oscillatory size changes will move back and forth relative to the arm mirrors. This motion shifts the effective optical path lengths seen by the two arms in opposite directions, generating a non-zero differential phase shift. Consequently, oscillations in fundamental constants such as those induced by an ULDM field can be converted into a measurable interferometric signal~\cite{Stadnik:2014tta}.

\paragraph{Connection to large-scale interferometers:} Modern Michelson interferometers with power-recycling and signal-recycling mirrors (e.g.\ those employed in GW detectors) enhance this basic principle by increasing the effective optical path length and circulating power~\cite{Grote:2019uvn,DeRocco:2018jwe}. The sensitivity to differential optical path changes is therefore extremely high, making such interferometers powerful probes of tiny length variations and, by
extension, of time-dependent fundamental constants.

In summary, optical interferometers measure differential optical path length variations between two nominally equal arms. When the optical components are freely suspended, time-varying changes in fundamental constants can induce relative motion of the beam splitter and mirrors, producing a phase difference that constitutes the
experimental signal.

\paragraph{Interferometric response to variations of fundamental constants:} Variations of fundamental constants induced by an oscillating ULDM field lead to measurable changes in the physical dimensions and optical properties of interferometer components. In particular, variations in the fine-structure constant $\alpha$ and the electron mass $m_e$ modify both the length of solid materials and their refractive indices.

For a solid material, the fractional change in length $\ell$ is given by
\begin{equation}
\frac{\delta \ell}{\ell}
=
- \left(
\frac{\delta \alpha}{\alpha}
+
\frac{\delta m_e}{m_e}
\right),
\end{equation}
reflecting the dependence of the Bohr radius on $\alpha$ and $m_e$. In addition, the refractive index $n$ of the material also varies according to\footnote{The numerical prefactor in \cref{eq:ligon}  is evaluated for fused silica at interferometer wavelengths relevant to LIGO; other materials yield different coefficients.}
\begin{equation}
\frac{\delta n}{n}
=
-5 \times 10^{-3}
\left(
2 \frac{\delta \alpha}{\alpha}
+
\frac{\delta m_e}{m_e}
\right).
\label{eq:ligon}
\end{equation}
In a Michelson-type interferometer, the observable is the differential optical path length between the two arms, $L_x$ and $L_y$. The variation of this quantity can be written as
\begin{equation}
\delta (L_x - L_y)
=
\sqrt{2}
\left[
\left( n - \frac{1}{2} \right) \delta \ell
+
\ell\, \delta n
\right]
\;\simeq\;
n\,\ell \, \bigl[ \delta_\alpha(a) + \delta_e(a) \bigr],
\end{equation}
where the approximation highlights that the dominant contribution arises from the oscillatory variations of $\alpha$ and $m_e$ induced by the DM field $a$.

\paragraph{Examples:} GEO\,600~\cite{Vermeulen:2021epa} is a modified Michelson interferometer in which the differential strain is measured as a function of frequency and used to set bounds on ALP couplings. The entire optimal frequency band of the detector, approximately $100~\mathrm{Hz}$ to $10~\mathrm{kHz}$, lies below the fundamental
frequency of the longitudinal oscillation mode of the interferometer, $\sim 37~\mathrm{kHz}$. As a result, GEO\,600 is sensitive to ALP masses in the range
\[
m_a \;\simeq\; 10^{-13} \text{--} 10^{-11}~\mathrm{eV}.
\]
Oscillations of fundamental constants can also be probed using Fabry--Perot interferometers such as LIGO~\cite{Morisaki:2018htj,Gottel:2024cfj}. The methodology is similar to that employed for GEO\,600; however, in LIGO the sensitivity is attenuated by the finite arm-cavity finesse, which is of order $\mathcal{O}(100)$. There is an additional contribution to
$\delta (L_x - L_y)$ arising from thickness variations of the mirrors mounted in the two arm cavities. This effect is subleading, since
\[
\delta (L_x - L_y) \;\propto\; \Delta t \;\sim\; 80~\mu\mathrm{m}.
\]
Current LIGO observations (LIGO--O3) have been used to set limits on ALP couplings in the mass range~\cite{Gottel:2024cfj,Fukusumi:2023kqd}
\[
m_a \;\simeq\; 10^{-14} \text{--} 10^{-11}~\mathrm{eV}.
\]

\subsubsection{Mechanical Resonators}

Mechanical resonators provide a complementary probe of oscillating dark matter through their sensitivity to time-dependent mechanical strain in macroscopic solid objects~\cite{Arvanitaki:2015iga,Manley:2019vxy}. Conceptually, they are closely related to optical cavities, but operate in the acoustic or elastic domain rather than the optical one.

A macroscopic elastic body consists of a large number of atoms whose equilibrium separation is set by the atomic length scale, which in turn is determined by the Bohr radius. Oscillations of fundamental constants , induced by an oscillating ALP field $a(t)$, therefore lead to coherent, time-dependent variations of atomic sizes. This manifests as a fractional change in the length of the solid, i.e., a mechanical strain.

The induced strain can be written as
\begin{equation}
h(t)
\;\equiv\;
\frac{\delta \ell(t)}{\ell}
\;=\;
- \left[ \delta_\alpha(a) + \delta_e(a) \right],
\end{equation}
where $\delta_\alpha(a)$ and $\delta_e(a)$ denote the ALP-induced fractional variations of the fine-structure constant $\alpha$ and the electron mass $m_e$, respectively. The minus sign reflects the fact that an increase in $\alpha$ or $m_e$ reduces the Bohr radius and hence the equilibrium interatomic spacing.

Mechanical resonators support a discrete set of acoustic or elastic normal modes, with characteristic resonance frequencies determined by the geometry and elastic properties of the body. If one of these acoustic modes has a resonance frequency $\omega_{\rm res}$ tuned close to the oscillation frequency (corresponding to a quadratically coupled ALP ULDM) of the strain signal,
\begin{equation}
\omega_{\rm res} \;\simeq\; 2 m_a,
\end{equation}
the response of the resonator is resonantly enhanced. In this regime, the small ALP-induced strain acts as a periodic driving force, leading to a coherent build-up of mechanical oscillations limited by the quality factor $Q$ of the resonator.

Mechanical resonators thus act as narrowband, resonantly enhanced detectors for ULDM. Compared to optical cavities, they typically operate at much lower frequencies (from Hz to MHz, depending on the device), making them particularly well suited for probing ALP masses corresponding to twice the mechanical resonance frequency. Their sensitivity is controlled by the achievable  factor, thermal noise, and readout precision, while the signal is directly linked to the same combination of fundamental constant variations that governs optical-cavity length changes.

\paragraph{AURIGA detector:} AURIGA~\cite{Branca:2016rez} is a cryogenic resonant-mass detector based on a macroscopic elastic bar of length $L_{\rm bar} \sim \mathcal{O}(\mathrm{m})$, operated at cryogenic temperatures in order to suppress thermal noise and achieve high mechanical quality factors.

The detector functions as a narrowband mechanical resonator, with optimal sensitivity concentrated in a small frequency window around the fundamental longitudinal acoustic mode of the bar.

\begin{figure}[htb!]
    \centering
    \includegraphics[width=0.75\linewidth]{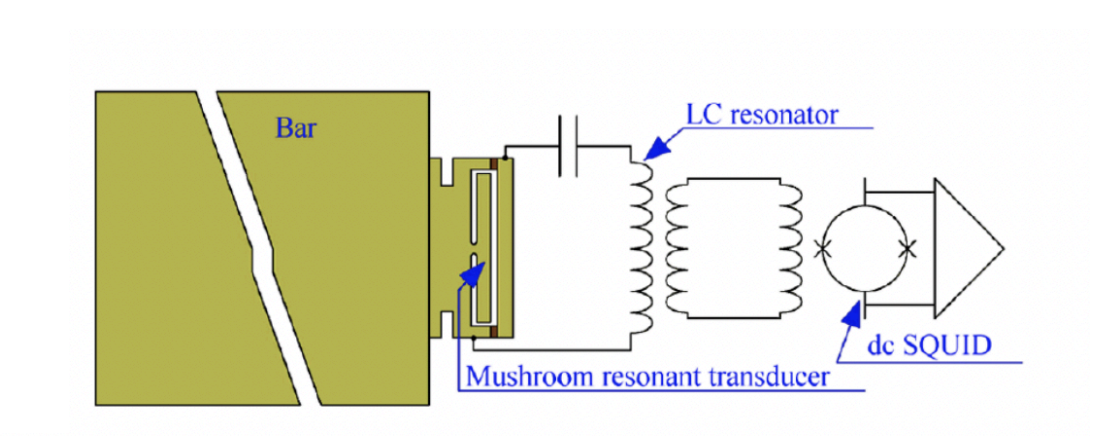}
    \caption{Schematic of AURIGA, illustrating mechanical excitation and readout via an electromechanical transducer and SQUID amplifier. Figure from Ref.~\cite{Branca:2016rez}.}
    \label{fig:auriga}
\end{figure}

\begin{figure}[htb!]
    \centering
    \includegraphics[width=0.65\linewidth]{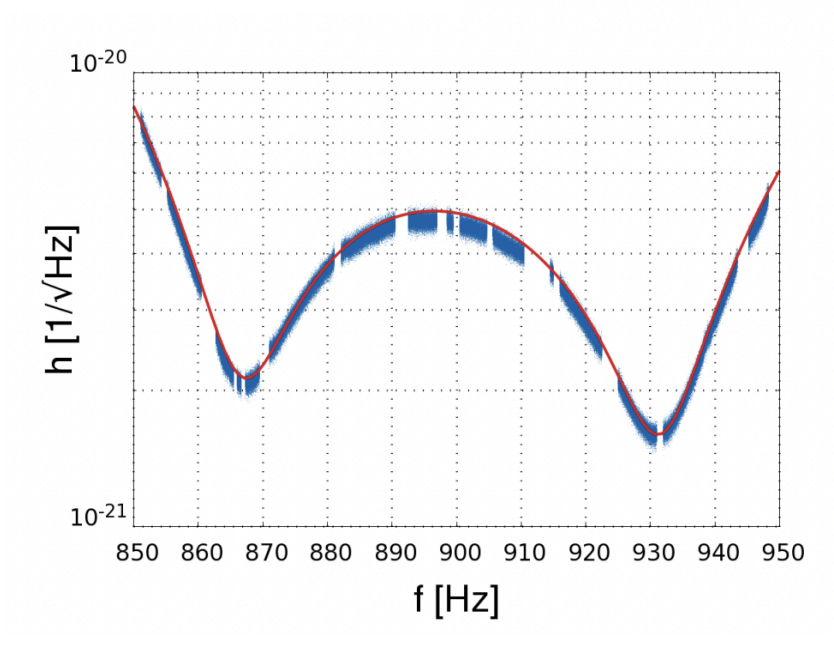}
    \caption{Strain response of AURIGA, showing the narrowband sensitivity around the mechanical resonance. Figure from Ref.~\cite{Branca:2016rez}.}
    \label{fig:aurigaresponse}
\end{figure}

AURIGA is sensitive over a narrow frequency band,
\[
f \simeq 850\text{--}950~\mathrm{Hz},
\]
which directly maps onto the ALP mass window through the quadratic coupling signal frequency $\omega_{\rm sig} \simeq 2 m_a$. This corresponds to an ALP mass range
\[
m_a \simeq 1.88\text{--}1.94~\mathrm{peV}.
\]
The mechanical motion of the bar is transduced via a resonant mechanical ``mushroom'' transducer, which converts the bar displacement into an electrical signal. This signal is then coupled to an LC resonator and read out using a dc~SQUID amplifier, providing ultra-sensitive displacement and strain measurements. The detector sensitivity is commonly expressed in terms of the strain noise spectral density, $h(f) \;\; [\mathrm{Hz}^{-1/2}]$, which exhibits a resonant enhancement within the narrow operational bandwidth. The sensitivity curve in \cref{fig:aurigaresponse} shows a characteristic minimum near the resonance frequency, with degraded sensitivity away from resonance, reflecting the narrowband nature of the instrument.

\subsubsection{Nuclear Clock: \texorpdfstring{$^{229}$Th}{Th-229}}

An especially promising platform for probing ULDM oscillation is provided by the nuclear clock based on
the isotope $^{229}\mathrm{Th}$~\cite{Seiferle:2019fbe,Caputo:2024doz}. The nucleus $^{229}\mathrm{Th}$ possesses an excited isomeric state, $^{229\mathrm{m}}\mathrm{Th}$, whose excitation energy is exceptionally low on nuclear scales, at the level of a few eVs. The currently measured
energy of the isomeric transition is~\cite{Sikorsky:2020peq}
\[
E_{\mathrm{iso}} \simeq 8.19(10)\ \mathrm{eV},
\]
corresponding to a transition wavelength of $\lambda \simeq 148.7\ \mathrm{nm}$, with an estimated lifetime of order $\tau \sim 5 \times 10^{3}\ \mathrm{s}$. This makes it the only known nuclear transition that is directly accessible to laser excitation and high-precision optical spectroscopy.

Typical nuclear binding energies are of order $E_{\mathrm{nuc}} \sim \mathcal{O}(\mathrm{MeV})$, while the isomeric transition energy is of order eV. The relative smallness of the transition energy compared to the characteristic nuclear scale can be expressed as
\[
\frac{E_{\mathrm{iso}}}{E_{\mathrm{nuc}}}
\;\sim\;
\frac{8\ \mathrm{eV}}{1\ \mathrm{MeV}}
\;\sim\; 10^{-5}.
\]
As a result, small fractional changes in the underlying nuclear energies can lead to enormously enhanced fractional shifts in the transition frequency.

\paragraph{Sensitivity to fundamental constants:} The isomeric transition energy arises from a delicate cancellation between large contributions to the nuclear binding energy, including Coulomb and strong interaction effects. Variations in fundamental parameters such as the fine-structure constant $\alpha$, light quark masses and the QCD scale $\Lambda_{\mathrm{QCD}}$ modify these large contributions differently in the ground and isomeric states. Because the observed transition energy is the small difference of two large quantities, the resulting sensitivity coefficients are strongly enhanced.

This leads to exceptionally large effective sensitivity coefficients, schematically,
\[
\frac{\delta \nu}{\nu}
\;=\;
k_\alpha \frac{\delta \alpha}{\alpha}
\;+\;
k_q \frac{\delta m_q}{m_q}
\;+\; \dots
\qquad
\text{with}
\quad
k_\alpha,\; k_q \sim \mathcal{O}(10^{4}\text{--}10^{5})~\cite{Caputo:2024doz},
\]
far exceeding those of atomic electronic transitions. Owing to these enormous enhancement factors,   $^{229}\mathrm{Th}$ would exhibit correspondingly large fractional frequency modulations, making it a uniquely powerful next-generation tool for testing fundamental physics and searching for ULDM.

\subsubsection{Atom Interferometers}

Another powerful next generation experiment are atom interferometers~\cite{Zhao:2021tie,Buchmueller:2023nll,Arvanitaki:2016fyj} which provide a complementary and highly versatile probe of ULDM by exploiting the phase coherence of matter waves and their sensitivity to time-varying atomic transition frequencies.

Atomic interferometers measure the relative phase accumulated between two spatially and temporally separated atomic wave packets that are coherently split and recombined using laser pulses (beam splitters and mirrors). The measured signal is the phase difference between the two interferometric paths, which arises from differences in the internal and external evolution of the atom along each path.

In the presence of an oscillating ULDM background, fundamental constants acquire a time-dependent component. This induces an oscillatory shift in the atomic electronic transition frequency,
\[
\omega_A(t,x) = \omega_A + \delta\omega_A(a),
\]
where the unperturbed transition frequency scales as
\[
\omega_A \propto m_e \alpha^{2+\xi},
\]
with $\xi$ encoding relativistic and many-body corrections specific to the atomic transition.

The fractional modulation of the transition frequency is therefore
\[
\frac{\delta\omega_A(a)}{\omega_A}
=
\delta_e(a) + (2+\xi)\,\delta_\alpha(a).
\]
For quadratic couplings of an ALP ULDM to Standard Model fields, this modulation is oscillatory,
\[
\frac{\delta\omega_A(a)}{\omega_A}
\simeq
\left[\delta_e + (2+\xi)\delta_\alpha\right]
\frac{\rho_{\rm DM}}{m_a^2 f^2}
\cos(2\omega_a t)
\;\equiv\;
\bar{\omega}_A \cos(2\omega_a t),
\]
where $\omega_a = m_a$ is the ALP Compton frequency.

\paragraph{Signal accumulation:} The phase accumulated by an atom while it is in the excited internal state between times $t_1$ and $t_2$ is given by
\[
\Phi^{t_2}_{t_1}
=
\int\limits_{t_1}^{t_2} \delta\omega_A(a)\, dt .
\]
In an atomic interferometer, the total signal phase is obtained by summing the contributions from all interferometric paths during which the atom occupies the excited state. Paths that remain in the ground state do not contribute to this internal-state phase shift.

The sensitivity of an atomic interferometer is maximised when the period of the ULDM-induced oscillation matches the total duration of the interferometric pulse sequence. In this regime, the oscillatory frequency shift adds coherently over the interferometer time, leading to an enhanced phase signal.

\paragraph{Gradiometers:} Gradiometers~\cite{Badurina:2021lwr} are based on the correlated operation of two or more interferometers separated by a finite spatial baseline. By forming a
differential signal between the interferometers, it is possible to cancel common-mode noise sources, most importantly the laser phase noise that limits single-interferometer sensitivity. This configuration is therefore especially well suited for probing ultralight dark matter signals that are spatially coherent over macroscopic distances.

Consider two atomic interferometers (AIs) separated by a distance $\Delta r$ along a common laser axis, sharing the same interrogation laser. Each interferometer individually accumulates a phase that contains both signal and noise contributions. By taking the difference of the two measured phases, the common laser phase noise cancels, while any spatially varying signal, such as one induced by a DM field, survives.

For an oscillating ALP background with Compton frequency $\omega_a = m_a$, the differential signal phase between two interferometers can be written as~\cite{Bauer:2024hfv}
\[
\Phi_s
=
\frac{4\,\bar{\omega}_A}{\omega_a}
\frac{\Delta r}{L}
\sin(\omega_a n L)\,
\sin(\omega_a T)\,
\sin\!\left[\omega_a\left(T + (n-1)L\right)\right].
\]
where $\bar{\omega}_A$ is the amplitude of the DM-induced oscillation of the atomic transition frequency, $\Delta r$ is the separation between the two atomic interferometers,  $L$ is the baseline length of the interferometer arms,  $T$ is the interrogation time, and  $n$ labels the number of light pulses in the interferometric sequence.

This expression explicitly shows how the signal depends on both temporal parameters (through $T$) and spatial parameters (through $L$ and $\Delta r$), reflecting the wave nature of the ULDM field.

In the limit where the geometry and timing are optimised, the signal phase can be approximated by
\[
\Phi_s
=
4\,\bar{\omega}_a\, n\, \Delta r \, \sin^2(m_a T),
\]
which highlights the key parametric dependences, such as linear scaling with the interferometer separation $\Delta r$, enhancement with the number of pulses $n$ and resonant sensitivity when the interrogation time $T$ is matched to the DM oscillation period.

A longer baseline directly increases the signal amplitude and therefore the sensitivity of the gradiometer. However, extending the baseline also introduces additional challenges such as the gravity-gradient noise becoming increasingly important at low frequencies, and therefore can limit sensitivity. Also, environmental and seismic noise must be carefully controlled, particularly for terrestrial long-baseline setups.

Several gradiometer concepts have been proposed and are under active
development~\cite{Badurina:2019hst,Abe_2021,AEDGE:2019nxb}:
\begin{itemize}
  \item \textbf{Compact gradiometers}: AION-10 and MAGIS, optimised for laboratory
        or near-term deployment with baselines of order tens of meters.
  \item \textbf{Long-baseline configurations}: ground-based setups with
        baselines of $\mathcal{O}(100\,\mathrm{m})$ up to kilometer scales,
        targeting enhanced sensitivity to lower DM frequencies.
  \item \textbf{Space-based gradiometers}: AEDGE, which exploits long baselines
        and low-noise space environments to access ultra-low frequency regimes.
\end{itemize}

\subsubsection{The Landscape of the Present and the Near Future }

Current precision experiments already probe an extensive region of ULDM parameter space, spanning many tens of orders of magnitude in mass and coupling strength, with sensitivity driven by advances in precision metrology, interferometry, and resonant mechanical systems.

\begin{figure}[htb!]
    \centering
    \includegraphics[width=0.85\linewidth]{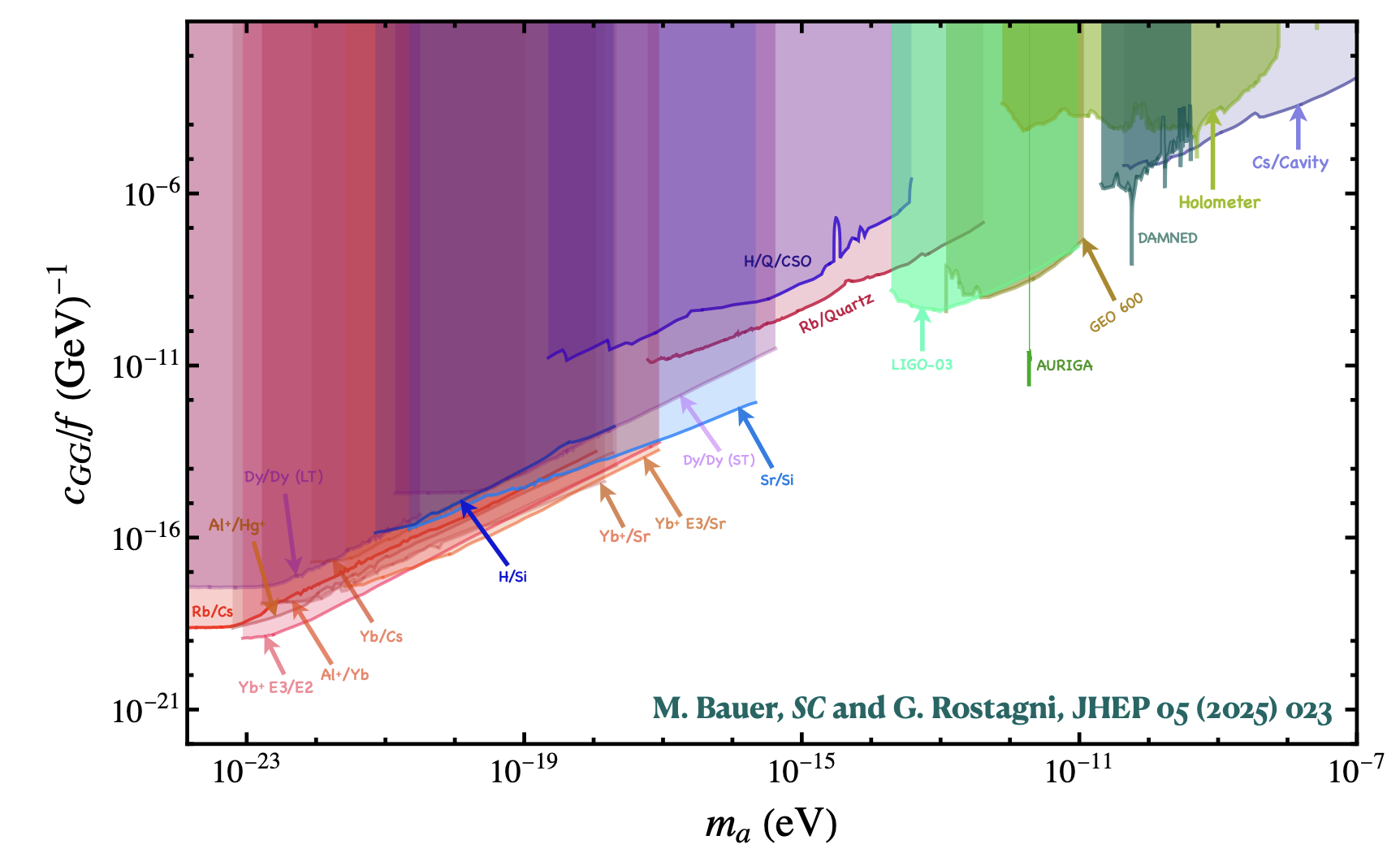}
    \caption{Present experimental coverage of quadratic ALP couplings induced radiatively by ALP-gluon interactions~\cite{Bauer:2024hfv}.}
    \label{fig:landscapecurrent}
\end{figure}

\cref{fig:landscapecurrent} summarises the current experimental landscape of constraints on ALP with quadratic couplings to SM fields. The results are presented in the plane spanned by the ALP mass \(m_a\) and the effective gluonic coupling \(c_{GG}/f\), illustrating how different experimental techniques probe complementary regions of parameter space.

Each experiment is sensitive to oscillatory signals within a characteristic frequency window, which maps directly onto a specific ALP mass range through the relation \(\omega \simeq 2 m_a\). As a result, the reach of a given experiment in mass is determined primarily by the timescales over which it can coherently
probe oscillations.

\paragraph{Low-mass regime: atomic clocks:} At very low ALP masses,
\[
m_a \lesssim 10^{-19}\text{--}10^{-18}~\mathrm{eV},
\]
the strongest constraints are provided by atomic clock comparisons. These include microwave clock comparisons such as Rb/Cs systems, optical clock comparisons such as Al\(^+\)/Hg\(^+\), Al\(^+\)/Yb, and Yb/Cs, as well as ultra-precise optical clock networks that monitor long-term frequency stability. Because these experiments operate with extremely long integration times, they are sensitive to ultra-low frequency oscillations of fundamental constants, leading to particularly strong bounds in the ultralight ALP mass regime.

\paragraph{Intermediate masses: clock--cavity comparisons:} In the intermediate mass range,
\[
m_a \sim 10^{-18}\text{--}10^{-15}~\mathrm{eV},
\]
the dominant constraints arise from clock--cavity and cavity--cavity comparison experiments. Representative examples include comparisons between optical lattice clocks and ultrastable optical cavities, such as Sr/Si systems, as well as comparisons involving hydrogen spectroscopy and silicon cavities that exploit different functional dependences on fundamental constants. These experiments benefit from the exceptional frequency stability achievable in the optical
domain and provide some of the most stringent constraints in this mass window.

\paragraph{Higher masses: interferometers and resonant detectors:} At larger ALP masses,
\[
m_a \gtrsim 10^{-14}~\mathrm{eV},
\]
experimental sensitivity shifts toward techniques capable of probing higher frequencies. This includes laser interferometers such as GEO--600 and the LIGO~O3 run, unequal-delay cavity interferometers such as DAMNED, resonant-mass detectors including AURIGA, and broadband interferometric setups such as the Holometer. These experiments are sensitive to mechanical strain or optical path length variations induced by oscillations of fundamental constants, allowing them to probe shorter oscillation periods than clock-based
measurements.


\begin{figure}[htb!]
    \centering
    \includegraphics[width=0.85\linewidth]{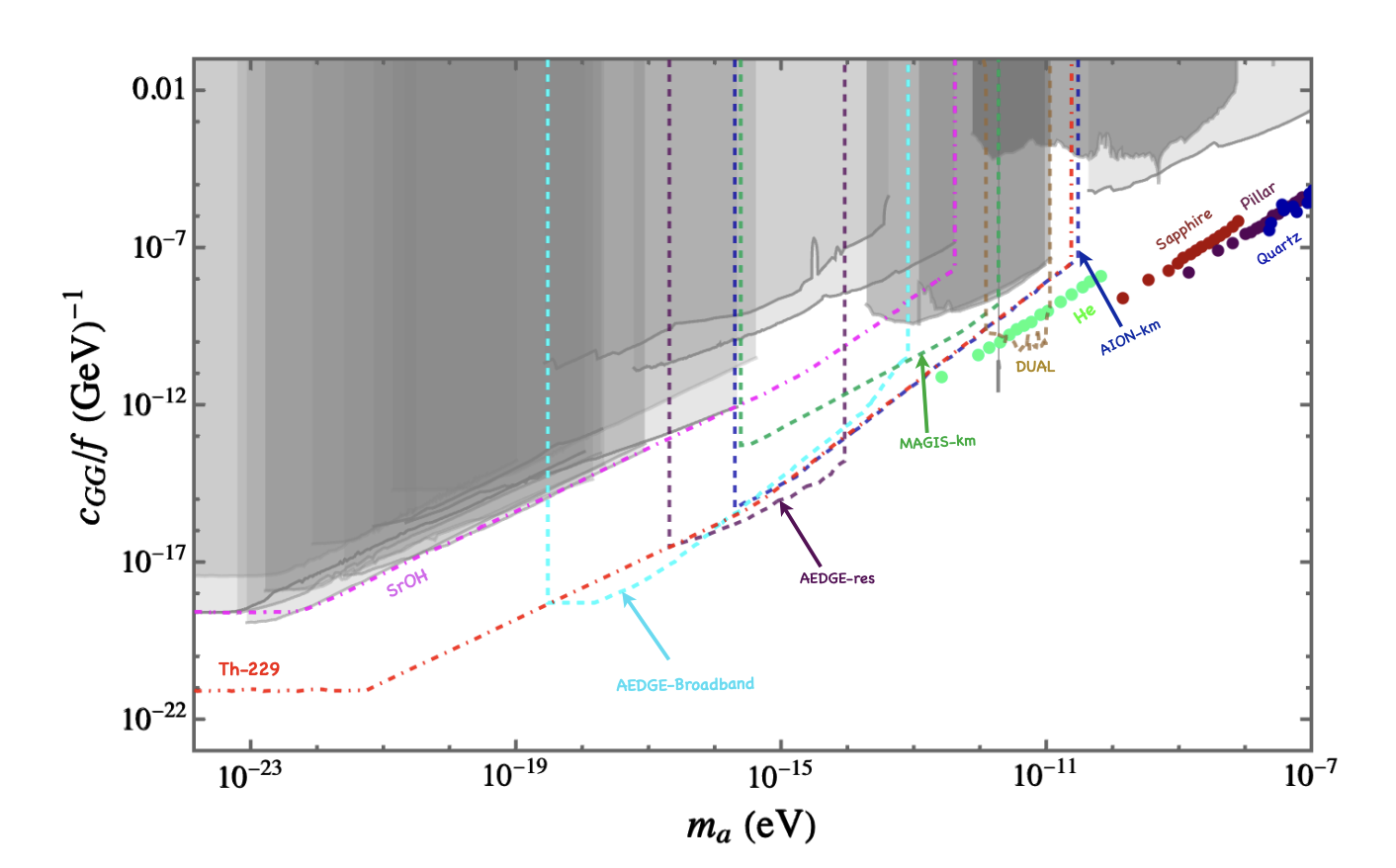}
    \caption{Future projections of experimental reach to probe quadratic ALP couplings induced radiatively by ALP-gluon interactions~\cite{Bauer:2024hfv}.}
    \label{fig:landscapefuture}
\end{figure}

\cref{fig:landscapefuture} summarises the expected future reach of laboratory experiments in the same $(c_{GG}/f$~vs. $m_a)$ plane. Existing precision constraints are indicated by shaded grey regions, while coloured lines denote projected sensitivities of future experiments.

At the lowest ALP masses, nuclear clock proposals based on the \(^{229}\mathrm{Th}\) are expected to probe couplings as small as
\(c_{GG}/f \sim 10^{-22}\,\mathrm{GeV}^{-1}\) for
\(m_a \lesssim 10^{-22}\,\mathrm{eV}\), making them uniquely sensitive to ultra-low frequency oscillations.

At intermediate masses, a sequence of atom interferometer proposals provides broad and continuous coverage. Broadband concepts such as AEDGE target \(m_a \sim 10^{-19}\text{--}10^{-16}\,\mathrm{eV}\), while resonant operation modes enhance sensitivity in narrower windows around \(m_a \sim 10^{-15}\,\mathrm{eV}\). Km-scale interferometers such as
MAGIS--km and AION--km extend the reach to higher masses, up to
\(m_a \sim 10^{-11}\,\mathrm{eV}\), by exploiting long baselines, extended interrogation times, and coherent phase accumulation. Dual-baseline concepts such as DUAL provide complementary sensitivity within this range.

At still higher ALP masses, solid--state and mechanical resonator technologies dominate the projected sensitivity. Helium--based resonators bridge the transition between atomic interferometers and solid devices, while quartz, pillar, and sapphire resonators extend sensitivity up to \(m_a \sim 10^{-7}\,\mathrm{eV}\). In this regime, acoustic and elastic modes respond directly to oscillations of fundamental constants induced by the ALP field.

Taken together, the figure illustrates how future laboratory experiments are expected to achieve broad and overlapping coverage across many decades in ALP mass, with different technologies optimised for different frequency regimes. This complementarity underscores the strong discovery potential of the next generation of precision experiments probing ULDM axions.


\section{Summary}
\label{sec:sum}

These lecture notes have surveyed the theoretical motivation, phenomenology, and experimental strategies underlying searches for ultralight ALPs using precision and quantum technologies. A central unifying theme has been the recognition that, in the ultralight regime, DM is most naturally described as a coherently oscillating classical field rather than as a collection of individual particles. This perspective fundamentally shapes both the form of the expected signals, which are typically coherent, time-dependent modulations, and the types of experiments that are best suited to detect them.

Starting from an EFT description of ALP interactions, we have shown how different couplings lead to qualitatively different experimental signatures. Dimension--5 axion-photon interactions enable conversion-based searches such as haloscopes and helioscopes, where resonant enhancement, magnetic fields, and coherence considerations determine sensitivity. Higher dimensional interactions, including quadratic couplings to Standard Model fields, induce oscillations of fundamental constants and material properties. These effects are probed using a large ensemble of precision experiments, including atomic and nuclear clocks, optical cavities, laser and unequal time-delay interferometers, and mechanical or solid-state resonators.

Throughout, emphasis has been placed on understanding the physical origin of the sensitivity of each platform and on how experimental noise, coherence times, and bandwidth shape the accessible mass ranges. Taken together, the material highlights the strong complementarity between different experimental approaches. Atomic clocks dominate sensitivity at the lowest masses, cavity and clock-based comparisons extend coverage to intermediate frequencies, and interferometric and resonant techniques probe higher mass regimes inaccessible to long-term averaging experiments. The resulting experimental landscape provides continuous coverage over many orders of magnitude in ALP mass and coupling strength, with future proposals promising substantial further gains. In addition, a systematic EFT prescription has been developed to connect high-scale ALP models to low-energy interactions. This demonstrates how minimal couplings at high energies can radiatively generate a rich set of effective operators at low energies, thereby enabling precision experiments to probe wide regions of parameter space that would otherwise remain inaccessible. By linking EFT to concrete observables, the notes develop a phenomenological framework that lays the groundwork for interpreting and designing future ultralight dark matter searches in non-minimal theories.


\section*{Acknowledgement}


This article is based on the work from COST Action COSMIC WISPers CA21106, supported by COST (European Cooperation in Science and Technology). The author also acknowledges support from the UKRI Future Leader Fellowship DarkMAP (Ref. no. MR/Y034112/1).

\bibliography{biblio}

\providecommand{\href}[2]{#2}\begingroup\raggedright\begin{thebibliography}{100}

\bibitem{Zwicky:1933gu}
F.~Zwicky, \emph{{Die Rotverschiebung von extragalaktischen Nebeln}}, \href{http://dx.doi.org/10.1007/s10714-008-0707-4}{\emph{Helv. Phys. Acta} {\bf 6} (1933) 110--127}.

\bibitem{Zwicky:1937zza}
F.~Zwicky, \emph{{On the Masses of Nebulae and of Clusters of Nebulae}}, \href{http://dx.doi.org/10.1086/143864}{\emph{Astrophys. J.} {\bf 86} (1937) 217--246}.

\bibitem{Rubin:1978kmz}
V.~C. Rubin, W.~K. Ford, Jr. and N.~Thonnard, \emph{{Extended rotation curves of high-luminosity spiral galaxies. IV. Systematic dynamical properties, Sa through Sc}}, \href{http://dx.doi.org/10.1086/182804}{\emph{Astrophys. J. Lett.} {\bf 225} (1978) L107--L111}.

\bibitem{Bartelmann:2010fz}
M.~Bartelmann, \emph{{Gravitational Lensing}}, \href{http://dx.doi.org/10.1088/0264-9381/27/23/233001}{\emph{Class. Quant. Grav.} {\bf 27} (2010) 233001}, [\href{https://arxiv.org/abs/1010.3829}{{\tt 1010.3829}}].

\bibitem{Hoekstra:2013via}
H.~Hoekstra, M.~Bartelmann, H.~Dahle, H.~Israel, M.~Limousin and M.~Meneghetti, \emph{{Masses of galaxy clusters from gravitational lensing}}, \href{http://dx.doi.org/10.1007/s11214-013-9978-5}{\emph{Space Sci. Rev.} {\bf 177} (2013) 75--118}, [\href{https://arxiv.org/abs/1303.3274}{{\tt 1303.3274}}].

\bibitem{Clowe:2006eq}
D.~Clowe, M.~Bradac, A.~H. Gonzalez, M.~Markevitch, S.~W. Randall, C.~Jones et~al., \emph{{A direct empirical proof of the existence of dark matter}}, \href{http://dx.doi.org/10.1086/508162}{\emph{Astrophys. J. Lett.} {\bf 648} (2006) L109--L113}, [\href{https://arxiv.org/abs/astro-ph/0608407}{{\tt astro-ph/0608407}}].

\bibitem{Sunyaev:1970eu}
R.~A. Sunyaev and Y.~B. Zeldovich, \emph{{Small scale fluctuations of relic radiation}}, {\emph{Astrophys. Space Sci.} {\bf 7} (1970) 3--19}.

\bibitem{Gelmini:2015zpa}
G.~B. Gelmini, \emph{{The hunt for dark matter.}},  in \emph{{Theoretical Advanced Study Institute in Elementary Particle Physics}: {Journeys Through the Precision Frontier: Amplitudes for Colliders}}, pp.~559--616, 2015.
\newblock \href{https://arxiv.org/abs/1502.01320}{{\tt 1502.01320}}.
\newblock \href{http://dx.doi.org/10.1142/9789814678766_0012}{DOI}.

\bibitem{PhysRevLett.49.1110}
A.~H. Guth and S.~Y. Pi, \emph{{Fluctuations in the New Inflationary Universe}}, \href{http://dx.doi.org/10.1103/PhysRevLett.49.1110}{\emph{Phys. Rev. Lett.} {\bf 49} (1982) 1110--1113}.

\bibitem{Bennett_2013}
{\scshape WMAP} collaboration, C.~L. Bennett, D.~Larson, J.~L. Weiland, N.~Jarosik, G.~Hinshaw, N.~Odegard et~al., \emph{{Nine-Year Wilkinson Microwave Anisotropy Probe (WMAP) Observations: Final Maps and Results}}, \href{http://dx.doi.org/10.1088/0067-0049/208/2/20}{\emph{Astrophys. J. Suppl.} {\bf 208} (2013) 20}, [\href{https://arxiv.org/abs/1212.5225}{{\tt 1212.5225}}].

\bibitem{Planck:2018vyg}
{\scshape Planck} collaboration, N.~Aghanim et~al., \emph{{Planck 2018 results. VI. Cosmological parameters}}, \href{http://dx.doi.org/10.1051/0004-6361/201833910}{\emph{Astron. Astrophys.} {\bf 641} (2020) A6}, [\href{https://arxiv.org/abs/1807.06209}{{\tt 1807.06209}}].

\bibitem{Bovy:2012tw}
J.~Bovy and S.~Tremaine, \emph{{On the local dark matter density}}, \href{http://dx.doi.org/10.1088/0004-637X/756/1/89}{\emph{Astrophys. J.} {\bf 756} (2012) 89}, [\href{https://arxiv.org/abs/1205.4033}{{\tt 1205.4033}}].

\bibitem{Bertone:2004pz}
G.~Bertone, D.~Hooper and J.~Silk, \emph{{Particle dark matter: Evidence, candidates and constraints}}, \href{http://dx.doi.org/10.1016/j.physrep.2004.08.031}{\emph{Phys. Rept.} {\bf 405} (2005) 279--390}, [\href{https://arxiv.org/abs/hep-ph/0404175}{{\tt hep-ph/0404175}}].

\bibitem{Kamionkowski_2010}
M.~Kamionkowski, S.~M. Koushiappas and M.~Kuhlen, \emph{{Galactic Substructure and Dark Matter Annihilation in the Milky Way Halo}}, \href{http://dx.doi.org/10.1103/PhysRevD.81.043532}{\emph{Phys. Rev. D} {\bf 81} (2010) 043532}, [\href{https://arxiv.org/abs/1001.3144}{{\tt 1001.3144}}].

\bibitem{Kolb:1990vq}
E.~Kolb and M.~Turner, \emph{{The Early Universe}}.
\newblock CRC Press, London, England, May, 2019.

\bibitem{RevModPhys.88.015004}
R.~H. Cyburt, B.~D. Fields, K.~A. Olive and T.-H. Yeh, \emph{{Big Bang Nucleosynthesis: 2015}}, \href{http://dx.doi.org/10.1103/RevModPhys.88.015004}{\emph{Rev. Mod. Phys.} {\bf 88} (2016) 015004}, [\href{https://arxiv.org/abs/1505.01076}{{\tt 1505.01076}}].

\bibitem{Baryakhtar:2022hbu}
M.~Baryakhtar et~al., \emph{{Dark Matter In Extreme Astrophysical Environments}},  in \emph{{Snowmass 2021}}, 3, 2022.
\newblock \href{https://arxiv.org/abs/2203.07984}{{\tt 2203.07984}}.

\bibitem{Drlica-Wagner:2022lbd}
A.~Drlica-Wagner et~al., \emph{{Report of the Topical Group on Cosmic Probes of Dark Matter for Snowmass 2021}},  in \emph{{Snowmass 2021}}, 9, 2022.
\newblock \href{https://arxiv.org/abs/2209.08215}{{\tt 2209.08215}}.

\bibitem{Hui:2016ltb}
L.~Hui, J.~P. Ostriker, S.~Tremaine and E.~Witten, \emph{{Ultralight scalars as cosmological dark matter}}, \href{http://dx.doi.org/10.1103/PhysRevD.95.043541}{\emph{Phys. Rev. D} {\bf 95} (2017) 043541}, [\href{https://arxiv.org/abs/1610.08297}{{\tt 1610.08297}}].

\bibitem{PhysRevD.28.1243}
M.~S. Turner, \emph{{Coherent Scalar Field Oscillations in an Expanding Universe}}, \href{http://dx.doi.org/10.1103/PhysRevD.28.1243}{\emph{Phys. Rev. D} {\bf 28} (1983) 1243}.

\bibitem{Preskill:1982cy}
J.~Preskill, M.~B. Wise and F.~Wilczek, \emph{{Cosmology of the Invisible Axion}}, \href{http://dx.doi.org/10.1016/0370-2693(83)90637-8}{\emph{Phys. Lett. B} {\bf 120} (1983) 127--132}.

\bibitem{Abbott:1982af}
L.~F. Abbott and P.~Sikivie, \emph{{A Cosmological Bound on the Invisible Axion}}, \href{http://dx.doi.org/10.1016/0370-2693(83)90638-X}{\emph{Phys. Lett. B} {\bf 120} (1983) 133--136}.

\bibitem{Dine:1982ah}
M.~Dine and W.~Fischler, \emph{{The Not So Harmless Axion}}, \href{http://dx.doi.org/10.1016/0370-2693(83)90639-1}{\emph{Phys. Lett. B} {\bf 120} (1983) 137--141}.

\bibitem{Dodelson:1993je}
S.~Dodelson and L.~M. Widrow, \emph{{Sterile-neutrinos as dark matter}}, \href{http://dx.doi.org/10.1103/PhysRevLett.72.17}{\emph{Phys. Rev. Lett.} {\bf 72} (1994) 17--20}, [\href{https://arxiv.org/abs/hep-ph/9303287}{{\tt hep-ph/9303287}}].

\bibitem{Shi:1998km}
X.-D. Shi and G.~M. Fuller, \emph{{A New dark matter candidate: Nonthermal sterile neutrinos}}, \href{http://dx.doi.org/10.1103/PhysRevLett.82.2832}{\emph{Phys. Rev. Lett.} {\bf 82} (1999) 2832--2835}, [\href{https://arxiv.org/abs/astro-ph/9810076}{{\tt astro-ph/9810076}}].

\bibitem{Bode:2000gq}
P.~Bode, J.~P. Ostriker and N.~Turok, \emph{{Halo formation in warm dark matter models}}, \href{http://dx.doi.org/10.1086/321541}{\emph{Astrophys. J.} {\bf 556} (2001) 93--107}, [\href{https://arxiv.org/abs/astro-ph/0010389}{{\tt astro-ph/0010389}}].

\bibitem{Bond:1983hb}
J.~R. Bond and A.~S. Szalay, \emph{{The Collisionless Damping of Density Fluctuations in an Expanding Universe}}, \href{http://dx.doi.org/10.1086/161460}{\emph{Astrophys. J.} {\bf 274} (1983) 443--468}.

\bibitem{PhysRevD.71.063534}
M.~Viel, J.~Lesgourgues, M.~G. Haehnelt, S.~Matarrese and A.~Riotto, \emph{{Constraining warm dark matter candidates including sterile neutrinos and light gravitinos with WMAP and the Lyman-alpha forest}}, \href{http://dx.doi.org/10.1103/PhysRevD.71.063534}{\emph{Phys. Rev. D} {\bf 71} (2005) 063534}, [\href{https://arxiv.org/abs/astro-ph/0501562}{{\tt astro-ph/0501562}}].

\bibitem{Bond:1982uy}
J.~R. Bond, A.~S. Szalay and M.~S. Turner, \emph{{Formation of Galaxies in a Gravitino Dominated Universe}}, \href{http://dx.doi.org/10.1103/PhysRevLett.48.1636}{\emph{Phys. Rev. Lett.} {\bf 48} (1982) 1636}.

\bibitem{Dayal:2023nwi}
P.~Dayal and S.~K. Giri, \emph{{Warm dark matter constraints from the JWST}}, \href{http://dx.doi.org/10.1093/mnras/stae176}{\emph{Mon. Not. Roy. Astron. Soc.} {\bf 528} (2024) 2784--2789}, [\href{https://arxiv.org/abs/2303.14239}{{\tt 2303.14239}}].

\bibitem{Gondolo:1990dk}
P.~Gondolo and G.~Gelmini, \emph{{Cosmic abundances of stable particles: Improved analysis}}, \href{http://dx.doi.org/10.1016/0550-3213(91)90438-4}{\emph{Nucl. Phys. B} {\bf 360} (1991) 145--179}.

\bibitem{Bergstrom:2012fi}
L.~Bergstrom, \emph{{Dark Matter Evidence, Particle Physics Candidates and Detection Methods}}, \href{http://dx.doi.org/10.1002/andp.201200116}{\emph{Annalen Phys.} {\bf 524} (2012) 479--496}, [\href{https://arxiv.org/abs/1205.4882}{{\tt 1205.4882}}].

\bibitem{Feng:2010gw}
J.~L. Feng, \emph{{Dark Matter Candidates from Particle Physics and Methods of Detection}}, \href{http://dx.doi.org/10.1146/annurev-astro-082708-101659}{\emph{Ann. Rev. Astron. Astrophys.} {\bf 48} (2010) 495--545}, [\href{https://arxiv.org/abs/1003.0904}{{\tt 1003.0904}}].

\bibitem{PhysRevLett.64.615}
K.~Griest and M.~Kamionkowski, \emph{{Unitarity Limits on the Mass and Radius of Dark Matter Particles}}, \href{http://dx.doi.org/10.1103/PhysRevLett.64.615}{\emph{Phys. Rev. Lett.} {\bf 64} (1990) 615}.

\bibitem{Kusenko:2001vu}
A.~Kusenko and P.~J. Steinhardt, \emph{{Q ball candidates for selfinteracting dark matter}}, \href{http://dx.doi.org/10.1103/PhysRevLett.87.141301}{\emph{Phys. Rev. Lett.} {\bf 87} (2001) 141301}, [\href{https://arxiv.org/abs/astro-ph/0106008}{{\tt astro-ph/0106008}}].

\bibitem{Ge:2019voa}
S.~Ge, K.~Lawson and A.~Zhitnitsky, \emph{{Axion quark nugget dark matter model: Size distribution and survival pattern}}, \href{http://dx.doi.org/10.1103/PhysRevD.99.116017}{\emph{Phys. Rev. D} {\bf 99} (2019) 116017}, [\href{https://arxiv.org/abs/1903.05090}{{\tt 1903.05090}}].

\bibitem{Deliyergiyev:2019vti}
M.~Deliyergiyev, A.~Del~Popolo, L.~Tolos, M.~Le~Delliou, X.~Lee and F.~Burgio, \emph{{Dark compact objects: an extensive overview}}, \href{http://dx.doi.org/10.1103/PhysRevD.99.063015}{\emph{Phys. Rev. D} {\bf 99} (2019) 063015}, [\href{https://arxiv.org/abs/1903.01183}{{\tt 1903.01183}}].

\bibitem{Hawking:1971ei}
S.~Hawking, \emph{{Gravitationally collapsed objects of very low mass}}, \href{http://dx.doi.org/10.1093/mnras/152.1.75}{\emph{Mon. Not. Roy. Astron. Soc.} {\bf 152} (1971) 75}.

\bibitem{Chapline:1975ojl}
G.~F. Chapline, \emph{{Cosmological effects of primordial black holes}}, \href{http://dx.doi.org/10.1038/253251a0}{\emph{Nature} {\bf 253} (1975) 251--252}.

\bibitem{Bird:2016dcv}
S.~Bird, I.~Cholis, J.~B. Mu{\~n}oz, Y.~Ali-Ha{\"\i}moud, M.~Kamionkowski, E.~D. Kovetz et~al., \emph{{Did LIGO detect dark matter?}}, \href{http://dx.doi.org/10.1103/PhysRevLett.116.201301}{\emph{Phys. Rev. Lett.} {\bf 116} (2016) 201301}, [\href{https://arxiv.org/abs/1603.00464}{{\tt 1603.00464}}].

\bibitem{Sasaki:2016jop}
M.~Sasaki, T.~Suyama, T.~Tanaka and S.~Yokoyama, \emph{{Primordial Black Hole Scenario for the Gravitational-Wave Event GW150914}}, \href{http://dx.doi.org/10.1103/PhysRevLett.117.061101}{\emph{Phys. Rev. Lett.} {\bf 117} (2016) 061101}, [\href{https://arxiv.org/abs/1603.08338}{{\tt 1603.08338}}].

\bibitem{Niikura:2017zjd}
H.~Niikura et~al., \emph{{Microlensing constraints on primordial black holes with Subaru/HSC Andromeda observations}}, \href{http://dx.doi.org/10.1038/s41550-019-0723-1}{\emph{Nature Astron.} {\bf 3} (2019) 524--534}, [\href{https://arxiv.org/abs/1701.02151}{{\tt 1701.02151}}].

\bibitem{PhysRevLett.39.165}
B.~W. Lee and S.~Weinberg, \emph{{Cosmological Lower Bound on Heavy Neutrino Masses}}, \href{http://dx.doi.org/10.1103/PhysRevLett.39.165}{\emph{Phys. Rev. Lett.} {\bf 39} (1977) 165--168}.

\bibitem{PhysRevD.31.3059}
M.~W. Goodman and E.~Witten, \emph{{Detectability of Certain Dark Matter Candidates}}, \href{http://dx.doi.org/10.1103/PhysRevD.31.3059}{\emph{Phys. Rev. D} {\bf 31} (1985) 3059}.

\bibitem{Akerib:2022ort}
D.~S. Akerib et~al., \emph{{Snowmass2021 Cosmic Frontier Dark Matter Direct Detection to the Neutrino Fog}},  in \emph{{Snowmass 2021}}, 3, 2022.
\newblock \href{https://arxiv.org/abs/2203.08084}{{\tt 2203.08084}}.

\bibitem{Abercrombie:2015wmb}
D.~Abercrombie et~al., \emph{{Dark Matter benchmark models for early LHC Run-2 Searches: Report of the ATLAS/CMS Dark Matter Forum}}, \href{http://dx.doi.org/10.1016/j.dark.2019.100371}{\emph{Phys. Dark Univ.} {\bf 27} (2020) 100371}, [\href{https://arxiv.org/abs/1507.00966}{{\tt 1507.00966}}].

\bibitem{Boveia:2022syt}
A.~Boveia et~al., \emph{{Snowmass 2021 Cross Frontier Report: Dark Matter Complementarity (Extended Version)}},  in \emph{{Snowmass 2021}}, 10, 2022.
\newblock \href{https://arxiv.org/abs/2210.01770}{{\tt 2210.01770}}.

\bibitem{Cirelli:2024ssz}
M.~Cirelli, A.~Strumia and J.~Zupan, \emph{{Dark Matter}},  \href{https://arxiv.org/abs/2406.01705}{{\tt 2406.01705}}.

\bibitem{Ando:2022kzd}
S.~Ando et~al., \emph{{Snowmass2021 Cosmic Frontier: Synergies between dark matter searches and multiwavelength/multimessenger astrophysics}},  in \emph{{Snowmass 2021}}, 3, 2022.
\newblock \href{https://arxiv.org/abs/2203.06781}{{\tt 2203.06781}}.

\bibitem{PhysRevLett.134.111802}
{\scshape XENON} collaboration, E.~Aprile, J.~Aalbers, K.~Abe, S.~Ahmed~Maouloud, L.~Althueser, B.~Andrieu et~al., \emph{{First Search for Light Dark Matter in the Neutrino Fog with XENONnT}}, \href{http://dx.doi.org/10.1103/PhysRevLett.134.111802}{\emph{Phys. Rev. Lett.} {\bf 134} (2025) 111802}, [\href{https://arxiv.org/abs/2409.17868}{{\tt 2409.17868}}].

\bibitem{Eberhardt:2025caq}
A.~Eberhardt and E.~G.~M. Ferreira, \emph{{Ultralight fuzzy dark matter review}},  \href{https://arxiv.org/abs/2507.00705}{{\tt 2507.00705}}.

\bibitem{PhysRevD.104.055037}
M.~Lisanti, M.~Moschella and W.~Terrano, \emph{{Stochastic properties of ultralight scalar field gradients}}, \href{http://dx.doi.org/10.1103/PhysRevD.104.055037}{\emph{Phys. Rev. D} {\bf 104} (2021) 055037}, [\href{https://arxiv.org/abs/2107.10260}{{\tt 2107.10260}}].

\bibitem{Abel:2020pzs}
C.~Abel et~al., \emph{{Measurement of the Permanent Electric Dipole Moment of the Neutron}}, \href{http://dx.doi.org/10.1103/PhysRevLett.124.081803}{\emph{Phys. Rev. Lett.} {\bf 124} (2020) 081803}, [\href{https://arxiv.org/abs/2001.11966}{{\tt 2001.11966}}].

\bibitem{GrillidiCortona:2015jxo}
G.~Grilli~di Cortona, E.~Hardy, J.~Pardo~Vega and G.~Villadoro, \emph{{The QCD axion, precisely}}, \href{http://dx.doi.org/10.1007/JHEP01(2016)034}{\emph{JHEP} {\bf 01} (2016) 034}, [\href{https://arxiv.org/abs/1511.02867}{{\tt 1511.02867}}].

\bibitem{Bauer:2021mvw}
M.~Bauer, M.~Neubert, S.~Renner, M.~Schnubel and A.~Thamm, \emph{{Flavor probes of axion-like particles}}, \href{http://dx.doi.org/10.1007/JHEP09(2022)056}{\emph{JHEP} {\bf 09} (2022) 056}, [\href{https://arxiv.org/abs/2110.10698}{{\tt 2110.10698}}].

\bibitem{DiLuzio:2020wdo}
L.~Di~Luzio, M.~Giannotti, E.~Nardi and L.~Visinelli, \emph{{The landscape of QCD axion models}}, \href{http://dx.doi.org/10.1016/j.physrep.2020.06.002}{\emph{Phys. Rept.} {\bf 870} (2020) 1--117}, [\href{https://arxiv.org/abs/2003.01100}{{\tt 2003.01100}}].

\bibitem{Shifman:1979if}
M.~A. Shifman, A.~I. Vainshtein and V.~I. Zakharov, \emph{{Can Confinement Ensure Natural CP Invariance of Strong Interactions?}}, \href{http://dx.doi.org/10.1016/0550-3213(80)90209-6}{\emph{Nucl. Phys. B} {\bf 166} (1980) 493--506}.

\bibitem{Kim:1979if}
J.~E. Kim, \emph{{Weak Interaction Singlet and Strong CP Invariance}}, \href{http://dx.doi.org/10.1103/PhysRevLett.43.103}{\emph{Phys. Rev. Lett.} {\bf 43} (1979) 103}.

\bibitem{Zhitnitsky:1980tq}
A.~R. Zhitnitsky, \emph{{On Possible Suppression of the Axion Hadron Interactions. (In Russian)}}, {\emph{Sov. J. Nucl. Phys.} {\bf 31} (1980) 260}.

\bibitem{Dine:1981rt}
M.~Dine, W.~Fischler and M.~Srednicki, \emph{{A Simple Solution to the Strong CP Problem with a Harmless Axion}}, \href{http://dx.doi.org/10.1016/0370-2693(81)90590-6}{\emph{Phys. Lett. B} {\bf 104} (1981) 199--202}.

\bibitem{Chadha-Day:2021szb}
F.~Chadha-Day, J.~Ellis and D.~J.~E. Marsh, \emph{{Axion dark matter: What is it and why now?}}, \href{http://dx.doi.org/10.1126/sciadv.abj3618}{\emph{Sci. Adv.} {\bf 8} (2022) abj3618}, [\href{https://arxiv.org/abs/2105.01406}{{\tt 2105.01406}}].

\bibitem{Svrcek:2006yi}
P.~Svrcek and E.~Witten, \emph{{Axions In String Theory}}, \href{http://dx.doi.org/10.1088/1126-6708/2006/06/051}{\emph{JHEP} {\bf 06} (2006) 051}, [\href{https://arxiv.org/abs/hep-th/0605206}{{\tt hep-th/0605206}}].

\bibitem{Choi:2009jt}
K.-S. Choi, H.~P. Nilles, S.~Ramos-Sanchez and P.~K.~S. Vaudrevange, \emph{{Accions}}, \href{http://dx.doi.org/10.1016/j.physletb.2009.04.028}{\emph{Phys. Lett. B} {\bf 675} (2009) 381--386}, [\href{https://arxiv.org/abs/0902.3070}{{\tt 0902.3070}}].

\bibitem{Arvanitaki:2009fg}
A.~Arvanitaki, S.~Dimopoulos, S.~Dubovsky, N.~Kaloper and J.~March-Russell, \emph{{String Axiverse}}, \href{http://dx.doi.org/10.1103/PhysRevD.81.123530}{\emph{Phys. Rev. D} {\bf 81} (2010) 123530}, [\href{https://arxiv.org/abs/0905.4720}{{\tt 0905.4720}}].

\bibitem{Adams:2022pbo}
C.~B. Adams et~al., \emph{{Axion Dark Matter}},  in \emph{{Snowmass 2021}}, 3, 2022.
\newblock \href{https://arxiv.org/abs/2203.14923}{{\tt 2203.14923}}.

\bibitem{Irastorza:2018dyq}
I.~G. Irastorza and J.~Redondo, \emph{{New experimental approaches in the search for axion-like particles}}, \href{http://dx.doi.org/10.1016/j.ppnp.2018.05.003}{\emph{Prog. Part. Nucl. Phys.} {\bf 102} (2018) 89--159}, [\href{https://arxiv.org/abs/1801.08127}{{\tt 1801.08127}}].

\bibitem{PhysRevD.32.2988}
P.~Sikivie, \emph{{Detection Rates for 'Invisible' Axion Searches}}, \href{http://dx.doi.org/10.1103/PhysRevD.36.974}{\emph{Phys. Rev. D} {\bf 32} (1985) 2988}.

\bibitem{Sikivie:1983ip}
P.~Sikivie, \emph{{Experimental Tests of the Invisible Axion}}, \href{http://dx.doi.org/10.1103/PhysRevLett.51.1415}{\emph{Phys. Rev. Lett.} {\bf 51} (1983) 1415--1417}.

\bibitem{ADMX:2018gho}
{\scshape ADMX} collaboration, N.~Du et~al., \emph{{A Search for Invisible Axion Dark Matter with the Axion Dark Matter Experiment}}, \href{http://dx.doi.org/10.1103/PhysRevLett.120.151301}{\emph{Phys. Rev. Lett.} {\bf 120} (2018) 151301}, [\href{https://arxiv.org/abs/1804.05750}{{\tt 1804.05750}}].

\bibitem{ADMX:2019uok}
{\scshape ADMX} collaboration, T.~Braine et~al., \emph{{Extended Search for the Invisible Axion with the Axion Dark Matter Experiment}}, \href{http://dx.doi.org/10.1103/PhysRevLett.124.101303}{\emph{Phys. Rev. Lett.} {\bf 124} (2020) 101303}, [\href{https://arxiv.org/abs/1910.08638}{{\tt 1910.08638}}].

\bibitem{PhysRevD.64.092003}
{\scshape ADMX} collaboration, S.~J. Asztalos et~al., \emph{{Large scale microwave cavity search for dark matter axions}}, \href{http://dx.doi.org/10.1103/PhysRevD.64.092003}{\emph{Phys. Rev. D} {\bf 64} (2001) 092003}.

\bibitem{ADMX:2021nhd}
{\scshape ADMX} collaboration, C.~Bartram et~al., \emph{{Search for Invisible Axion Dark Matter in the 3.3{\textendash}4.2{\,}{\,}{\ensuremath{\mu}}eV Mass Range}}, \href{http://dx.doi.org/10.1103/PhysRevLett.127.261803}{\emph{Phys. Rev. Lett.} {\bf 127} (2021) 261803}, [\href{https://arxiv.org/abs/2110.06096}{{\tt 2110.06096}}].

\bibitem{HAYSTAC:2020kwv}
{\scshape HAYSTAC} collaboration, K.~M. Backes et~al., \emph{{A quantum-enhanced search for dark matter axions}}, \href{http://dx.doi.org/10.1038/s41586-021-03226-7}{\emph{Nature} {\bf 590} (2021) 238--242}, [\href{https://arxiv.org/abs/2008.01853}{{\tt 2008.01853}}].

\bibitem{HAYSTAC:2023cam}
{\scshape HAYSTAC} collaboration, M.~J. Jewell et~al., \emph{{New results from HAYSTAC{\textquoteright}s phase II operation with a squeezed state receiver}}, \href{http://dx.doi.org/10.1103/PhysRevD.107.072007}{\emph{Phys. Rev. D} {\bf 107} (2023) 072007}, [\href{https://arxiv.org/abs/2301.09721}{{\tt 2301.09721}}].

\bibitem{refId0}
{Majorovits, Béla}, \emph{The search of axion dark matter with a dielectric halo-scope: Madmax}, \href{http://dx.doi.org/10.1051/epjconf/202328201008}{\emph{EPJ Web Conf.} {\bf 282} (2023) 01008}.

\bibitem{Li:2021oV}
X.~Li, \emph{{MADMAX: A Dielectric Haloscope Experiment}}, \href{http://dx.doi.org/10.22323/1.390.0645}{\emph{PoS} {\bf ICHEP2020} (2021) 645}.

\bibitem{c749-419q}
{\scshape MADMAX} collaboration, B.~A. d.~S. Garcia et~al., \emph{{First Search for Axion Dark Matter with a MADMAX Prototype}}, \href{http://dx.doi.org/10.1103/c749-419q}{\emph{Phys. Rev. Lett.} {\bf 135} (2025) 041001}, [\href{https://arxiv.org/abs/2409.11777}{{\tt 2409.11777}}].

\bibitem{PhysRevLett.118.091801}
{\scshape MADMAX Working Group} collaboration, A.~Caldwell, G.~Dvali, B.~Majorovits, A.~Millar, G.~Raffelt, J.~Redondo et~al., \emph{{Dielectric Haloscopes: A New Way to Detect Axion Dark Matter}}, \href{http://dx.doi.org/10.1103/PhysRevLett.118.091801}{\emph{Phys. Rev. Lett.} {\bf 118} (2017) 091801}, [\href{https://arxiv.org/abs/1611.05865}{{\tt 1611.05865}}].

\bibitem{Berlin:2019ahk}
A.~Berlin, R.~T. D'Agnolo, S.~A.~R. Ellis, C.~Nantista, J.~Neilson, P.~Schuster et~al., \emph{{Axion Dark Matter Detection by Superconducting Resonant Frequency Conversion}}, \href{http://dx.doi.org/10.1007/JHEP07(2020)088}{\emph{JHEP} {\bf 07} (2020) 088}, [\href{https://arxiv.org/abs/1912.11048}{{\tt 1912.11048}}].

\bibitem{Giaccone:2022pke}
B.~Giaccone et~al., \emph{{Design of axion and axion dark matter searches based on ultra high Q SRF cavities}},  \href{https://arxiv.org/abs/2207.11346}{{\tt 2207.11346}}.

\bibitem{PhysRevLett.123.021801}
Z.~Bogorad, A.~Hook, Y.~Kahn and Y.~Soreq, \emph{{Probing Axionlike Particles and the Axiverse with Superconducting Radio-Frequency Cavities}}, \href{http://dx.doi.org/10.1103/PhysRevLett.123.021801}{\emph{Phys. Rev. Lett.} {\bf 123} (2019) 021801}, [\href{https://arxiv.org/abs/1902.01418}{{\tt 1902.01418}}].

\bibitem{vanBibber:1988ge}
K.~van Bibber, P.~M. McIntyre, D.~E. Morris and G.~G. Raffelt, \emph{{A Practical Laboratory Detector for Solar Axions}}, \href{http://dx.doi.org/10.1103/PhysRevD.39.2089}{\emph{Phys. Rev. D} {\bf 39} (1989) 2089}.

\bibitem{Redondo:2013wwa}
J.~Redondo, \emph{{Solar axion flux from the axion-electron coupling}}, \href{http://dx.doi.org/10.1088/1475-7516/2013/12/008}{\emph{JCAP} {\bf 12} (2013) 008}, [\href{https://arxiv.org/abs/1310.0823}{{\tt 1310.0823}}].

\bibitem{CAST:2009jdc}
{\scshape CAST} collaboration, S.~Andriamonje et~al., \emph{{Search for 14.4-keV solar axions emitted in the M1-transition of Fe-57 nuclei with CAST}}, \href{http://dx.doi.org/10.1088/1475-7516/2009/12/002}{\emph{JCAP} {\bf 12} (2009) 002}, [\href{https://arxiv.org/abs/0906.4488}{{\tt 0906.4488}}].

\bibitem{DiLuzio:2021qct}
L.~Di~Luzio et~al., \emph{{Probing the axion{\textendash}nucleon coupling with the next generation of~axion helioscopes}}, \href{http://dx.doi.org/10.1140/epjc/s10052-022-10061-1}{\emph{Eur. Phys. J. C} {\bf 82} (2022) 120}, [\href{https://arxiv.org/abs/2111.06407}{{\tt 2111.06407}}].

\bibitem{RevModPhys.54.767}
J.~N. Bahcall, W.~F. Huebner, S.~H. Lubow, P.~D. Parker and R.~K. Ulrich, \emph{{Standard Solar Models and the Uncertainties in Predicted Capture Rates of Solar Neutrinos}}, \href{http://dx.doi.org/10.1103/RevModPhys.54.767}{\emph{Rev. Mod. Phys.} {\bf 54} (1982) 767}.

\bibitem{Vogel:2023rfa}
J.~K. Vogel and I.~G. Irastorza, \emph{{Solar Production of Ultralight Bosons}},  in \emph{{The Search for Ultralight Bosonic Dark Matter}} (D.~F.~J. Kimball and K.~van Bibber, eds.), pp.~141--171.
\newblock Springer, 2023.
\newblock \href{http://dx.doi.org/10.1007/978-3-030-95852-7_5}{DOI}.

\bibitem{Barth:2013sma}
K.~Barth et~al., \emph{{CAST constraints on the axion-electron coupling}}, \href{http://dx.doi.org/10.1088/1475-7516/2013/05/010}{\emph{JCAP} {\bf 05} (2013) 010}, [\href{https://arxiv.org/abs/1302.6283}{{\tt 1302.6283}}].

\bibitem{Irastorza:2011gs}
I.~G. Irastorza et~al., \emph{{Towards a new generation axion helioscope}}, \href{http://dx.doi.org/10.1088/1475-7516/2011/06/013}{\emph{JCAP} {\bf 06} (2011) 013}, [\href{https://arxiv.org/abs/1103.5334}{{\tt 1103.5334}}].

\bibitem{Abbon:2007ug}
P.~Abbon et~al., \emph{{The Micromegas detector of the CAST experiment}}, \href{http://dx.doi.org/10.1088/1367-2630/9/6/170}{\emph{New J. Phys.} {\bf 9} (2007) 170}, [\href{https://arxiv.org/abs/physics/0702190}{{\tt physics/0702190}}].

\bibitem{Kuster:2007ue}
M.~Kuster et~al., \emph{{The X-ray Telescope of CAST}}, \href{http://dx.doi.org/10.1088/1367-2630/9/6/169}{\emph{New J. Phys.} {\bf 9} (2007) 169}, [\href{https://arxiv.org/abs/physics/0702188}{{\tt physics/0702188}}].

\bibitem{Autiero:2007uf}
D.~Autiero et~al., \emph{{The CAST Time Projection Chamber}}, \href{http://dx.doi.org/10.1088/1367-2630/9/6/171}{\emph{New J. Phys.} {\bf 9} (2007) 171}, [\href{https://arxiv.org/abs/physics/0702189}{{\tt physics/0702189}}].

\bibitem{CAST:2011rjr}
{\scshape CAST} collaboration, S.~Aune et~al., \emph{{CAST search for sub-eV mass solar axions with 3He buffer gas}}, \href{http://dx.doi.org/10.1103/PhysRevLett.107.261302}{\emph{Phys. Rev. Lett.} {\bf 107} (2011) 261302}, [\href{https://arxiv.org/abs/1106.3919}{{\tt 1106.3919}}].

\bibitem{CAST:2013bqn}
{\scshape CAST} collaboration, M.~Arik et~al., \emph{{Search for Solar Axions by the CERN Axion Solar Telescope with $^3$He Buffer Gas: Closing the Hot Dark Matter Gap}}, \href{http://dx.doi.org/10.1103/PhysRevLett.112.091302}{\emph{Phys. Rev. Lett.} {\bf 112} (2014) 091302}, [\href{https://arxiv.org/abs/1307.1985}{{\tt 1307.1985}}].

\bibitem{CAST:2004gzq}
{\scshape CAST} collaboration, K.~Zioutas et~al., \emph{{First results from the CERN Axion Solar Telescope (CAST)}}, \href{http://dx.doi.org/10.1103/PhysRevLett.94.121301}{\emph{Phys. Rev. Lett.} {\bf 94} (2005) 121301}, [\href{https://arxiv.org/abs/hep-ex/0411033}{{\tt hep-ex/0411033}}].

\bibitem{CAST:2007jps}
{\scshape CAST} collaboration, S.~Andriamonje et~al., \emph{{An Improved limit on the axion-photon coupling from the CAST experiment}}, \href{http://dx.doi.org/10.1088/1475-7516/2007/04/010}{\emph{JCAP} {\bf 04} (2007) 010}, [\href{https://arxiv.org/abs/hep-ex/0702006}{{\tt hep-ex/0702006}}].

\bibitem{CAST:2008ixs}
{\scshape CAST} collaboration, E.~Arik et~al., \emph{{Probing eV-scale axions with CAST}}, \href{http://dx.doi.org/10.1088/1475-7516/2009/02/008}{\emph{JCAP} {\bf 02} (2009) 008}, [\href{https://arxiv.org/abs/0810.4482}{{\tt 0810.4482}}].

\bibitem{CAST:2015qbl}
{\scshape CAST} collaboration, M.~Arik et~al., \emph{{New solar axion search using the CERN Axion Solar Telescope with $^4$He filling}}, \href{http://dx.doi.org/10.1103/PhysRevD.92.021101}{\emph{Phys. Rev. D} {\bf 92} (2015) 021101}, [\href{https://arxiv.org/abs/1503.00610}{{\tt 1503.00610}}].

\bibitem{CAST:2017uph}
{\scshape CAST} collaboration, V.~Anastassopoulos et~al., \emph{{New CAST Limit on the Axion-Photon Interaction}}, \href{http://dx.doi.org/10.1038/nphys4109}{\emph{Nature Phys.} {\bf 13} (2017) 584--590}, [\href{https://arxiv.org/abs/1705.02290}{{\tt 1705.02290}}].

\bibitem{Zioutas:1998cc}
K.~Zioutas et~al., \emph{{A Decommissioned LHC model magnet as an axion telescope}}, \href{http://dx.doi.org/10.1016/S0168-9002(98)01442-9}{\emph{Nucl. Instrum. Meth. A} {\bf 425} (1999) 480--489}, [\href{https://arxiv.org/abs/astro-ph/9801176}{{\tt astro-ph/9801176}}].

\bibitem{Giannotti:2016drd}
M.~Giannotti, J.~Ruz and J.~K. Vogel, \emph{{IAXO, next-generation of helioscopes}}, \href{http://dx.doi.org/10.22323/1.282.0195}{\emph{PoS} {\bf ICHEP2016} (2016) 195}, [\href{https://arxiv.org/abs/1611.04652}{{\tt 1611.04652}}].

\bibitem{Armengaud:2014gea}
E.~Armengaud et~al., \emph{{Conceptual Design of the International Axion Observatory (IAXO)}}, \href{http://dx.doi.org/10.1088/1748-0221/9/05/T05002}{\emph{JINST} {\bf 9} (2014) T05002}, [\href{https://arxiv.org/abs/1401.3233}{{\tt 1401.3233}}].

\bibitem{IAXO:2019mpb}
{\scshape IAXO} collaboration, E.~Armengaud et~al., \emph{{Physics potential of the International Axion Observatory (IAXO)}}, \href{http://dx.doi.org/10.1088/1475-7516/2019/06/047}{\emph{JCAP} {\bf 06} (2019) 047}, [\href{https://arxiv.org/abs/1904.09155}{{\tt 1904.09155}}].

\bibitem{IAXO:2025ltd}
{\scshape IAXO} collaboration, A.~Arcusa et~al., \emph{{The International Axion Observatory (IAXO): case, status and plans. Input to the European Strategy for Particle Physics}},  \href{https://arxiv.org/abs/2504.00079}{{\tt 2504.00079}}.

\bibitem{axionlimits}
C.~O'Hare, ``Axionlimits.'' https://cajohare.github.io/AxionLimits/.

\bibitem{Bauer:2020jbp}
M.~Bauer, M.~Neubert, S.~Renner, M.~Schnubel and A.~Thamm, \emph{{The Low-Energy Effective Theory of Axions and ALPs}}, \href{http://dx.doi.org/10.1007/JHEP04(2021)063}{\emph{JHEP} {\bf 04} (2021) 063}, [\href{https://arxiv.org/abs/2012.12272}{{\tt 2012.12272}}].

\bibitem{Bauer:2024hfv}
M.~Bauer, S.~Chakraborti and G.~Rostagni, \emph{{Axion bounds from quantum technology}}, \href{http://dx.doi.org/10.1007/JHEP05(2025)023}{\emph{JHEP} {\bf 05} (2025) 023}, [\href{https://arxiv.org/abs/2408.06412}{{\tt 2408.06412}}].

\bibitem{Chala:2020wvs}
M.~Chala, G.~Guedes, M.~Ramos and J.~Santiago, \emph{{Running in the ALPs}}, \href{http://dx.doi.org/10.1140/epjc/s10052-021-08968-2}{\emph{Eur. Phys. J. C} {\bf 81} (2021) 181}, [\href{https://arxiv.org/abs/2012.09017}{{\tt 2012.09017}}].

\bibitem{Kim:2023pvt}
H.~Kim, A.~Lenoci, G.~Perez and W.~Ratzinger, \emph{{Probing an ultralight QCD axion with electromagnetic quadratic interaction}}, \href{http://dx.doi.org/10.1103/PhysRevD.109.015030}{\emph{Phys. Rev. D} {\bf 109} (2024) 015030}, [\href{https://arxiv.org/abs/2307.14962}{{\tt 2307.14962}}].

\bibitem{Hoferichter:2015tha}
M.~Hoferichter, J.~Ruiz~de Elvira, B.~Kubis and U.-G. Mei{\ss}ner, \emph{{Matching pion-nucleon Roy-Steiner equations to chiral perturbation theory}}, \href{http://dx.doi.org/10.1103/PhysRevLett.115.192301}{\emph{Phys. Rev. Lett.} {\bf 115} (2015) 192301}, [\href{https://arxiv.org/abs/1507.07552}{{\tt 1507.07552}}].

\bibitem{Kim:2022ype}
H.~Kim and G.~Perez, \emph{{Oscillations of atomic energy levels induced by QCD axion dark matter}}, \href{http://dx.doi.org/10.1103/PhysRevD.109.015005}{\emph{Phys. Rev. D} {\bf 109} (2024) 015005}, [\href{https://arxiv.org/abs/2205.12988}{{\tt 2205.12988}}].

\bibitem{Beadle:2023flm}
C.~Beadle, S.~A.~R. Ellis, J.~Quevillon and P.~N. Hoa~Vuong, \emph{{Quadratic coupling of the axion to photons}}, \href{http://dx.doi.org/10.1103/PhysRevD.110.035019}{\emph{Phys. Rev. D} {\bf 110} (2024) 035019}, [\href{https://arxiv.org/abs/2307.10362}{{\tt 2307.10362}}].

\bibitem{Arvanitaki:2014faa}
A.~Arvanitaki, J.~Huang and K.~Van~Tilburg, \emph{{Searching for dilaton dark matter with atomic clocks}}, \href{http://dx.doi.org/10.1103/PhysRevD.91.015015}{\emph{Phys. Rev. D} {\bf 91} (2015) 015015}, [\href{https://arxiv.org/abs/1405.2925}{{\tt 1405.2925}}].

\bibitem{Flambaum:2004tm}
V.~V. Flambaum, D.~B. Leinweber, A.~W. Thomas and R.~D. Young, \emph{{Limits on the temporal variation of the fine structure constant, quark masses and strong interaction from quasar absorption spectra and atomic clock experiments}}, \href{http://dx.doi.org/10.1103/PhysRevD.69.115006}{\emph{Phys. Rev. D} {\bf 69} (2004) 115006}, [\href{https://arxiv.org/abs/hep-ph/0402098}{{\tt hep-ph/0402098}}].

\bibitem{Karshenboim:2004tx}
S.~G. Karshenboim, V.~V. Flambaum and E.~Peik, \emph{{Atomic clocks and constraints on variations of fundamental constants}},  \href{https://arxiv.org/abs/physics/0410074}{{\tt physics/0410074}}.

\bibitem{Wynands_2005}
R.~Wynands and S.~Weyers, \emph{Atomic fountain clocks}, \href{http://dx.doi.org/10.1088/0026-1394/42/3/S08}{\emph{Metrologia} {\bf 42} (2005) S64}.

\bibitem{ludlow2015opticalatomicclocks}
A.~D. Ludlow, M.~M. Boyd, J.~Ye, E.~Peik and P.~O. Schmidt, \emph{Optical atomic clocks},  \href{https://arxiv.org/abs/1407.3493}{{\tt 1407.3493}}.

\bibitem{Peik:2020cwm}
E.~Peik, T.~Schumm, M.~S. Safronova, A.~P{\'a}lffy, J.~Weitenberg and P.~G. Thirolf, \emph{{Nuclear clocks for testing fundamental physics}}, \href{http://dx.doi.org/10.1088/2058-9565/abe9c2}{\emph{Quantum Sci. Technol.} {\bf 6} (2021) 034002}, [\href{https://arxiv.org/abs/2012.09304}{{\tt 2012.09304}}].

\bibitem{1966IEEEP..54..221A}
D.~W. {Allan}, \emph{{Statistics of atomic frequency standards}}, \href{http://dx.doi.org/10.1109/PROC.1966.4634}{\emph{IEEE Proceedings} {\bf 54} (1966) 221--230}.

\bibitem{5570702}
J.~A. Barnes, A.~R. Chi, L.~S. Cutler, D.~J. Healey, D.~B. Leeson, T.~E. McGunigal et~al., \emph{Characterization of frequency stability}, \href{http://dx.doi.org/10.1109/TIM.1971.5570702}{\emph{IEEE Transactions on Instrumentation and Measurement} {\bf IM-20} (1971) 105--120}.

\bibitem{804271}
W.~Riley and D.~Howe, \emph{Handbook of Frequency Stability Analysis}.
\newblock Special Publication (NIST SP), National Institute of Standards and Technology, Gaithersburg, MD, July, 2008.

\bibitem{Kessler:2013nez}
E.~M. Kessler, P.~K{\'o}m{\'a}r, M.~Bishof, L.~Jiang, A.~S. S{\o}rensen, J.~Ye et~al., \emph{{Heisenberg-Limited Atom Clocks Based on Entangled Qubits}}, \href{http://dx.doi.org/10.1103/PhysRevLett.112.190403}{\emph{Phys. Rev. Lett.} {\bf 112} (2014) 190403}, [\href{https://arxiv.org/abs/1310.6043}{{\tt 1310.6043}}].

\bibitem{Flambaum:2006ip}
V.~V. Flambaum and A.~F. Tedesco, \emph{{Dependence of nuclear magnetic moments on quark masses and limits on temporal variation of fundamental constants from atomic clock experiments}}, \href{http://dx.doi.org/10.1103/PhysRevC.73.055501}{\emph{Phys. Rev. C} {\bf 73} (2006) 055501}, [\href{https://arxiv.org/abs/nucl-th/0601050}{{\tt nucl-th/0601050}}].

\bibitem{Barontini:2021mvu}
G.~Barontini et~al., \emph{{Measuring the stability of fundamental constants with a network of clocks}}, \href{http://dx.doi.org/10.1140/epjqt/s40507-022-00130-5}{\emph{EPJ Quant. Technol.} {\bf 9} (2022) 12}, [\href{https://arxiv.org/abs/2112.10618}{{\tt 2112.10618}}].

\bibitem{Hees:2016gop}
A.~Hees, J.~Gu{\'e}na, M.~Abgrall, S.~Bize and P.~Wolf, \emph{{Searching for an oscillating massive scalar field as a dark matter candidate using atomic hyperfine frequency comparisons}}, \href{http://dx.doi.org/10.1103/PhysRevLett.117.061301}{\emph{Phys. Rev. Lett.} {\bf 117} (2016) 061301}, [\href{https://arxiv.org/abs/1604.08514}{{\tt 1604.08514}}].

\bibitem{Sherrill:2023zah}
N.~Sherrill et~al., \emph{{Analysis of atomic-clock data to constrain variations of fundamental constants}}, \href{http://dx.doi.org/10.1088/1367-2630/aceff6}{\emph{New J. Phys.} {\bf 25} (2023) 093012}, [\href{https://arxiv.org/abs/2302.04565}{{\tt 2302.04565}}].

\bibitem{Kobayashi_2022}
T.~Kobayashi, A.~Takamizawa, D.~Akamatsu, A.~Kawasaki, A.~Nishiyama, K.~Hosaka et~al., \emph{{Search for Ultralight Dark Matter from Long-Term Frequency Comparisons of Optical and Microwave Atomic Clocks}}, \href{http://dx.doi.org/10.1103/PhysRevLett.129.241301}{\emph{Phys. Rev. Lett.} {\bf 129} (2022) 241301}, [\href{https://arxiv.org/abs/2212.05721}{{\tt 2212.05721}}].

\bibitem{Filzinger:2023zrs}
M.~Filzinger, S.~D{\"o}rscher, R.~Lange, J.~Klose, M.~Steinel, E.~Benkler et~al., \emph{{Improved Limits on the Coupling of Ultralight Bosonic Dark Matter to Photons from Optical Atomic Clock Comparisons}}, \href{http://dx.doi.org/10.1103/PhysRevLett.130.253001}{\emph{Phys. Rev. Lett.} {\bf 130} (2023) 253001}, [\href{https://arxiv.org/abs/2301.03433}{{\tt 2301.03433}}].

\bibitem{BACON:2020ubh}
{\scshape BACON} collaboration, K.~Beloy et~al., \emph{{Frequency ratio measurements at 18-digit accuracy using an optical clock network}}, \href{http://dx.doi.org/10.1038/s41586-021-03253-4}{\emph{Nature} {\bf 591} (2021) 564--569}, [\href{https://arxiv.org/abs/2005.14694}{{\tt 2005.14694}}].

\bibitem{Banerjee:2023bjc}
A.~Banerjee, D.~Budker, M.~Filzinger, N.~Huntemann, G.~Paz, G.~Perez et~al., \emph{{Oscillating Nuclear Charge Radii as Sensors for Ultralight Dark Matter}}, \href{http://dx.doi.org/10.1103/37vw-gc1r}{\emph{Phys. Rev. Lett.} {\bf 135} (2025) 22}, [\href{https://arxiv.org/abs/2301.10784}{{\tt 2301.10784}}].

\bibitem{PhysRevLett.123.031304}
A.~A. Geraci, C.~Bradley, D.~Gao, J.~Weinstein and A.~Derevianko, \emph{{Searching for Ultralight Dark Matter with Optical Cavities}}, \href{http://dx.doi.org/10.1103/PhysRevLett.123.031304}{\emph{Phys. Rev. Lett.} {\bf 123} (2019) 031304}, [\href{https://arxiv.org/abs/1808.00540}{{\tt 1808.00540}}].

\bibitem{Antypas:2022asj}
D.~Antypas et~al., \emph{{New Horizons: Scalar and Vector Ultralight Dark Matter}},  \href{https://arxiv.org/abs/2203.14915}{{\tt 2203.14915}}.

\bibitem{Kennedy:2020bac}
C.~J. Kennedy, E.~Oelker, J.~M. Robinson, T.~Bothwell, D.~Kedar, W.~R. Milner et~al., \emph{{Precision Metrology Meets Cosmology: Improved Constraints on Ultralight Dark Matter from Atom-Cavity Frequency Comparisons}}, \href{http://dx.doi.org/10.1103/PhysRevLett.125.201302}{\emph{Phys. Rev. Lett.} {\bf 125} (2020) 201302}, [\href{https://arxiv.org/abs/2008.08773}{{\tt 2008.08773}}].

\bibitem{Tretiak:2022ndx}
O.~Tretiak, X.~Zhang, N.~L. Figueroa, D.~Antypas, A.~Brogna, A.~Banerjee et~al., \emph{{Improved Bounds on Ultralight Scalar Dark Matter in the Radio-Frequency Range}}, \href{http://dx.doi.org/10.1103/PhysRevLett.129.031301}{\emph{Phys. Rev. Lett.} {\bf 129} (2022) 031301}, [\href{https://arxiv.org/abs/2201.02042}{{\tt 2201.02042}}].

\bibitem{Savalle:2020vgz}
E.~Savalle, A.~Hees, F.~Frank, E.~Cantin, P.-E. Pottie, B.~M. Roberts et~al., \emph{{Searching for Dark Matter with an Optical Cavity and an Unequal-Delay Interferometer}}, \href{http://dx.doi.org/10.1103/PhysRevLett.126.051301}{\emph{Phys. Rev. Lett.} {\bf 126} (2021) 051301}, [\href{https://arxiv.org/abs/2006.07055}{{\tt 2006.07055}}].

\bibitem{Grote:2019uvn}
H.~Grote and Y.~V. Stadnik, \emph{{Novel signatures of dark matter in laser-interferometric gravitational-wave detectors}}, \href{http://dx.doi.org/10.1103/PhysRevResearch.1.033187}{\emph{Phys. Rev. Res.} {\bf 1} (2019) 033187}, [\href{https://arxiv.org/abs/1906.06193}{{\tt 1906.06193}}].

\bibitem{Stadnik:2014tta}
Y.~V. Stadnik and V.~V. Flambaum, \emph{{Searching for dark matter and variation of fundamental constants with laser and maser interferometry}}, \href{http://dx.doi.org/10.1103/PhysRevLett.114.161301}{\emph{Phys. Rev. Lett.} {\bf 114} (2015) 161301}, [\href{https://arxiv.org/abs/1412.7801}{{\tt 1412.7801}}].

\bibitem{DeRocco:2018jwe}
W.~DeRocco and A.~Hook, \emph{{Axion interferometry}}, \href{http://dx.doi.org/10.1103/PhysRevD.98.035021}{\emph{Phys. Rev. D} {\bf 98} (2018) 035021}, [\href{https://arxiv.org/abs/1802.07273}{{\tt 1802.07273}}].

\bibitem{Vermeulen:2021epa}
S.~M. Vermeulen et~al., \emph{{Direct limits for scalar field dark matter from a gravitational-wave detector}},  \href{https://arxiv.org/abs/2103.03783}{{\tt 2103.03783}}.

\bibitem{Morisaki:2018htj}
S.~Morisaki and T.~Suyama, \emph{{Detectability of ultralight scalar field dark matter with gravitational-wave detectors}}, \href{http://dx.doi.org/10.1103/PhysRevD.100.123512}{\emph{Phys. Rev. D} {\bf 100} (2019) 123512}, [\href{https://arxiv.org/abs/1811.05003}{{\tt 1811.05003}}].

\bibitem{Gottel:2024cfj}
A.~S. G{\"o}ttel, A.~Ejlli, K.~Karan, S.~M. Vermeulen, L.~Aiello, V.~Raymond et~al., \emph{{Searching for Scalar Field Dark Matter with LIGO}}, \href{http://dx.doi.org/10.1103/PhysRevLett.133.101001}{\emph{Phys. Rev. Lett.} {\bf 133} (2024) 101001}, [\href{https://arxiv.org/abs/2401.18076}{{\tt 2401.18076}}].

\bibitem{Fukusumi:2023kqd}
K.~Fukusumi, S.~Morisaki and T.~Suyama, \emph{{Upper limit on scalar field dark matter from LIGO-Virgo third observation run}}, \href{http://dx.doi.org/10.1103/PhysRevD.108.095054}{\emph{Phys. Rev. D} {\bf 108} (2023) 095054}, [\href{https://arxiv.org/abs/2303.13088}{{\tt 2303.13088}}].

\bibitem{Arvanitaki:2015iga}
A.~Arvanitaki, S.~Dimopoulos and K.~Van~Tilburg, \emph{{Sound of Dark Matter: Searching for Light Scalars with Resonant-Mass Detectors}}, \href{http://dx.doi.org/10.1103/PhysRevLett.116.031102}{\emph{Phys. Rev. Lett.} {\bf 116} (2016) 031102}, [\href{https://arxiv.org/abs/1508.01798}{{\tt 1508.01798}}].

\bibitem{Manley:2019vxy}
J.~Manley, R.~Stump, D.~Wilson, D.~Grin and S.~Singh, \emph{{Searching for scalar dark matter with compact mechanical resonators}}, \href{http://dx.doi.org/10.1103/PhysRevLett.124.151301}{\emph{Phys. Rev. Lett.} {\bf 124} (2020) 151301}, [\href{https://arxiv.org/abs/1910.07574}{{\tt 1910.07574}}].

\bibitem{Branca:2016rez}
A.~Branca et~al., \emph{{Search for an Ultralight Scalar Dark Matter Candidate with the AURIGA Detector}}, \href{http://dx.doi.org/10.1103/PhysRevLett.118.021302}{\emph{Phys. Rev. Lett.} {\bf 118} (2017) 021302}, [\href{https://arxiv.org/abs/1607.07327}{{\tt 1607.07327}}].

\bibitem{Seiferle:2019fbe}
B.~Seiferle et~al., \emph{{Energy of the$^{229}$Th nuclear clock transition}}, \href{http://dx.doi.org/10.1038/s41586-019-1533-4}{\emph{Nature} {\bf 573} (2019) 243--246}, [\href{https://arxiv.org/abs/1905.06308}{{\tt 1905.06308}}].

\bibitem{Caputo:2024doz}
A.~Caputo, D.~Gazit, H.-W. Hammer, J.~Kopp, G.~Paz, G.~Perez et~al., \emph{{Sensitivity of nuclear clocks to new physics}}, \href{http://dx.doi.org/10.1103/l29n-gt5j}{\emph{Phys. Rev. C} {\bf 112} (2025) L031302}, [\href{https://arxiv.org/abs/2407.17526}{{\tt 2407.17526}}].

\bibitem{Sikorsky:2020peq}
T.~Sikorsky et~al., \emph{{Measurement of the $^{229}$Th isomer energy with a magnetic micro-calorimeter}}, \href{http://dx.doi.org/10.1103/PhysRevLett.125.142503}{\emph{Phys. Rev. Lett.} {\bf 125} (2020) 142503}, [\href{https://arxiv.org/abs/2005.13340}{{\tt 2005.13340}}].

\bibitem{Zhao:2021tie}
W.~Zhao, D.~Gao, J.~Wang and M.~Zhan, \emph{{Investigating the environmental dependence of ultralight scalar dark matter with atom interferometers}}, \href{http://dx.doi.org/10.1007/s10714-022-02925-4}{\emph{Gen. Rel. Grav.} {\bf 54} (2022) 41}, [\href{https://arxiv.org/abs/2102.02391}{{\tt 2102.02391}}].

\bibitem{Buchmueller:2023nll}
O.~Buchmueller, J.~Ellis and U.~Schneider, \emph{{Large-scale atom interferometry for fundamental physics}}, \href{http://dx.doi.org/10.1080/00107514.2023.2239008}{\emph{Contemp. Phys.} {\bf 64} (2023) 93--110}, [\href{https://arxiv.org/abs/2306.17726}{{\tt 2306.17726}}].

\bibitem{Arvanitaki:2016fyj}
A.~Arvanitaki, P.~W. Graham, J.~M. Hogan, S.~Rajendran and K.~Van~Tilburg, \emph{{Search for light scalar dark matter with atomic gravitational wave detectors}}, \href{http://dx.doi.org/10.1103/PhysRevD.97.075020}{\emph{Phys. Rev. D} {\bf 97} (2018) 075020}, [\href{https://arxiv.org/abs/1606.04541}{{\tt 1606.04541}}].

\bibitem{Badurina:2021lwr}
L.~Badurina, D.~Blas and C.~McCabe, \emph{{Refined ultralight scalar dark matter searches with compact atom gradiometers}}, \href{http://dx.doi.org/10.1103/PhysRevD.105.023006}{\emph{Phys. Rev. D} {\bf 105} (2022) 023006}, [\href{https://arxiv.org/abs/2109.10965}{{\tt 2109.10965}}].

\bibitem{Badurina:2019hst}
L.~Badurina et~al., \emph{{AION: An Atom Interferometer Observatory and Network}}, \href{http://dx.doi.org/10.1088/1475-7516/2020/05/011}{\emph{JCAP} {\bf 05} (2020) 011}, [\href{https://arxiv.org/abs/1911.11755}{{\tt 1911.11755}}].

\bibitem{Abe_2021}
{\scshape MAGIS-100} collaboration, M.~Abe, P.~Adamson, M.~Borcean, D.~Bortoletto, K.~Bridges, S.~P. Carman et~al., \emph{{Matter-wave Atomic Gradiometer Interferometric Sensor (MAGIS-100)}}, \href{http://dx.doi.org/10.1088/2058-9565/abf719}{\emph{Quantum Sci. Technol.} {\bf 6} (2021) 044003}, [\href{https://arxiv.org/abs/2104.02835}{{\tt 2104.02835}}].

\bibitem{AEDGE:2019nxb}
{\scshape AEDGE} collaboration, Y.~A. El-Neaj et~al., \emph{{AEDGE: Atomic Experiment for Dark Matter and Gravity Exploration in Space}}, \href{http://dx.doi.org/10.1140/epjqt/s40507-020-0080-0}{\emph{EPJ Quant. Technol.} {\bf 7} (2020) 6}, [\href{https://arxiv.org/abs/1908.00802}{{\tt 1908.00802}}].

\end{thebibliography}\endgroup
\bibliographystyle{jhep}
\end{document}